\documentclass[a4paper,11pt]{article}
\usepackage{amsfonts}
\usepackage{mathrsfs}
\usepackage{amsmath,amssymb,amsthm}
\usepackage{latexsym}
\usepackage{indentfirst}

\usepackage[top=1in, bottom=1in, left=1in, right=1in]{geometry}
\usepackage{color}
\usepackage{footmisc}
\usepackage{endnotes}
\usepackage{algorithm}
\usepackage{float}
\usepackage{graphicx}
\usepackage{epstopdf}
\usepackage{longtable}
\usepackage{authblk}
 \usepackage[numbers]{natbib}
\usepackage{bm}
\usepackage{natbib}
\usepackage{graphicx}
\usepackage{float}
\usepackage{subfig}
\usepackage{url}
\usepackage[font={small}]{caption}

 \DeclareMathOperator*{\var}{var}
\DeclareMathOperator*{\bias}{bias}
\DeclareMathOperator*{\dimm}{dim}

\DeclareMathOperator*{\dett}{det}
\DeclareMathOperator*{\tr}{tr}

\newcommand{\ud}{\mathrm{d}}
\newcommand{\T}{\top}

\DeclareMathOperator*{\argmin}{arg\,min}
\DeclareMathOperator*{\argmax}{arg\,max}

\newtheorem{theorem}{Theorem}
\newtheorem{assumption}{Assumption}
\newtheorem{lemma}{Lemma}

\newtheorem{proposition}{Proposition}

\theoremstyle{definition}
\newtheorem{remark}{Remark}

\allowdisplaybreaks

\begin{document}

\title{\textbf{Simple, Scalable and Accurate Posterior Interval Estimation}}

\author[1]{Cheng Li \thanks{stalic@nus.edu.sg}}
\author[2]{Sanvesh Srivastava \thanks{sanvesh-srivastava@uiowa.edu}}
\author[3]{David B. Dunson \thanks{dunson@duke.edu}}

\affil[1]{Department of Statistics and Applied Probability, National University of Singapore}
\affil[2]{Department of Statistics and Actuarial Science, The University of Iowa}
\affil[3]{Department of Statistical Science, Duke University}

\date{}
\maketitle

\begin{abstract}
There is a lack of simple and scalable algorithms for uncertainty quantification.  Bayesian methods quantify uncertainty through posterior and predictive distributions, but it is difficult to rapidly estimate summaries of these distributions, such as quantiles and intervals.  Variational Bayes approximations are widely used, but may badly underestimate posterior covariance.  Typically, the focus of Bayesian inference is on point and interval estimates for one-dimensional functionals of interest.  In small scale problems, Markov chain Monte Carlo algorithms remain the gold standard, but such algorithms face major problems in scaling up to big data.  Various modifications have been proposed based on parallelization and approximations based on subsamples, but such approaches are either highly complex or lack theoretical support and/or good performance outside of narrow settings.  We propose a simple and general posterior interval estimation algorithm, which is based on running Markov chain Monte Carlo in parallel for subsets of the data and averaging quantiles estimated from each subset.  We provide strong theoretical guarantees and illustrate performance in several applications.
\end{abstract}

Key words: Bayesian; Big data; Credible interval; Embarrassingly parallel; Markov chain Monte Carlo; Quantile estimation; Wasserstein barycenter.

\section{Introduction}

We propose a posterior interval estimation algorithm for uncertainty quantification in massive data settings in which usual Bayesian sampling algorithms are too slow.  Bayesian models quantify uncertainty via the joint posterior distribution of the model parameters and predictive distributions of new observations.  As joint posteriors and predictives are difficult to visualize and use in practice, the focus is almost always on posterior summaries of one-dimensional functionals.  For example, it is typical to report 95\% posterior credible intervals for a variety of one-dimensional functionals of interest. In practice, by far the most common approach to estimate credible intervals relies on running a Markov chain Monte Carlo algorithm to obtain samples from the joint posterior, based on which estimating intervals for different one-dimensional functionals is trivial.  Traditional Markov chain Monte Carlo algorithms are too slow to be practically useful in massive data applications.  However, given their rich history and broad use, it would be appealing to be able to incorporate a simple fix-up, which would allow trivial modifications of existing code, solve the computational bottleneck, and enable provably accurate estimation of posterior quantiles for any one-dimensional functional of interest.

Current classes of analytic approximations, such as Gaussian/Laplace, variational Bayes \citep{Hofetal13,Broetal13,TanNot14}, and expectation propagation \citep{Xuetal14}, clearly do not provide a generally useful alternative to sampling methods in terms of accurate estimation of posterior credible intervals.  Hence, in comparing with the literature, we focus on scalable sampling algorithms.  There has been a recent interest in scaling up Bayesian sampling in general and Markov chain Monte Carlo algorithms in particular, with many different threads considered.  Three of the most successful include (i) approximating expensive Markov chain Monte Carlo transition kernels with easier to sample surrogates; (ii) running Markov chain Monte Carlo on a single machine but with different subsets of the data used as sampling proceeds \citep{WelTeh11,Macetal14}; and (iii) running Markov chain Monte Carlo in parallel for different data subsets and then combining \citep{Scoetal16, NeiWanXin13, Minetal14, Srietal15, Wangetal15}.  Motivated by our goal of defining a very simple and theoretically supported algorithm, we focus on embarassingly parallel Markov chain Monte Carlo following strategy (iii).

The key question in embarassingly parallel Markov chain Monte Carlo is how to combine samples from the different subset posteriors.  If each subset posterior were approximately Gaussian, then weighted averaging is well justified, motivating the consensus Monte Carlo algorithm \citep{Scoetal16}. Outside of this restrictive setting, one can instead rely on the product equation representation to combine using kernel smoothing \citep{NeiWanXin13} or multi-scale histograms \citep{Wangetal15}.  Such approaches have theory support in terms of accuracy as the number of samples increases, but rely heavily on the accuracy of density estimators for the subset posteriors, suffering badly when subset posteriors have even slightly non-overlapping supports.  Moreover, the product equation representation obtained by splitting the prior is not invariant to model reparameterization. An alternative approach is to use data subsamples to define noisy approximations to the full data posterior, and then take an appropriate notion of geometric center, such as geometric median \citep{Minetal14} or mean \citep{Srietal15} of these approximations.  These later approaches are invariant to model reparameterization, but they require a somewhat conceptually and computationally complex combining algorithm.

In this article, we propose a new scalable algorithm for posterior interval estimation. Our algorithm first runs Markov chain Monte Carlo or any alternative posterior sampling algorithm in parallel for each subset posterior, with the subset posteriors proportional to the prior multiplied by the subset likelihood raised to the full data sample size divided by the subset sample size. To obtain an accurate estimate of a posterior quantile for any one-dimensional functional of interest, we simply calculate the quantile estimates in parallel for each subset posterior and then average these estimates.  Hence, our combining step is completely trivial conceptually and computationally.
We also provide theory justifying the performance of the quantile estimates. We emphasize that we are not proposing a new Markov chain Monte Carlo algorithm, but we are instead developing a simple approach to scale up existing algorithms to datasets with large numbers of observations.

Our approach is related to the frequentist Bag of Little Bootstraps \citep{Kleetal14} and provides a Bayesian interpretation. Bag of Little Bootstraps divides massive data into small subsets and obtains bootstrap confidence intervals for a one-dimensional parameter on every subset from weighted bootstrap samples. Then the confidence interval of the one-dimensional parameter based on the whole data is constructed by averaging lower and upper bounds of the bootstrap confidence intervals across all subsets.  Similarly, our algorithm averages quantiles from all subset posteriors. Our theory leads to new insights into Bag of Little Bootstraps, showing that its confidence intervals correspond to the confidence intervals of the Wasserstein barycenter of bootstrap distributions across all subsets.

\section{Preliminaries}
\subsection{Wasserstein Distance and Barycenter}
Our algorithm is related to the concept of Wasserstein barycenter of subset posteriors \citep{Srietal15}, which depends on the notions of Wasserstein distance and Wasserstein barycenter. Suppose $\Theta\in \mathcal{R}^d$ and $\|\theta_1-\theta_2\|$ is the Euclidean distance between any $\theta_1,\theta_2\in \Theta$. For any two measures $\nu_1,\nu_2$ on $\Theta$, their Wasserstein-2 distance is defined as
\begin{align}
W_2(\nu_1,\nu_2) = \left\{\inf_{\gamma \in \Gamma(\nu_1,\nu_2)}\int_{\Theta\times \Theta} \|\theta_1-\theta_2\|^2 \ud \gamma( \nu_1,\nu_2 ) \right\}^{1/2},\nonumber
\end{align}
where $\Gamma(\nu_1,\nu_2)$ is the set of all probability measures on $\Theta\times \Theta$ with marginals $\nu_1$ and $\nu_2$, respectively. If we let $\mathcal{P}_2(\Theta)=\left\{\nu:\int_{\Theta} \|\theta\|^2\ud \nu(\theta) <\infty\right\}$, then the $W_2$ distance is well defined for every pair of measures in $\mathcal{P}_2(\Theta)$. The topological space $\{ \Theta,\mathcal{P}_2(\Theta)\}$ is a Polish space, and the $W_2$ distance metricizes the weak convergence of measures on $\mathcal{P}_2(\Theta)$. Convergence in $W_2$ distance on $\mathcal{P}_2(\Theta)$ is equivalent to weak convergence plus convergence of the second moment; see for example, Lemma 8.3 in \citep{BikFre81}. Given $N$ different measures $\nu_1,\ldots,\nu_N$ in $\mathcal{P}_2(\Theta)$, their Wasserstein barycenter is defined as the solution to the following optimization problem \citep{AguCar11}:
\begin{align}\label{wasp0}
\overline \nu = \argmin_{\mu \in \mathcal{P}_2(\Theta)} \sum_{j=1}^N W_2^2 \left(\mu, \nu_j \right),
\end{align}
which can be viewed as the geometric center of the $N$ measures $\nu_1,\ldots,\nu_N$.

\subsection{Wasserstein Posterior and Posterior Interval Estimation}
\label{wasp-pie}
Consider $n$ observations that are conditionally independent given model parameters and can be
partitioned into $K$ non-overlapping subsets. For ease of presentation, we assume that all subsets have the same sample size $m$, such that $n=Km$. The data in the $j$th subset are denoted $X_j=\{X_{1j},X_{2j},\ldots,X_{mj}\}$ for $j=1,\ldots,K$, and the whole dataset is denoted $X=\cup_{j=1}^K X_j$. The model $P(x\mid  \theta)$, or for short $P_{\theta}$, describes the distribution of $X$, with parameter $\theta\in \Theta \subseteq \mathcal{R}^d$, where $d$ is the dimension of $\theta$. Suppose $P(x\mid  \theta)$ is absolutely continuous with respect to dominating measure $\lambda$ such that $\ud P(x\mid \theta) = p(x\mid \theta) \ud \lambda(x)$. For theory development, we assume the existence of a true parameter $\theta_0\in \Theta$, such that the data $X$ are generated from $P_{\theta_0}$. Given a prior distribution $\Pi(\theta)$ over $\Theta$ with density $\pi(\theta)$, define the overall posterior density of $\theta$ given $X$ and the $j$th subset posterior density of $\theta$ given $X_j$, $j=1,\ldots,K$, as
\begin{align}\label{subsetpost}
\pi_n \left(\theta\mid  X\right) &= \frac{ \prod_{j=1}^K\prod_{i=1}^m p(X_{ij}\mid  \theta)\pi(\theta) \ud \theta} {\int_{\Theta} \prod_{j=1}^K\prod_{i=1}^m p(X_{ij}\mid  \theta) \pi(\theta) \ud \theta} \nonumber \\
\pi_m  \left(\theta|X_j\right) &= \frac{\left\{\prod_{i=1}^m p(X_{ij}\mid  \theta)\right\}^K \pi(\theta) \ud \theta} {\int_{\Theta} \left\{\prod_{i=1}^m p(X_{ij}\mid  \theta)\right\}^K \pi(\theta) \ud \theta},
\end{align}
and we denote their corresponding distribution functions as $\Pi_n(\theta\mid X)$ and $\Pi_m(\theta\mid X_j)$, respectively. In the definition of subset posterior density $\pi_m(\theta\mid  X_j)$, we have raised the subset likelihood function to the $K$th power. As a stochastic approximation to the overall posterior $\pi_n \left(\theta\mid  X\right)$, this modification rescales the variance of each subset posterior given $X_j$ to be roughly of the same order as the variance of the overall posterior $\Pi_n(\theta\mid  X)$, as in \citep{Minetal14} and \citep{Srietal15}.  Based on \eqref{subsetpost}, \citep{Srietal15} runs Markov chain Monte Carlo algorithms on the $K$ subsets in parallel, producing draws from each $\Pi_m(\theta\mid  X_j)$, $j=1,\ldots,K$. Empirical estimates of $\Pi_m(\theta\mid  X_j)$ for all $K$ subsets are obtained from the Markov chain Monte Carlo draws, their Wasserstein barycenter is estimated via a linear program, and used as an approximation of the overall posterior $\Pi_n(\theta\mid  X)$.

Suppose we are interested in a scalar parameter $\xi=h(\theta) \in \Xi$ with $h:\Theta \mapsto \Xi\subseteq \mathcal{R}$. We denote the overall posterior for $\xi$ by $\Pi_n(\xi\mid  X)$ and the $j$th subset posterior for $\xi$ by $\Pi_m(\xi\mid  X_j)$. For theory development, we mainly focus on the linear functional $\xi=h(\theta)=a^\T \theta +b$ for some fixed $a\in \mathcal{R}^d$ and $b\in \mathcal{R}$, which includes the individual components in $\theta$ as special cases. We can define the $W_2$ distance and the set of measures $\mathcal{P}_2(\Xi)$ on the univariate space $\Xi$. If $\Pi_m \left(\xi \mid  X_j \right) \in \mathcal{P}_2(\Xi)$ for all $j=1,\ldots,K$, then the one-dimensional Wasserstein posterior $\overline \Pi_n(\xi\mid  X)$ is defined as the Wasserstein barycenter of $\Pi_m \left(\xi \mid  X_j \right)$ as in \eqref{wasp0}:
\begin{align}\label{wasp}
\overline \Pi_n(\xi\mid  X) = \argmin_{\mu \in \mathcal{P}_2(\Xi)} \sum_{j=1}^K W_2^2 \left\{ \mu, \Pi_m \left(\xi \mid  X_j \right)  \right\}.
\end{align}
In the one-dimensional case, the Wasserstein posterior has an explicit relation with the $K$ subset posteriors. Let $F^{-1}(u)=\inf\{x:F(x)\geq u\}$ be the quantile function of a generic univariate distribution function $F(x)$. Let $F_1$ and $F_2$ be two univariate distributions in $\mathcal{P}_2(\Xi)$, with quantile functions $F_1^{-1}(u)$ and $F_2^{-1}(u)$, for any $u\in (0,1)$, respectively. Then the $W_2$ distance between $F_1$ and $F_2$ has an explicit expression by Lemma 8.2 of \citep{BikFre81}:
$$W_2(F_1,F_2) = \left[ \int_0^1 \left\{ F_1^{-1}(u)-F_2^{-1}(u)\right\}^2 \ud u\right]^{1/2}.$$
Therefore, $\overline \Pi_n(\xi\mid  X)$ in \eqref{wasp} is explicitly related to the subset posteriors $\Pi_m \left(\xi \mid  X_j \right) $ by
\begin{align*}
\overline \Pi_n^{-1} (u\mid  X) = \frac{1}{K} \sum_{j=1}^K \Pi_m^{-1} \left(u\mid  X_j \right),
\end{align*}
where $\Pi_m^{-1} \left(u\mid  X_j \right)$ and $\overline \Pi_n^{-1} (u\mid  X)$ are the quantile functions of $\Pi_m(\xi\mid  X_j)$ and $\overline \Pi_n(\xi\mid  X)$, respectively. This expression for the one-dimensional $W_2$ barycenter has been derived in \citep{AguCar11} from an optimal transport perspective. The relation indicates that for a scalar functional $\xi$, the average of subset posterior quantiles produces another quantile function that corresponds exactly to the one-dimensional Wasserstein posterior. Therefore, in our algorithm, to evaluate the Wasserstein posterior of $\xi$, we simply take the empirical quantiles based on posterior draws from each $\Pi_m(\xi\mid  X_j)$ and then average them over $j=1,\ldots,K$. Our algorithm is summarized in Algorithm \ref{pie-algo}.

\begin{algorithm}[!h]
\caption{Posterior Interval Estimation} \label{pie-algo}
\vspace*{-12pt}
\begin{tabbing}
   \enspace {\bf Input}: $K$ subsets of data $X_1,\ldots,X_K$, each with sample size $m$. \\
   \enspace {\bf Output}: Posterior credible intervals $[\overline q_{\alpha/2},\overline q_{1-\alpha/2}]$, for $\alpha\in (0,1)$.\\
   \enspace For $j=1$ to $K$  \qquad \# Parallel in $K$ subsets \\
   \qquad For $t=1$ to $T$ \\\
   \qquad\qquad  Draw $\theta_{tj}$ from $\Pi_m \left(\theta \mid  X_j \right)$, using an appropriate posterior sampler. \\
   \qquad\qquad  Calculate $\xi_{tj}=h(\theta_{tj})$. \\
   \qquad End for \\
   \qquad Sort $\left\{\xi_{1j},\ldots,\xi_{Tj}\right\}$ into $\left\{\xi_{(1)j}\leq \ldots \leq \xi_{(T)j}\right\}$; \\
   \qquad Obtain the empirical $\alpha/2$ and $1-\alpha/2$ quantiles
   $q_{\alpha/2,j}=\xi_{(\lfloor T\alpha/2\rfloor)j}$ \\
   \qquad and $q_{1-\alpha/2,j}=\xi_{(\lfloor T(1-\alpha/2)\rfloor)j}$, where $\lfloor x \rfloor$ denotes the integer part of $x$.\\
   \enspace End for \\
   \enspace  Set $\overline q_{\alpha/2} = \frac{1}{K}\sum_{j=1}^K q_{\alpha/2,j}$ and $\overline q_{1-\alpha/2} = \frac{1}{K}\sum_{j=1}^K q_{1-\alpha/2,j}$. \\
    \enspace {\bf Return}: $[\overline q_{\alpha/2},\overline q_{1-\alpha/2}]$.
\end{tabbing}
\end{algorithm}


\section{Main Results}
In this section, we develop theory supporting our approach.  Under mild regularity conditions, we show that the one-dimensional Wasserstein posterior  $\overline\Pi_n(\xi\mid  X)$ is an accurate approximation to the overall posterior $\Pi_n(\xi\mid  X)$.  Essentially, as the subset sample size $m$ increases, the $W_2$ distance between them diminishes at a faster than parametric rate $o_p(m^{-1/2})$. Their biases, variances and quantiles are only different in high orders of $m$. This rate can be improved to $o_p(n^{-1/2})$ when the maximum likelihood estimator of $\xi$ is unbiased. Our results are improved relative to previous papers relying on combining subset posteriors, such as \citep{Minetal14} and \citep{Srietal15}, with more detailed description of the limiting behavior of the estimated posterior and weaker restrictions on the growth rate of the number of subsets $K$.

Our theory relies on the parametric Bernstein-von Mises theorem. The consensus Monte Carlo algorithm in \citep{Scoetal16} also leverages approximate normality in their asymptotic justification and can be viewed as a different way of averaging subset posteriors. They used weighted averages of subset posterior samples as an approximate sample from the true posterior, where the weights were taken as the inverse covariance matrices based on each subset posterior samples. Their weighting strategy relies more heavily on the normality assumption than our strategy of averaging quantiles. In contrast to the heuristic arguments in \citep{Scoetal16}, we provide formal justification for using normal approximations on a large number of subsets, and quantify the asymptotic orders of the induced approximation errors.

We first define some useful notation. Let $\ell_j(\theta) = \sum_{i=1}^m \log p(X_{ij}\mid  \theta)$ be the log-likelihood in the $j$ subset, and $\ell(\theta)=\sum_{j=1}^K \ell_j(\theta)$ be the overall log-likelihood. Let $\ell_j'(\theta) = \partial \ell_j(\theta)/\partial \theta $ and $\ell_j''(\theta) = -\partial^2 \ell_j(\theta)/\partial\theta\partial\theta^\T$ be the first and second derivatives of $\ell_j(\theta)$ with respect to $\theta$. Let $\hat \theta_j =\argmax_{\theta\in \Theta} \ell_j(\theta)$ be the maximum likelihood estimator of $\theta$ based on the $j$th subset $X_j$, $j=1,\ldots,K$. Similarly let $\hat \theta =\argmax_{\theta\in \Theta} \ell(\theta)$ be the maximum likelihood estimator of $\theta$ based on the full dataset $X$. Let $\overline \theta = \sum_{j=1}^K \hat \theta_j /K$ denote the average of maximum likelihood estimators across subsets.

We make the following assumptions on the data generating process, the prior and the posterior.

\begin{assumption}\label{a1}
$\theta_0$ is an interior point of $\Theta \in \mathcal{R}^d$, where $d$ is a fixed positive integer and does not depend on $n$. $P_{\theta}=P_{\theta_0}$ almost everywhere if and only if $\theta=\theta_0$. $X$ contains independent and identically distributed observations generated from $P_{\theta_0}$.
\end{assumption}
\begin{assumption}\label{a2}
The support of $p(x\mid  \theta)$ is the same for all $\theta\in \Theta$.
\end{assumption}
\begin{assumption}\label{a3}
$\log p(x\mid  \theta)$ is three times differentiable with respect to $\theta$ in a neighborhood $B_{\delta_0}(\theta_0)\equiv \{\theta\in \Theta:\|\theta-\theta_0\|\leq \delta_0\}$ of $\theta_0$, for some constant $\delta_0>0$. $E_{P_{\theta_0}}\left\{p'(X\mid  \theta_0)/p(X\mid  \theta_0)\right\}=0$.
    Furthermore, there exists an envelope function $M(x)$ such that
    $\sup_{\theta\in B_{\delta_0}(\theta_0)}\left|\partial \log p(x\mid  \theta) / \partial \theta_{l_1}\right|\leq M(x)$,
      $\sup_{\theta\in B_{\delta_0}(\theta_0)}\left|\partial^2 \log p(x\mid  \theta) / \partial \theta_{l_1}\partial \theta_{l_2}\right|\leq M(x)$,
      $\sup_{\theta\in B_{\delta_0}(\theta_0)}\left|\partial^3 \log p(x\mid  \theta) / \partial \theta_{l_1}\partial\theta_{l_2} \partial\theta_{l_3}\right|\leq M(x)$ for all $l_1,l_2,l_3 = 1,\ldots,d$, for all values of $x$, and $E_{P_{\theta_0}} M(X)^4 <\infty$.
\end{assumption}
\begin{assumption}\label{a4}
$I(\theta)=E_{P_{\theta_0}}\{ -\partial^2 p(X\mid  \theta)/\partial\theta\partial\theta^\T \}
    = E_{P_{\theta_0}}\left[ \{ \partial p(X\mid  \theta)/\partial\theta\} \{ \partial p(X\mid  \theta)/\partial\theta\}^\T\right]$.
    $-\ell_1''(\theta)/m$ is positive definite with eigenvalues bounded from below and above by constants, for all $\theta\in \Theta$, all values of $X_1$, and all sufficiently large $m$.
\end{assumption}
\begin{assumption}\label{a5}
For any $\delta>0$, there exists an $\epsilon>0$ such that\\
 $\lim_{m\to\infty} P_{\theta_0}\left[ \sup_{|\theta-\theta_0|\geq \delta}\left\{ \ell_1(\theta)-\ell_1(\theta_0)\right\}/m\leq -\epsilon\right]=1$.
\end{assumption}
\begin{assumption}\label{a6}
The prior density $\pi(\theta)$ is continuous, bounded from above in $\Theta$ and bounded below at $\theta_0$. The prior has finite second moment $\int_{\Theta} \|\theta\|^2 \pi(\theta)\ud \theta < \infty$.
\end{assumption}
\begin{assumption}\label{a7}
Let $\psi(X_1) = E_{\Pi_m(\theta\mid  X_1)} Km\|\theta - \hat \theta_1 \|^2 $, where $E_{\Pi_m(\theta\mid  X_1)}$ is the expectation with respect to $\theta$ under the posterior $\Pi_m(\theta\mid  X_1)$. Then there exists an integer $m_0\geq 1$, such that $\left\{\psi(X_1):m\geq m_0, K\geq 1\right\}$ is uniformly integrable under $P_{\theta_0}$. In other words, \\ $\lim_{C\to+\infty}\sup_{m\geq m_0,K\geq 1}E_{P_{\theta_0}}\psi(X_1)I\{ \psi(X_1)\geq C\} = 0$, where $I(\cdot)$ is the indicator function.
\end{assumption}

Assumptions \ref{a1}-\ref{a5} are standard and mild regularity conditions on the model $P(x\mid  \theta)$, which are similar to the assumptions of Theorem 8.2 in Chapter 6 of \citep{LehCas98} and Theorem 4.2 in \citep{Ghosh06} for showing the asymptotic normality of posteriors. Assumption \ref{a6} requires the prior to have a finite second moment, such that with high probability all the posterior distributions are in the $\mathcal{P}_2(\Theta)$ space and the $W_2$ distance is well defined.
In models with heavy tailed priors, such as our example in Section D.1, one can replace Assumption \ref{a6} by assuming that the posterior distribution conditional on a fixed number of initial observations has finite second moment; see Example 8.5 in Chapter 6 of \citep{LehCas98} and our Proposition 3 in the Appendix. The uniform integrability of subset posteriors in Assumption \ref{a7} is an extra mild technical assumption that helps us to generalize the usual Bernstein-von Mises result on the subsets from the convergence in probability to the convergence in $L_1$ distance.  We verify Assumption \ref{a7} for normal linear models and some exponential family distributions in the Appendix. A stronger condition that can replace Assumption \ref{a7} is $\sup_{m\geq m_0,K\geq 1}E_{X_1}E_{\Pi_m(\theta\mid  X_1)} Km\|\theta - \hat \theta_1 \|^2<+\infty$.  The following theorems hold for the one-dimensional Wasserstein posterior defined in \eqref{wasp}.

\begin{theorem}\label{w2main}
Suppose Assumptions \ref{a1}--\ref{a7} hold and $\xi=a^\T \theta+b$ for some fixed $a\in \mathcal{R}^d$ and $b\in \mathcal{R}$. Let $I_{\xi}(\theta_0) = \left\{ a^\T I^{-1}(\theta_0)a\right\}^{-1}$. Let $\overline \xi = a^\T \overline \theta + b$, $\hat \xi= a^\T \hat \theta+b$. Let $\Phi(\cdot;\mu,\Sigma)$ be the normal distribution with mean $\mu$ and variance $\Sigma$. \\
\noindent (i) As $m\to \infty$,
\begin{align*}
& n^{1/2}~W_2\left(\overline \Pi_{n}\left(\xi \mid  X\right),\Phi\left[\xi;\overline\xi,\left\{ nI_{\xi}(\theta_0)\right\}^{-1}\right]\right) \to 0,\\
&n^{1/2}~W_2\left(\Pi_{n}\left(\xi \mid  X\right),\Phi\left[\xi;\hat \xi,\left\{ nI_{\xi}(\theta_0)\right\}^{-1}\right]\right) \to 0,\\
& m^{1/2}~W_2\left\{ \overline \Pi_{n}\left(\xi \mid  X\right),\Pi_{n}\left(\xi \mid  X\right)\right\} \to 0,
\end{align*}
where the convergence is in $P_{\theta_0}$-probability.\\
\noindent (ii) If $\hat\theta_1$ is an unbiased estimator for $\theta$, so $E_{P_{\theta_0}} \hat \theta_1 = \theta_0$, then as $m\to\infty$,
\begin{align*}
&n^{1/2}~ W_2\left\{ \overline \Pi_{n}\left(\xi \mid  X\right),\Pi_{n}\left(\xi \mid  X\right)\right\} \to 0 \qquad \text{in $P_{\theta_0}$-probability.}
\end{align*}
\end{theorem}

Theorem \ref{w2main} shows that both the one-dimensional Wasserstein posterior of $\xi$ from combining $K$ subset posteriors and the overall posterior of $\xi$ based on the full dataset are asymptotically close in the $W_2$ distance to their respective limiting normal distributions, with slightly different means and the same variance. Such convergence in the $W_2$ distance implies weak convergence and convergence of the second moment. Furthermore, the $W_2$ distance between the Wasserstein and full posteriors converges to zero in probability with rates $m^{1/2}$ and $n^{1/2}$, depending on the behavior of the maximum likelihood estimator $\hat\theta_1$.

Previous asymptotic justifications for embarrassingly parallel Markov chain Monte Carlo approaches focus on consistency \citep{Srietal15} or convergence rates \citep{Minetal14}, while the above theorem is stronger in providing a limiting distribution.  In addition, our conditions are much weaker in only requiring the subset sample size $m$ to increase, while imposing no restrictions on the growth rates of $m$ and $K$.  Hence, the number of subsets $K$ can grow polynomially in $n$, mimicking the case in which many computers are available but computational resources per computer are limited.  For example, the theorem allows $K=O(n^c)$, $m=O(n^{1-c})$ for any $c\in (0,1)$.  Under this setup, the one-dimensional Wasserstein posterior, the overall posterior and their normal limits will all converge to $\theta_0$ at the same rate of $O_p(n^{-1/2})$, and their mutual difference is of order $o_p(m^{-1/2})$.

When the maximum likelihood estimator $\hat\theta_1$ is unbiased, Part (ii) of the theorem provides a sharper convergence rate of $O_p(n^{-1/2})$ compared to the $O_p(m^{-1/2})$ rate in Part (i), still with no explicit restrictions on the growth rates of $m$ and $K$. When $K$ increases very fast, for example $K\approx n^{1/2}$ and $m\approx n^{1/2}$, the $O_p(n^{-1/2})$ rate in Part (ii) is much faster than the $O_p(n^{-1/4})$ rate from Part (i). Moreover, $O_p(m^{-1/2})$ is suboptimal since it is the parametric rate based on only the subset data with size $m$, while $O_p(n^{-1/2})$ is the optimal parametric rate based on the full data with size $n$. The reason for the improvement in Part (ii) lies in the high order difference between the two means $\overline \xi$ and $\hat\xi$ of the limiting normal distributions of the one-dimensional Wasserstein posterior and the overall posterior. When the unbiasedness assumption does not hold and $K$ increases with $n$, the difference between the averaged maximum likelihood estimator $\overline \xi$ and the overall maximum likelihood estimator $\hat\xi$ is typically of order $o_p(m^{-1/2})$, which does not scale in the number of subsets $K$. However, when all subset maximum likelihood estimators are unbiased, this difference is reduced by a factor of $K^{1/2}$ due to the averaging effect over $K$ subset posteriors and decreases faster as $o_p(n^{-1/2})$. Hence, in models having unbiased maximum likelihood estimators, the one-dimensional Wasserstein posterior achieves high order accuracy in approximating the overall posterior with a difference $o_p(n^{-1/2})$.


Independently, \citep{ShaChe15} has considered a nonparametric generalized linear model and shown a related Bernstein-von Mises theorem. Besides the difference between the form of models, we emphasize that our result in Theorem \ref{w2main} does not rely on the strong requirement of a uniform normal approximation for all subset posteriors, as used in Shang and Cheng's paper. Instead, to show Theorem \ref{w2main}, it is only necessary for the normal approximation to work well on average among all subset posteriors. As a result, we have no explicit constraint on the growth rate on the number of subsets $K$, while their paper needs to control $K$ explicitly depending on the posterior convergence rate.

\begin{theorem}\label{biasvar}
Suppose Assumptions \ref{a1}--\ref{a7} hold. Let $\xi_0=a^\T\theta_0+b$ and $\hat\xi$ be the same as defined in Theorem \ref{w2main}. For a generic distribution $F$ on $\Xi$, let $\bias(F) = E_F(\xi)-\xi_0$ and $\var(F)$ be the variance of $F$. Let $u_1$ and $u_2$ be two arbitrary fixed numbers such that $0<u_1<u_2<1$. Then the following relations hold:
\begin{flalign*}
(i) &~ \bias\left\{\overline \Pi_{n}(\xi\mid  X)\right\}= \overline \xi-\xi_0 + o_p\left(n^{-1/2}\right), ~~ \bias\left\{\Pi_{n}(\xi\mid  X)\right\}= \hat \xi-\xi_0 + o_p\left(n^{-1/2}\right);&&\\
(ii) &~ \var\left\{\overline \Pi_{n}(\xi\mid  X)\right\} = \frac{1}{n}I_{\xi}^{-1}(\theta_0) + o_p\left(n^{-1}\right),~~
\var\left\{\Pi_{n}(\xi\mid  X)\right\}= \frac{1}{n}I_{\xi}^{-1}(\theta_0) + o_p\left(n^{-1}\right); &&\\
(iii) &~~ \sup_{u\in [u_1,u_2]}\left|\overline \Pi_n^{-1}(u\mid  X) - \Pi_n^{-1}(u\mid  X)\right| = o_p\left(m^{-1/2}\right),&&
%
\end{flalign*}
where $O_p$ and $o_p$ are in $P_{\theta_0}$-probability.  Furthermore, if $\hat\theta_1$ is an unbiased estimator of $\theta_0$, then
\begin{align*}
& \bias\left\{\overline \Pi_{n}(\xi\mid  X)\right\} - \bias\left\{\Pi_{n}(\xi\mid  X)\right\} =  o_p\left(n^{-1/2}\right);\\
& \sup_{u\in [u_1,u_2]}\left|\overline \Pi_n^{-1}(u\mid  X) - \Pi_n^{-1}(u\mid  X)\right| = o_p\left(n^{-1/2}\right).
\end{align*}
\end{theorem}

Theorem \ref{biasvar} provides the order for the differences for the bias, the variance and the quantiles between the one-dimensional Wasserstein posterior and the overall posterior. Essentially the one-dimensional Wasserstein posterior has an asymptotic bias $\overline \xi-\hat \xi$ from the overall posterior, which is generally of order $o_p(m^{-1/2})$ and has higher order $o_p(n^{-1/2})$ when the subset maximum likelihood estimators are unbiased. The variances of the one-dimensional Wasserstein posterior and the overall posterior agree in the leading order. Similar to the biases, the difference between their quantiles has order $o_p(m^{-1/2})$ in the general case, and improves to a higher order $o_p(n^{-1/2})$ when the subset maximum likelihood estimators are unbiased. In our algorithm, when we take $K$ different subset posterior credible intervals and average them, the averages of the lower and upper quantiles are asymptotically close to the quantiles from the overall posterior in the leading order. Therefore, Theorem \ref{biasvar} also validates our algorithm in the sense of posterior uncertainty quantification. We can also account for Monte Carlo errors in approximating subset posteriors using samples under mild mixing conditions on the subset Markov chains; see Theorem 3 in the Appendix.

\section{Experiments}

We applied the proposed algorithm in a variety of cases, using consensus Monte Carlo \citep{Scoetal16}, Wasserstein posterior \citep{Srietal15}, semiparametric density product \citep{NeiWanXin13}, and variational Bayes as our competitors. Posterior summaries from Markov chain Monte Carlo applied to the full data served as the benchmark for all the comparisons. As our theory guarantees good performance for very large samples, we focused on simulations with moderate sample sizes. All Markov chain Monte Carlo algorithms were run for 10,000 iterations. After discarding the first 5000 samples as burn-in, we retained every fifth sample in all the chains; convergence diagnostics suggested that every chain had converged to its stationary distribution. We used the combination step implemented in R package parallelMCMCcombine \citep{MirCon14} for consensus Monte Carlo and semiparametric density product methods. We implemented the combination step of our algorithm in R and of Srivastava et al.'s algorithm in Matlab. All experiments were run on an Oracle Grid Engine cluster with 2.6GHz 16 core compute nodes. Memory resources were capped at 8GB for all the methods, except for Markov chain Monte Carlo based on the full data, which had a maximum memory limit of 16GB.

The accuracy of a density $q (\theta \mid  X)$ approximating $\pi_n(\theta \mid  X)$ was evaluated using the metric
\begin{align}
  \text{accuracy} \left\{ q (\theta\mid  X)\right\} = 1 - \tfrac{1}{2} \int_{\Theta} \left|q(\theta\mid  X) - \pi_n(\theta \mid  X)\right| \ud \theta. \label{acc}
\end{align}
This accuracy metric lies in $[0,1]$, with larger values indicating better performance of $q$ in approximating $\pi_n$ \citep{Faeetal12}. In our experiments, we first estimated $q (\theta\mid  X)$ and $\pi_n(\theta \mid  X)$ based on the posterior samples using the bkde or bkde2D functions in R package {KernSmooth}, with automatic bandwidth selection via dpik  \citep{Wan15}. The density estimates were used to compute a numerical approximation of the integral in \eqref{acc}.

\subsection{Linear model with varying dimension}

We first evaluated the performance of our proposed algorithm under varying sample size, dimension, and number of subsets in Bayesian linear models. Let the response, design matrix, regression coefficients, and random error be denoted as $y$, $X$, $\beta$, and $\epsilon$, where $y, \epsilon \in \mathcal{R}^n$, $\beta \in \mathcal{R}^{p \times 1}$, and $X \in \mathcal{R}^{n \times p}$. The model assumes that
\begin{align}
  \label{eq:vs1}
  y = X \beta + \epsilon, \; \epsilon \sim \mathcal{N}(0_{n \times 1}, \sigma^2 I_n),\; \beta \sim \text{gdP},  \;
  \sigma \sim \text{Half-}t,
\end{align}
where $\text{gdP}$ denotes the generalized double Pareto shrinkage prior of \citep{ArmDunLee13} and Half-$t$ is chosen to be weakly-informative \citep{Gel06}. See Section D.1 in the Appendix for detailed specifications. The priors on $\beta$ and $\sigma$ in \eqref{eq:vs1} are both heavy-tailed with infinite second moments, and therefore do not satisfy Assumption 6. However, one can verify that conditional on the initial $m_0$ observations with $m_0\geq p+4$, every subset posterior has finite second moments for both $\beta$ and $\sigma$. The result is summarized in Proposition 3 in the Appendix.

We applied our approach for inference on $\beta$  in \eqref{eq:vs1} compared with an asymptotic normal approximation. We calculated the accuracy of approximations using a full data Gibbs sampler as the benchmark (Table \ref{tab:varsel}). The first $10\%$ of entries of $\beta$ were set to $\pm 1$ with the remaining 0.
The entries of $X$ were randomly set to $\pm 1$ and $\sigma^2$ was fixed at 1. We ran 10 replications for $n \in \{10^4, 10^5\}$ and $p \in \{10, 100, 200, 300, 400, 500\}$. We varied $K\in \{10, 20\}$ and applied Algorithm \ref{pie-algo} after running a modification of the Gibbs sampler in \eqref{subsetpost} for each subset.  We considered two versions of normal approximations for the full posterior. The first version used $\mathcal N(\widehat m, \widehat V)$ to approximate the posterior of $\beta$, where $\widehat m$ and $ \widehat V$ are the maximum likelihood estimates of $\beta$ and its estimated asymptotic covariance matrix in \eqref{eq:vs1}. For the second version, we first obtained the asymptotic normal approximation of the $j$th subset posterior as $\mathcal N$($\widehat m_j$, $\widehat V_j$), where $\widehat m_j$ and $\widehat V_j$ ($j=1,\ldots,K$) are the maximum likelihood estimates of $\beta$ and its estimated asymptotic covariance matrix for the $j$th subset. Then we found the $W_2$ barycenter of the $K$ subset normal approximations, which is again a normal distribution $\mathcal N(m^*, V^*)$  \citep{AguCar11}. This provides an empirical verification of Theorem \ref{w2main}. See Section D.1 in the Appendix for details of the Gibbs sampler and the form of $m^*$ and $V^*$.

The performance of all the approaches was fairly similar across all simulations and agreed with our asymptotic theory (Table \ref{tab:varsel}). The results in Table \ref{tab:varsel} show that the proposed algorithm closely matched the Gibbs sampling results for the full data in terms of uncertainty quantification. It also performed better than the asymptotic normal approximations in some cases. When the subset sample size was too small compared to the dimension, such as when $n=10^4,p=400,K=20$ which has a subset size of only $m=500$, we observe poor performance for both the asymptotic approximations and the proposed approach.

\begin{table}[ht]
\def~{\hphantom{0}}
\caption{\it Accuracy of approximate posteriors for the non-zero and zero elements of $\beta$ in \eqref{eq:vs1}. The accuracies are averaged over 10 simulation replications. Normal, the asymptotic normal approximation based on the full data; PIE, our posterior interval estimation algorithm; NB, the $W_2$ barycenter of $K$ asymptotic normal approximations of subset posteriors.}
{
\begin{tabular}{r|cc|cc|cc|cc|cc|cc}
  \hline
  & \multicolumn{4}{|c|}{$p=10$}   & \multicolumn{4}{|c|}{$p=100$} &   \multicolumn{4}{|c}{$p=200$} \\
  \hline
  & \multicolumn{2}{|c|}{$n=10^4$} & \multicolumn{2}{|c|}{$n=10^5$} & \multicolumn{2}{|c|}{$n=10^4$} & \multicolumn{2}{|c|}{$n=10^5$} & \multicolumn{2}{|c|}{$n=10^4$} & \multicolumn{2}{|c}{$n=10^5$}\\
  \hline
  & 0s & non-0s & 0s & non-0s & 0s & non-0s & 0s & non-0s & 0s & non-0s & 0s & non-0s\\
  \hline
  Normal & 0.95 & 0.89 & 0.96 & 0.96 & 0.95 & 0.90 & 0.96 & 0.95 & 0.95 & 0.89 & 0.96 & 0.95 \\
  NB (K=10) & 0.94 & 0.91 & 0.96 & 0.95 & 0.89 & 0.87 & 0.95 & 0.94 & 0.84 & 0.83 & 0.94 & 0.94 \\
  PIE (K=10) & 0.95 & 0.97 & 0.97 & 0.97 & 0.90 & 0.92 & 0.96 & 0.96 & 0.85 & 0.85 & 0.95 & 0.95 \\
  NB (K=20) & 0.93 & 0.92 & 0.96 & 0.95 & 0.84 & 0.84 & 0.94 & 0.93 & 0.75 & 0.76 & 0.92 & 0.92 \\
  PIE (K=20) & 0.94 & 0.97 & 0.97 & 0.97 & 0.85 & 0.87 & 0.95 & 0.95 & 0.77 & 0.78 & 0.93 & 0.93 \\
   \hline
  & \multicolumn{4}{|c|}{$p=300$}   & \multicolumn{4}{|c|}{$p=400$} &   \multicolumn{4}{|c}{$p=500$} \\
  \hline
  & \multicolumn{2}{|c|}{$n=10^4$} & \multicolumn{2}{|c|}{$n=10^5$} & \multicolumn{2}{|c|}{$n=10^4$} & \multicolumn{2}{|c|}{$n=10^5$} & \multicolumn{2}{|c|}{$n=10^4$} & \multicolumn{2}{|c}{$n=10^5$}\\
  \hline
  & 0s & non-0s & 0s & non-0s & 0s & non-0s & 0s & non-0s & 0s & non-0s & 0s & non-0s\\
  \hline
  Normal & 0.95 & 0.89 & 0.96 & 0.95 & 0.94 & 0.89 & 0.96 & 0.95 & 0.94 & 0.89 & 0.96 & 0.95 \\
  NB (K=10) & 0.80 & 0.79 & 0.93 & 0.93 & 0.75 & 0.75 & 0.93 & 0.92 & 0.71 & 0.71 & 0.92 & 0.91 \\
  PIE (K=10) & 0.82 & 0.81 & 0.94 & 0.94 & 0.77 & 0.78 & 0.93 & 0.93 & 0.73 & 0.74 & 0.93 & 0.93 \\
  NB (K=20) & 0.65 & 0.67 & 0.91 & 0.91 & 0.51 & 0.52 & 0.90 & 0.90 & - & - & 0.89 & 0.88 \\
  PIE (K=20) & 0.67 & 0.68 & 0.92 & 0.91 & 0.52 & 0.53 & 0.91 & 0.91 & 0.31 & 0.31 & 0.90 & 0.89 \\
  \hline
\end{tabular}
}
\label{tab:varsel}
\end{table}
\vspace{.5cm}

\subsection{Linear mixed effects model}
\label{sim-lme}

Linear mixed effects models are widely used to characterize dependence in longitudinal and nested data structures. Let $n_i$ be the number of observations associated with the $i$th individual, for $i=1,\ldots,s$. Let $y_i \in \mathcal{R}^{n_i}$ be the responses of the $i$th individual, $X_i \in \mathcal{R}^{n_i \times p}$ and $Z_i \in \mathcal{R}^{n_i \times q}$ be matrices including predictors having coefficients that are fixed across individuals and varying across individuals, respectively.  Let $\beta \in \mathcal{R}^p$ and $u_i \in \mathcal{R}^q$, respectively, represent the fixed effects and $i$th random effect. The linear mixed effects model lets
\begin{align}
  y_i \sim \mathcal{N}(X_i \beta + Z_i u_i, \sigma^2 I_{n_i}),\quad u_i \sim \mathcal{N}(0_q, \Sigma) ,\quad i = 1, \ldots, s. \label{lme}
\end{align}
Many software packages are available for Markov chain Monte Carlo-based Bayesian inference in \eqref{lme}, but current implementations become intractable for data with large $s$ and $n = \sum_{i=1}^s n_i$.


We applied our algorithm for inference on $\beta$ and $\Sigma$ in \eqref{lme} and compared its performance with maximum likelihood, consensus Monte Carlo, semiparametric density product, Wasserstein posterior, and variational Bayes. We set $s=5000$, $n_i=20$ for $i=1,\ldots,s$, $n=10^5$, $p=4$, $q=3$, $\beta = (-1, 1, -1, 1)^\T$, and $\sigma=1$. The random effects covariance $\Sigma$ had $\Sigma_{ii}= i, i=1, 2, 3$, $\Sigma_{12}=-$ 0.56, $\Sigma_{31} = $ 0.52, and $\Sigma_{23}=$ 0.0025. This matrix included negative, positive, and small to moderate strength correlations \citep{Kimetal13}. The simulation was replicated 10 times. The approximate posterior distributions were obtained using consensus Monte Carlo, semiparametric density product, Wasserstein posterior, and our algorithm in three steps. First, full data were randomly partitioned into 20 subsets such that data for each individual were in the same subset. Second, the Markov chain Monte Carlo sampler for $\beta$ and $\Sigma$ in \eqref{lme} was modified following \eqref{subsetpost} and Equation (2) in \citep{Scoetal16} and implemented in Stan (Version 2.5.0). Finally, the posterior samples from all the subsets were combined. We used the streamlined algorithm for variational Bayes \citep{LeeWan16}. Maximum likelihood produced a point estimate and asymptotic covariance for $\beta$, and only a point estimate for $\Sigma$.



\begin{figure}[ht]
  \centering
  \includegraphics[width=0.9\textwidth]{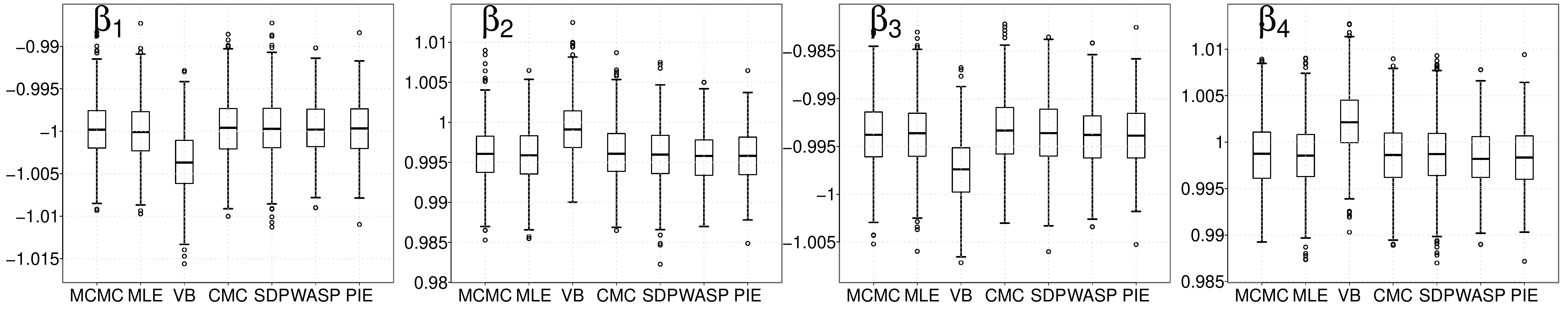}
  \captionsetup{font=small}
\caption{Boxplots of posterior samples for fixed effects in simulation for linear mixed effects model. MCMC, Markov chain Monte Carlo based on the full data;  ML, maximum likelihood estimator; VB, variational Bayes; CMC, consensus Monte Carlo; SDP, semiparametric density product; WASP, the algorithm in \citet{Srietal15}; PIE, our posterior interval estimation algorithm.}
\label{fig:simfixef}
\end{figure}

\begin{figure}[ht]
  \includegraphics[width=\textwidth]{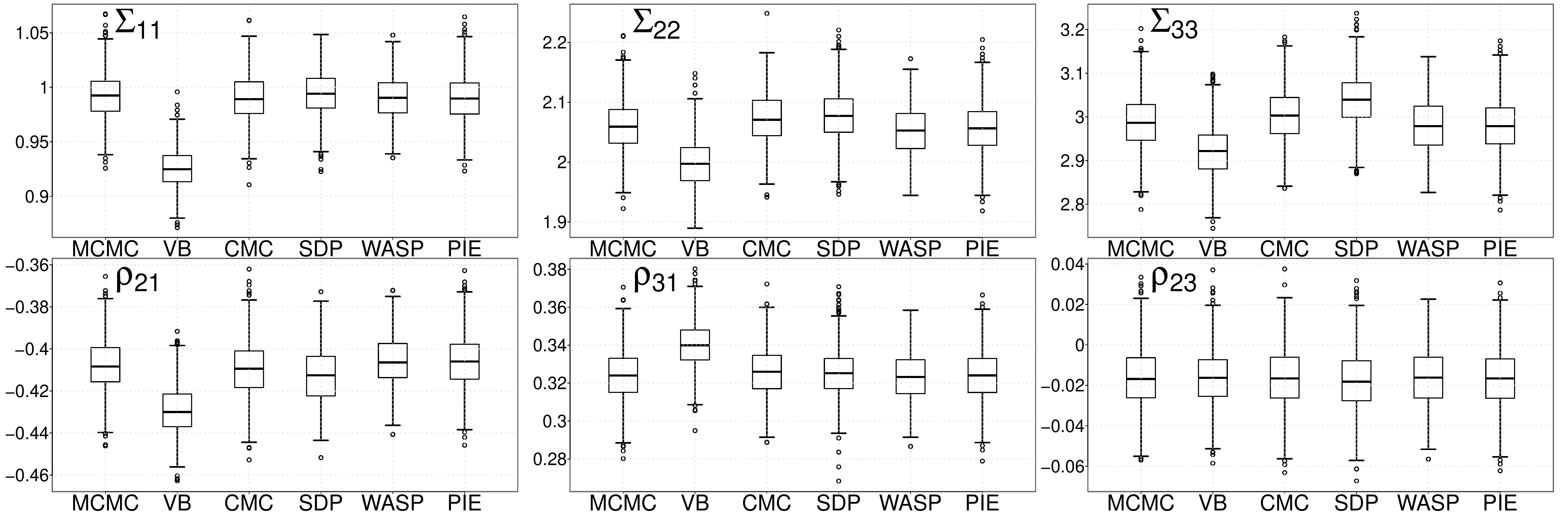}
  \captionsetup{font=small}
  \caption{Boxplots of posterior samples for variances and correlations of random effects in simulation for the linear mixed effects model. MCMC, Markov chain Monte Carlo based on the full data;  ML, maximum likelihood estimator; VB, variational Bayes; CMC, consensus Monte Carlo; SDP, semiparametric density product; WASP, the algorithm in \citet{Srietal15}; PIE, our posterior interval estimation algorithm.}
\label{fig:simranef}
\end{figure}

We compared the performance of the seven methods for inference on the fixed effects $\beta$, the variances of random effects $\Sigma_{ii}$ ($i=1,2,3$), and the correlations of random effects $\rho_{ij}=\Sigma_{ij}/(\Sigma_{ii}\Sigma_{jj})^{1/2}$ ($1\leq i<j\leq 3$). The correlations are nonlinear functionals of the model parameters $\Sigma$. Maximum likelihood estimator, consensus Monte Carlo, semiparametric density product, Wasserstein posterior, and our algorithm had excellent performance in estimation of $\beta$ (Figure \ref{fig:simfixef}), as well as the variances and correlations (Tables \ref{tab:sim-ran-ci} and \ref{tab:sim-ran-acc} and Figure \ref{fig:simranef}). Uncertainty quantification using consensus Monte Carlo, semiparametric density product, Wasserstein posterior, and our algorithm closely agreed with Markov chain Monte Carlo based on the full data.
As shown in Figure 1 of the Appendix, variational Bayes was computationally most efficient, but it showed poor accuracy in approximating the posterior of $\beta$ and the variances, with underestimation of posterior uncertainty.

\begin{table}[ht]
\def~{\hphantom{0}}
\caption{\it 90\% credible intervals for variances and correlations of random effects in simulation for linear mixed effects model. The upper and lower bounds are averaged over 10 replications.MLE, maximum likelihood estimator; MCMC, Markov chain Monte Carlo based on the full data; VB, variational Bayes; CMC, consensus Monte Carlo; SDP, semiparametric density product; WASP, the algorithm in \citet{Srietal15}; PIE, our posterior interval estimation algorithm.}{
\begin{tabular}{r|cccccc}
  \hline
  & $\Sigma_{11}$ & $\Sigma_{22}$ &  $\Sigma_{33}$ & $\rho_{12}$ & $\rho_{13}$ &  $\rho_{23}$   \\
  \hline
  \large{\small{MLE}} & 0.99 & 2.00 & 3.00 & -0.40 & 0.30 & 0.00 \\
  \large{\small{MCMC}} & (0.96, 1.03) & (1.94, 2.07) & (2.90, 3.10) & (-0.42, -0.38) & (0.28, 0.32) & (-0.03, 0.02) \\
  \large{\small{VB}}  &  (0.90, 0.96) & (1.88, 2.01) & (2.84, 3.04) & (-0.44, -0.41) & (0.29, 0.34) & (-0.03, 0.02) \\
  \large{\small{CMC}} & (0.96, 1.03) & (1.94, 2.08) & (2.91, 3.13) & (-0.42, -0.38) & (0.28, 0.32) & (-0.02, 0.02) \\
  \large{\small{SDP}} & (0.97, 1.03) & (1.95, 2.09) & (2.95, 3.14) & (-0.42, -0.38) & (0.28, 0.32) & (-0.03, 0.02) \\
  \large{\small{WASP}} & (0.96, 1.03) & (1.94, 2.07) & (2.90, 3.10) & (-0.42, -0.38) & (0.28, 0.32) & (-0.03, 0.02) \\
  \large{\small{PIE}} & (0.96, 1.03) & (1.94, 2.07) & (2.90, 3.10) & (-0.42, -0.38) & (0.28, 0.32) & (-0.03, 0.02) \\
   \hline
\end{tabular}}
\label{tab:sim-ran-ci}
\end{table}

\begin{table}[ht]
\def~{\hphantom{0}}
\caption{\it Accuracy of approximate posteriors for variances and correlations of random effects in simulation for linear mixed effects model. The standard deviation of accuracy across 10 replications is in parentheses. VB, variational Bayes; CMC, consensus Monte Carlo; SC, semiparametric density product; WASP, the algorithm in \citet{Srietal15}; PIE, our posterior interval estimation algorithm.}
{
  \begin{tabular}{r|cccccc}
    \hline
    & $\Sigma_{11}$ & $\Sigma_{22}$ &  $\Sigma_{33}$ & $\rho_{12}$ & $\rho_{13}$ &  $\rho_{23}$   \\
    \hline
    \large{\small{VB}} & 0.11 (0.02) & 0.38 (0.02) & 0.59 (0.03) & 0.45 (0.02) & 0.96 (0.01) & 0.61 (0.02) \\
    \large{\small{CMC}} & 0.92 (0.03) & 0.95 (0.03) & 0.93 (0.03) & 0.92 (0.04) & 0.96 (0.01) & 0.90 (0.03) \\
    \large{\small{SDP}} & 0.90 (0.04) & 0.92 (0.04) & 0.89 (0.07) & 0.85 (0.09) & 0.95 (0.03) & 0.74 (0.07) \\
    \large{\small{WASP}} & 0.95 (0.02) & 0.94 (0.02) & 0.95 (0.02) & 0.94 (0.02) & 0.96 (0.01) & 0.95 (0.01) \\
    \large{\small{PIE}} & 0.95 (0.02) & 0.95 (0.02) & 0.96 (0.02) & 0.96 (0.02) & 0.97 (0.01) & 0.96 (0.01) \\
    \hline
  \end{tabular}
}
\label{tab:sim-ran-acc}
\end{table}

\subsection{United States natality data}

We applied our algorithm to United States natality data on birth weight of infants and variables related to their mothers' health \citep{Abe2006}. Linear mixed effects models were used for the covariance in birth weights among siblings. Following the example in \citep{LeeWan16}, we selected the data for mothers who smoked, had two infants, and had some college education but not a college degree. Detailed information about the variables are in the Appendix.  The data set contained $s=3809$ mothers and $n=7618$ births.  There were 13 variables related to mother's health. All these covariates and an intercept were used as fixed effects in \eqref{lme}, so $p = 14$. The random effects included mother's age, gestation period, and number of living infants, so $q=3$. We performed 10 fold cross-validation and randomly split the data into 10 data sets such that data for siblings belonged to the same training data.  We estimated fixed effects and covariance matrix for random effects as in Section \ref{sim-lme} using $K=20$.

\begin{figure}[ht]
  \includegraphics[width=\textwidth]{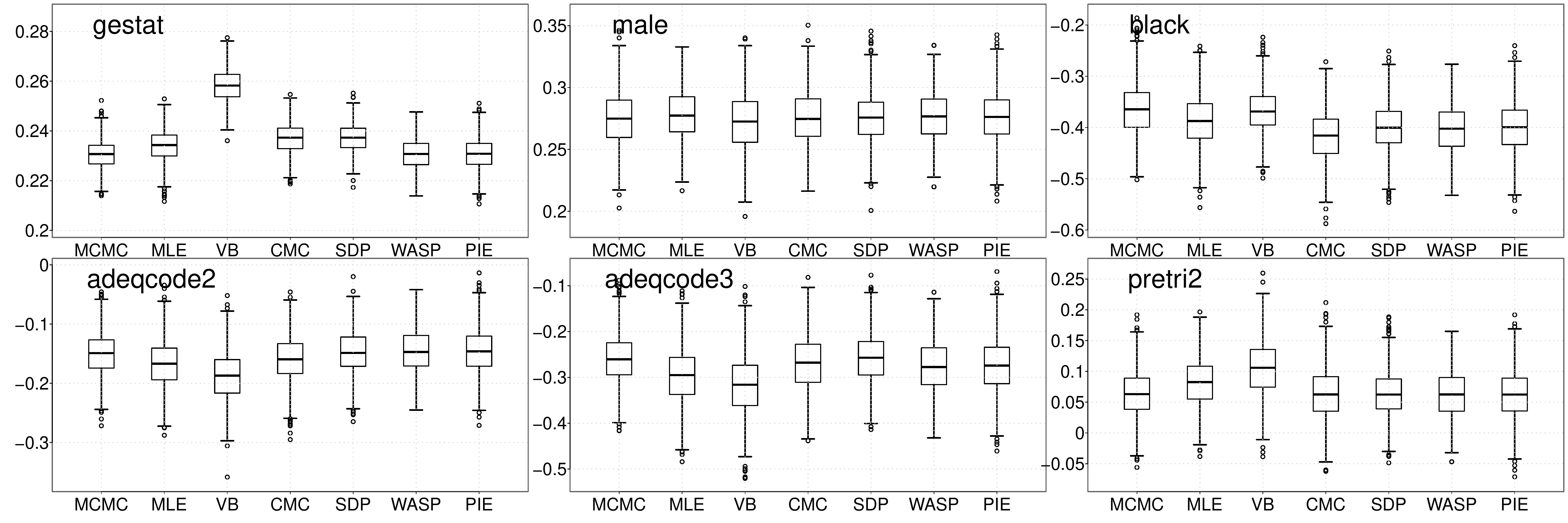}
  \captionsetup{font=small}
  \caption{Boxplots of posterior samples for six fixed effects in the United States natality data analysis. MCMC, Markov chain Monte Carlo based on the full data;  ML, maximum likelihood estimator; VB, variational Bayes; CMC, consensus Monte Carlo; SDP, semiparametric density product; WASP, the algorithm in \citet{Srietal15}; PIE, our posterior interval estimation algorithm.}
  \label{fig:abesimef}
\end{figure}

\begin{figure}[ht]
  \includegraphics[width=\textwidth]{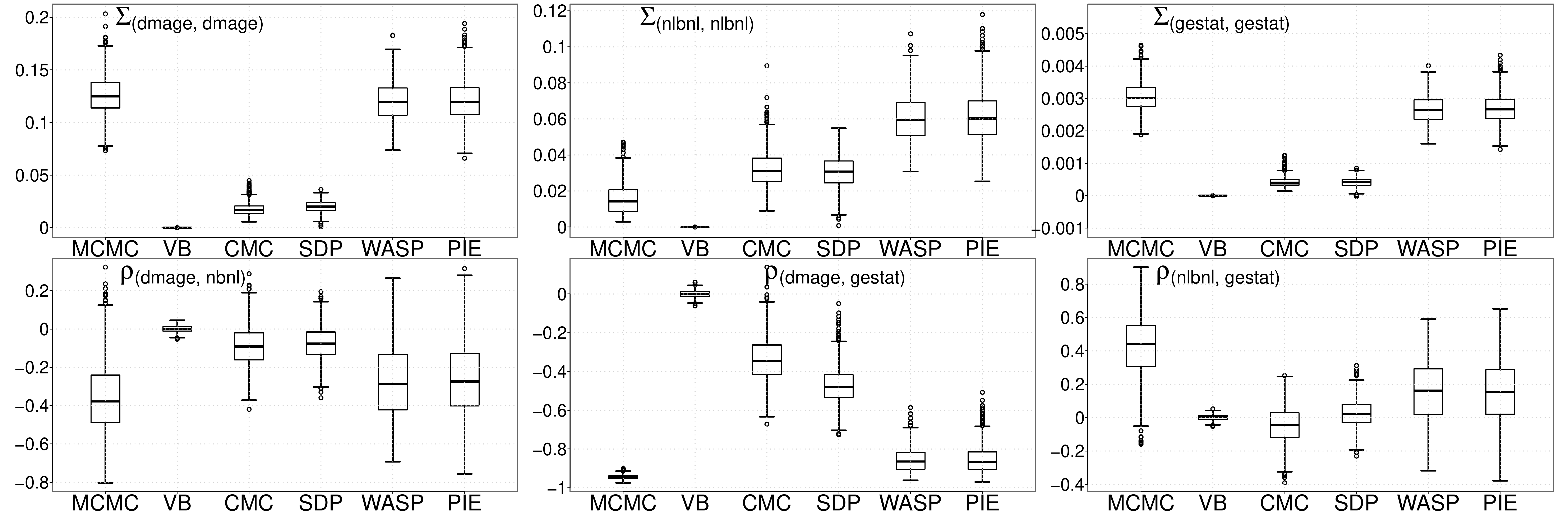}
  \captionsetup{font=small}
  \caption{Boxplots of posterior samples for variances and correlations of random effects in the United States natality data analysis. MCMC, Markov chain Monte Carlo based on the full data;  ML, maximum likelihood estimator; VB, variational Bayes; CMC, consensus Monte Carlo; SDP, semiparametric density product; WASP, the algorithm in \citet{Srietal15}; PIE, our posterior interval estimation algorithm.}
  \label{fig:aberanef}
\end{figure}

The seven methods in the previous section generally agreed in the inference on fixed effects (Figure \ref{fig:abesimef}), with variational Bayes deviating the furthest. Our algorithm and the algorithm of \citep{Srietal15} differed significantly from variational Bayes, consensus Monte Carlo, and semiparametric density product in the inference on variances and correlations of random effects (Tables \ref{tab:abe-cov-ci} and \ref{tab:abe-cov-acc} and Figure \ref{fig:aberanef}). Our algorithm and the algorithm of \citep{Srietal15} showed better agreement with Markov chain Monte Carlo based on the full data in estimating the correlations. The 90\% credible intervals from our algorithm included the maximum likelihood estimates of correlations. Variational Bayes posterior concentrated very close to 0 for every element of the covariance matrix and significantly underestimated posterior uncertainty. Consensus Monte Carlo and semiparametric density product methods performed poorly in the inference on random effects but were better than variational Bayes. Markov chain Monte Carlo based on the full data was extremely slow compared to the other methods (see Figure 1 in the Appendix). Taking into account both the approximation accuracy and the computational efficiency, we concluded that our proposed algorithm performs better than the competing algorithms in estimating the covariance matrix of random effects.


\subsection{Extension to multi-dimensional parameters}

Although Algorithm \ref{pie-algo} only applies to one-dimensional functionals, we provide a simple extension to the multi-dimensional case with a numerical illustration. Suppose our goal is to find the joint posterior of the $d$-dimensional parameter $\theta$. First, we center and scale the posterior samples of $\theta$ in every subset. Let $\widehat m_j$ and $\widehat V_j$ be the empirical mean and covariance matrix for the $j$th subset posterior samples $\{\theta_{1j},\ldots,\theta_{Tj}\}$. Let $\widehat m = K^{-1} \sum_{j=1}^K \widehat m_j$, $\widehat V^{-1} = K^{-1}\sum_{j=1}^K \widehat V_j^{-1}$. We transform every subset draw $\theta_{ij}$ to $\theta'_{ij}=\widehat V^{-1/2} (\theta_{ij} - \widehat m)$. If every subset posterior of $\theta$ is asymptotically normal, then the centered and rescaled version $\theta'$ will be asymptotically standard normal with approximately independent components, since $T$ is large in practice. For every component of $\theta'$, we apply Algorithm \ref{pie-algo} to combine its $K$ subset posterior samples and obtain approximations of posterior quantiles for a fine grid of $[0,1]$. This leads to accurate approximations of the marginal posteriors of $\theta'$; we repeatedly draw samples from these marginals, and then transform back to the original parameter using $\theta= \widehat V^{1/2} \theta' + \widehat m$. This yields approximate samples from the full posterior of $\theta$, and credible regions can be estimated based on these samples.

\begin{figure}[ht]
  \includegraphics[width=\textwidth]{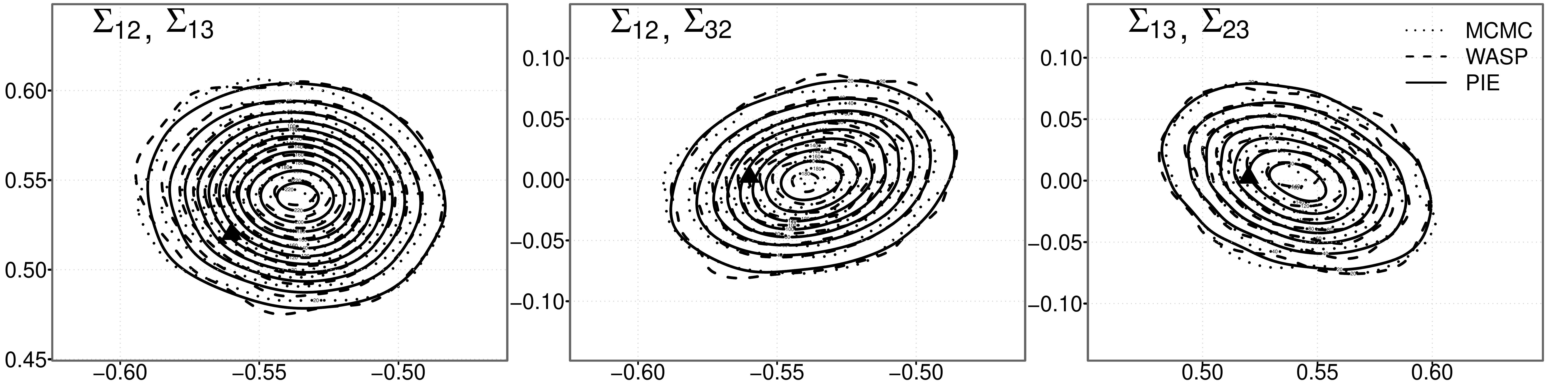}
  \captionsetup{font=small}
  \caption{{Kernel density estimates of the posterior densities for all the covariance pairs in \eqref{lme} and their true values (black triangle). var1, var2 represents the two-dimensional posterior density of (var1, var2). MCMC, Markov chain Monte Carlo based on the full data; WASP, the algorithm in \citet{Srietal15}; PIE, our posterior interval estimation algorithm.}%
  }
  \label{fig:2d}
\end{figure}

We implemented this generalized algorithm for combining subset posterior samples of all pairs of variances and covariances in the simulation from Section D.2, and compared the results with consensus Monte Carlo, semiparametric density product, Wasserstein posterior, and variational Bayes. The accuracies of our algorithm and the algorithm of \citep{Srietal15} were higher than the accuracies of the other three methods for all pairs of variances and covariances (Table \ref{tab:2d-cov-acc}). Variational Bayes performed poorly in the estimation of posterior distributions for all the pairs of variances. We obtained kernel density estimates of the three pairs of covariances in \eqref{lme} using the combined posterior samples and the bkde2D function in the KernSmooth R package with a bandwidth of 0.01 (Figure \ref{fig:2d}). The kernel density estimates centered very close to the true values of the covariance pairs. Compared to the algorithm of \citep{Srietal15}, our algorithm was more efficient, easier to implement, and robust to the grid-size of quantiles, while having similar accuracy and stability across all simulation replications.

\begin{table}[ht]
  \caption{\it 90\% credible intervals for variances and correlations of random effects in United States natality data analysis. The upper and lower bounds are averaged over 10 folds of cross-validation. MLE, maximum likelihood estimator; MCMC, Markov chain Monte Carlo based on the full data; VB, variational Bayes; CMC, consensus Monte Carlo; SDP, semiparametric density product; WASP, the algorithm in \citet{Srietal15}; PIE, our posterior interval estimation algorithm.}
{
    \begin{tabular}{r|cccccc}
      \hline
      & \scriptsize{(\texttt{dmage}, \texttt{dmage})} &  \scriptsize{(\texttt{nlbnl}, \texttt{nlbnl})} & \scriptsize{(\texttt{gestat}, \texttt{gestat})} & \scriptsize{(\texttt{dmage}, \texttt{nlbnl})} & \scriptsize{(\texttt{dmage}, \texttt{gestat})} & \scriptsize{(\texttt{nlbnl}, \texttt{gestat})} \\
      \hline
 \large{\small{MLE}} & 0.135 & 0.006 & 0.004 & -0.628 & -0.955 & 0.750 \\
 \large{\small{MCMC}} & (0.086, 0.152) & (0.003, 0.021) & (0.002, 0.004) & (-0.637, 0.283) & (-0.959, -0.912) & (-0.194, 0.728) \\
 \large{\small{VB}} & (0.000, 0.000) & (0.000, 0.000) & (0.000, 0.000) & (-0.027, 0.028) & (-0.028, 0.028) & (-0.027, 0.027) \\
 \large{\small{CMC}} & (0.010, 0.029) & (0.019, 0.051) & (0.000, 0.001) & (-0.292, 0.043) & (-0.574, -0.252) & (-0.181, 0.159) \\
 \large{\small{SDP}} & (0.015, 0.032) & (0.018, 0.049) & (0.000, 0.001) & (-0.298, -0.029) & (-0.656, -0.372) & (-0.054, 0.228) \\
 \large{\small{WASP}} & (0.100, 0.163) & (0.042, 0.088) & (0.002, 0.004) & (-0.526, 0.066) & (-0.928, -0.688) & (-0.145, 0.464) \\
 \large{\small{PIE}} & (0.098, 0.163) & (0.042, 0.088) & (0.002, 0.004) & (-0.526, 0.073) & (-0.927, -0.691) & (-0.145, 0.465) \\
 \hline
    \end{tabular}
  }
  \label{tab:abe-cov-ci}
\end{table}


\begin{table}[ht]
  \caption{\it Accuracy of approximate posteriors for variances and correlations of random effects in United States natality data analysis. The standard deviation of accuracy across 10 folds of cross-validation is in parentheses. VB, variational Bayes; CMC, consensus Monte Carlo; SDP, semiparametric density product; WASP, the algorithm in \citet{Srietal15}; PIE, our posterior interval estimation algorithm.}
{
    \begin{tabular}{r|cccccc}
      \hline
      & \scriptsize{(\texttt{dmage}, \texttt{dmage})} &  \scriptsize{(\texttt{nlbnl}, \texttt{nlbnl})} & \scriptsize{(\texttt{gestat}, \texttt{gestat})} & \scriptsize{(\texttt{dmage}, \texttt{nlbnl})} & \scriptsize{(\texttt{dmage}, \texttt{gestat})} & \scriptsize{(\texttt{nlbnl}, \texttt{gestat})} \\
      \hline
      \large{\small{VB}} & 0.00 (0.00) & 0.00 (0.01) & 0.00 (0.00) & 0.08 (0.02) & 0.00 (0.00) & 0.07 (0.02) \\
      \large{\small{CMC}} & 0.00 (0.00) & 0.15 (0.10) & 0.00 (0.00) & 0.42 (0.07) & 0.05 (0.16) & 0.33 (0.11) \\
      \large{\small{SDP}} & 0.00 (0.00) & 0.16 (0.11) & 0.00 (0.00) & 0.39 (0.08) & 0.06 (0.18) & 0.33 (0.08) \\
      \large{\small{WASP}} & 0.72 (0.21) & 0.03 (0.04) & 0.78 (0.15) & 0.72 (0.11) & 0.22 (0.14) & 0.63 (0.14) \\
      \large{\small{PIE}} & 0.73 (0.21) & 0.03 (0.04) & 0.78 (0.15) & 0.73 (0.11) & 0.22 (0.14) & 0.63 (0.14) \\
      \hline
    \end{tabular}
  }
\label{tab:abe-cov-acc}
\end{table}


\begin{table}[ht]
  \caption{\it Accuracy of approximate posteriors for all pairs of variances and covariances in \eqref{lme}. The standard deviation of accuracy across 10 folds of cross-validation is in parentheses. VB, variational Bayes; CMC, consensus Monte Carlo; SDP, semiparametric density product; WASP, the algorithm in \citet{Srietal15}; PIE, our posterior interval estimation algorithm.}
{
    \begin{tabular}{r|cccccc}
      \hline
      & ($\Sigma_{11}$, $\Sigma_{22}$) & ($\Sigma_{11}$, $\Sigma_{33}$) & ($\Sigma_{22}$, $\Sigma_{33}$) &
        ($\Sigma_{12}$, $\Sigma_{13}$) & ($\Sigma_{12}$, $\Sigma_{23}$) & ($\Sigma_{13}$, $\Sigma_{23}$) \\
      \hline
      \large{\small{VB}} & 0.09 (0.01) & 0.10 (0.01) & 0.36 (0.02) & 0.89 (0.01) & 0.90 (0.01) & 0.90 (0.01) \\
      \large{\small{CMC}} & 0.86 (0.04) & 0.84 (0.03) & 0.84 (0.03) & 0.83 (0.05) & 0.84 (0.05) & 0.88 (0.03) \\
      \large{\small{SDP}} & 0.82 (0.05) & 0.73 (0.10) & 0.74 (0.11) & 0.82 (0.05) & 0.85 (0.05) & 0.83 (0.06) \\
      \large{\small{WASP}} &  0.92 (0.02) & 0.90 (0.01) & 0.88 (0.01) & 0.93 (0.02) & 0.94 (0.01) & 0.93 (0.01) \\
      \large{\small{PIE}} & 0.91 (0.02) & 0.91 (0.01) & 0.92 (0.01) & 0.91 (0.02) & 0.92 (0.02) & 0.91 (0.01) \\
      \hline
    \end{tabular}
  }
\label{tab:2d-cov-acc}
\end{table}

\section{Conclusion}
We have proposed a simple posterior interval estimation algorithm to rapidly and accurately estimate quantiles of the posterior distributions for different one-dimensional functionals of interest.  The algorithm is simple and efficient relative to existing competitors, just averaging quantile estimates for each subset posterior based on applying existing sampling algorithms in an embarrassingly parallel manner.  There is a fascinating mathematical relationship with the Wasserstein barycenter of subset posteriors: our algorithm calculates quantiles of the Wasserstein posterior without the optimization step in \citep{Srietal15}. The credible intervals from our algorithm asymptotically approximate those from the full posterior in the leading parametric order. The quality of approximation is the same even if the subset sample size increases slowly and the number of subsets increases polynomially fast. Our experiments have demonstrated excellent performance for linear mixed effects models and linear models with varying dimension.

Although our current theory focuses on parametric models and one-dimensional linear functionals, the proposed algorithm can be practically implemented for general one-dimensional functionals for semiparametric and nonparametric models. For example, in simulations not shown in the paper, we found that our algorithm shows excellent performance for Dirichlet process mixture models for multivariate categorical data \citep{DunXin09}, and Gaussian process nonparametric regression. Furthermore, we have provided an extension to the multi-dimensional case. It would be appealing to develop theory justification in these more complex settings, and to develop guarantees on approximation accuracy for fixed subset sizes and growing numbers of subsets.
Also of interest in future work is to consider algorithms that do not require non-overlapping subsets, potentially relying on subsampling.  Although such modifications can be implemented trivially, our proof techniques for the combining step in Theorem \ref{w2main} do not apply directly.  Other important extensions include optimal design of subsampling algorithms and extensions beyond product likelihoods.


\newpage
\appendix

\setcounter{equation}{0}
\setcounter{lemma}{0}
\renewcommand{\theequation}{A.\arabic{equation}}

\noindent {\bf \LARGE Appendix}
\vspace{.5cm}

In Section A we provide the detailed technical proofs of Theorem 1 and Theorem 2 in the main paper. In Section B, we present a theorem that quantifies the Monte Carlo errors in subset posterior sampling. In Section C, we verify Assumption 7 in the main paper for the normal linear model and some exponential family distributions. Section D includes further details about the data analysis in the main paper. In particular, for the heavy tailed priors used in Example 1 in the main paper, we verify a relaxed version of Assumption 6 in Section D.1.

\section{Proofs of Theorem 1 and Theorem 2}



\subsection{Technical Lemmas}
\begin{lemma}\label{vilthm}
(Villani \citep{Vil08} Theorem 6.15) For two measures $P_1,P_2\in \mathcal{P}_2(\Theta)$, or similarly $\mathcal{P}_2(\Xi)$,
$$W_2^2(P_1,P_2) \leq 2~TV_2(P_1,P_2),$$
where the total variation of moments distance \citep{CheHon03, Vil08} is defined as
$$TV_2(P_1,P_2) = \int_{\Theta} (1+\|\theta\|^2)\ud|P_1(\theta)-P_2(\theta)|.$$
\end{lemma}

\begin{lemma}\label{subsetbound}
Suppose that Assumptions 1--7 hold. Let $\hat\theta_j$ be a weakly consistent estimator of $\theta_0$ based on the subset $X_j$ such that it solves the score equation $\ell_j'(\hat\theta_j)=0$; $\hat\theta_j \to \theta_0$ in $P_{\theta_0}$-probability. Let $\hat\theta$ be a weakly consistent estimator of $\theta_0$ based on the whole $X$ such that it solves the score equation $\ell'(\hat\theta)=0$; $\hat\theta \to \theta_0$ in $P_{\theta_0}$-probability. Let $t=n^{1/2}(\theta-\hat \theta_j)$ be the local parameter for the $j$th subset, and $s=n^{1/2}(\theta-\hat \theta)$.  Let $\Pi_{m,t}(t\mid X_j)$ be the $j$th subset posterior induced by $\Pi_{m}(\theta\mid X_j)$ and $\Pi_{n,s}(s\mid X)$ be the posterior of $s$ induced by the overall posterior $\Pi_{n}(\theta\mid X)$. Then
\begin{align}
\label{tj} &\lim_{m\to \infty} E_{P_{\theta_0}} TV_2\left[ \Pi_{m,t}(t\mid X_j),\Phi\left\{ t; 0,I^{-1}(\theta_0)\right\} \right] =0, \\
\label{tn} &\lim_{m\to \infty} E_{P_{\theta_0}} TV_2\left[ \Pi_{n,s}(s\mid X),\Phi\left\{ s; 0,I^{-1}(\theta_0)\right\} \right] =0.
\end{align}
\end{lemma}
\begin{remark}
In comparison, the usual parametric Bernstein-von Mises theorem on the subset $X_j$ without raising the likelihood to the $K$th power gives
$$\lim_{m\to \infty} TV_2\left[ \Pi_{m}(z\mid X_j),\Phi\left\{z; 0,I^{-1}(\theta_0)\right\}\right] = 0,$$
in $P_{\theta_0}$-probability, where $z=m^{1/2}(\theta-\hat\theta_j)$. See, for example, Theorem 8.2 in \citep{LehCas98} and Theorem 4.2 in \citep{Ghosh06}.
\end{remark}

\noindent {\bf Proof of Lemma 2:}\\
The relation \eqref{tn} in Lemma 2 is the usual Bernstein-von Mises theorem for the overall posterior $\Pi_n(\theta\mid X)$.  The proof of \eqref{tn} follows a related line to the proof of Theorem 4.2 in \citep{Ghosh06}, and can be treated as a special case of \eqref{tj} with $m=n$ and $K=1$. In the following we focus on the proof of \eqref{tj} in Lemma 2.

Given the independent and identically distributed assumption, we only need to show the result for a fixed index $j$. To emphasize the different roles played by the subset sample size $m$ and the number of subsets $K$, in the following proofs we will write the total sample size $n$ as $Km$. We complete the proof in 3 steps. For a generic matrix $A$ or a 3-dimensional array $A$, we use $\|A\|$ to denote its Frobenius norm.\\

\noindent {Step 1:} Show the existence of weakly consistent estimator $\hat\theta_j$ for $\theta_0$ that solves $\ell_j'(\hat\theta_j)=0$. Given Assumption 3, $\ell_j'(\theta)$ is continuously differentiable in an open neighborhood of $\theta_0$ and $E_{P_{\theta_0}}\left\{p'(X \mid \theta_0)/p(X\mid \theta_0)\right\}=0$. Therefore, with large $P_{\theta_0}$-probability, there exists a root for the equation $\ell_j'(\theta)=0$ inside the neighborhood that attains the maximum of $\ell(\theta)$. Denote the root as $\hat\theta_j$ and then $\hat\theta_j\to \theta_0$ in $P_{\theta_0}$-probability is a clear consequence of Assumption 5.

\vspace{.4cm}

\noindent {Step 2:} Show the following convergence as $m\to \infty$ in $P_{\theta_0}$-probability:
\begin{align}\label{astv2}
TV_2\left[\Pi_{m,t}(t\mid X_j),\Phi\left\{t; 0,I^{-1}(\theta_0)\right\}\right] \to 0.
\end{align}
We prove this result for a fixed subset $X_j$, since the data are independent and identically distributed, and the conclusion is identical for any $j=1,\ldots,K$. Define the following quantities
\begin{align*}
w(t) &= \ell_j\left\{\hat \theta_j +\frac{t}{(Km)^{1/2}}\right\} - \ell_j(\hat \theta_j) \\
C_m & = \int e^{Kw(z)} \pi\left\{\hat \theta_j +\frac{z}{(K m)^{1/2}}\right\} \ud z.
\end{align*}
Then based on the expression of $\Pi_m(\theta\mid X_j)$, with the likelihood raised to the $K$th power,
$$\pi_m \left(\theta\mid X_j\right) = \frac{\exp\left\{K \ell_j(\theta)\right\} \pi(\theta) \ud \theta} {\int_{\Theta} \exp\left\{K \ell_j(\theta)\right\} \pi(\theta) \ud \theta}.$$
The induced posterior density on $t=(K m)^{1/2}(\theta-\hat \theta_j)$ can be written as
\begin{align*}
\pi_m(t\mid X_j) & =  \frac{\exp\{K w(t)\}\pi\left\{\hat \theta_j + \frac{t}{(K m)^{1/2}}\right\}}{C_m}.
\end{align*}
Let $\mathcal{T} = \{t=(K m)^{1/2}(\theta-\hat\theta_j):\theta\in\Theta\}$. Define
\begin{align*}
g_m(t) &= \left(1+\|t\|^2\right) \left[ e^{K w(t)} \pi\left\{\hat \theta_j +\frac{t}{(K m)^{1/2}}\right\} - \exp\left\{-\frac{1}{2}t^\T I(\theta_0)t\right\}\pi(\theta_0) \right].
\end{align*}
If we can show that $\int_{\mathcal{T}} |g_m(z)|\ud z  \xrightarrow{P_{\theta_0}} 0$ as $m\to \infty$ in $P_{\theta_0}$-probability, then
$$ C_m \to \int_{\mathcal{R}^d}\exp\left\{-\frac{1}{2}z^\T I(\theta_0)z\right\}\pi(\theta_0)\ud z  = (2\pi)^{d/2} \left\{\dett{I(\theta_0)}\right\}^{-1/2} \pi(\theta_0)$$
as $m\to \infty$ in $P_{\theta_0}$-probability. Hence, for the difference in \eqref{astv2}, we obtain that
\begin{align}\label{tv2exp}
& TV_2\left[\Pi_{m,t}(t\mid X_j),\Phi\left\{t; 0,I^{-1}(\theta_0)\right\}\right] \nonumber \\
={}& \int_{\mathcal{T}} \left(1+\|z\|^2\right) \left| \frac{ e^{K w(z)} \pi\left\{\hat \theta_j +\frac{z}{(K m)^{1/2}}\right\}}{C_m} - \frac{1}{(2\pi)^{d/2} \left\{\dett{I(\theta_0)}\right\}^{-1/2} }\exp\left\{-\frac{1}{2}z^\T I(\theta_0)z\right\}\right| \ud z \nonumber \\
\leq{}& \frac{1}{C_m} \int_{\mathcal{T}} |g_m(z)| \ud z +  \left|\frac{(2\pi)^{d/2} \left\{\dett{I(\theta_0)}\right\}^{-1/2} \pi(\theta_0)}{C_m}-1\right|\times \nonumber\\
& \int_{\mathcal{R}^d} \frac{\left(1+\|z\|^2\right)}{(2\pi)^{d/2} \left\{\dett{I(\theta_0)}\right\}^{-1/2} }\exp\left\{-\frac{1}{2}z^\T I(\theta_0)z\right\} \ud z \to 0
\end{align}
as $m\to\infty$ in $P_{\theta_0}$-probability and \eqref{astv2} is proved. Therefore it suffices to show that $\int_{\mathcal{T}} |g_m(z)|\ud z \to 0$ as $m\to\infty$ in $P_{\theta_0}$-probability.\\

Divide the domain of the integral into 3 parts: $A_1=\{z: \|z\|\geq \delta_1 (K m)^{1/2}\}$, $A_2=\{z:\delta_2\leq \|z\|< \delta_1(K m)^{1/2}\}$, $A_3=\{z:\|z\|<\delta_2\}$, where the constants $\delta_1,\delta_2$ will be chosen later. Then
\begin{align}\label{gm}
\int_{\mathcal{T}} |g_m(z)|\ud z \leq \int_{A_1} |g_m(z)|\ud z + \int_{A_2} |g_m(z)|\ud z +  \int_{A_3} |g_m(z)|\ud z.
\end{align}
We have that
\begin{align}\label{g1a1}
& \int_{A_1} |g_m(z)|\ud z \nonumber \\
&\leq \int_{A_1} \left( 1+\|z\|^2\right) e^{K w(z)} \pi\left\{\hat \theta_j +\frac{z}{(K m)^{1/2}}\right\} \ud z +  \int_{A_1} \left( 1+\|z\|^2\right)e^{-\frac{1}{2}z^\T I(\theta_0)z}\pi(\theta_0) \ud z,
\end{align}
and
$$\int_{A_1} \left( 1+\|z\|^2\right)e^{-\frac{1}{2}z^\T I(\theta_0)z}\pi(\theta_0) \ud z =
\pi(\theta_0)\int_{\|z\|\geq \delta_1(K m)^{1/2}} \left( 1+\|z\|^2\right)e^{-\frac{1}{2}z^\T I(\theta_0)z} \ud z \to 0$$
as $m\to\infty$, because the integral on the whole $z\in \mathcal{R}^d$ is finite, $\pi(\theta_0)$ is bounded from above according to Assumption 6, and $K\geq 1$.\\

Next we bound the first term in \eqref{g1a1}. By Assumption 5 and the weak consistency of $\hat \theta_j$, there exists a constant $\epsilon_1$ that depends on $\delta_1$, such that for any $z\in A_1$ and all sufficiently large $m$, with $P_{\theta_0}$-probability approaching 1,
$$\ell_j\{\hat\theta_j+ z/(K m)^{1/2}\} - \ell_j(\hat\theta_j)  \leq - m\epsilon_1.$$
Furthermore, the weak consistency of $\hat \theta_j$ implies that for all sufficiently large $m$, with $P_{\theta_0}$-probability approaching 1, $\|\hat\theta_j\|\leq \|\hat\theta_j-\theta_0\| +\|\theta_0\|\leq \delta_0+\|\theta_0\|$. Therefore, as $m\to\infty$ in $P_{\theta_0}$-probability,
\begin{align}\label{prior2moment}
& \int_{A_1} \left( 1+\|z\|^2\right)e^{K w(z)} \pi\left\{\hat \theta_j +\frac{z}{(K m)^{1/2}}\right\} \ud z \nonumber \\
\leq {}& \exp(-K m\epsilon_1) \int \left( 1+\|z\|^2\right) \pi\left\{\hat \theta_j +\frac{z}{(K m)^{1/2}}\right\} \ud z \nonumber \\
\leq {}& \exp(-K m\epsilon_1) \left\{1+(Km)^{d/2}\int_{\Theta}2(\|\theta\|^2+\|\hat\theta_j\|^2)\pi(\theta)\ud \theta  \right\} \nonumber \\
\leq {}& \exp(-K m\epsilon_1) \left[1+2(Km)^{d/2}\left\{2\|\theta_0\|^2 + 2\delta_0^2+ \int_{\Theta}\|\theta\|^2\pi(\theta)\ud \theta \right\} \right]\to  0,
\end{align}
where we have used the finite second moment of $\pi(\theta)$ from Assumption 6 in the last step.
Hence, we have proved that the first integral in \eqref{gm} goes to zero in $P_{\theta_0}$-probability. \\

For the second integral in \eqref{gm}, by the Taylor series expansion and $\ell_j'(\hat\theta_j)=0$,
\begin{align}\label{wA2}
&w(z)=\ell_j\left\{\hat \theta_j +\frac{z}{(K m)^{1/2}}\right\} - \ell_j(\hat \theta_j) = -\frac{1}{2K } z^\T I(\hat\theta_j)z + R_m(z) \\
&R_m(z) \equiv \frac{1}{6} \frac{\partial^3 \ell_j(\tilde \theta)}{\partial \theta^3} \left\{\frac{z}{(K m)^{1/2}}, \frac{z}{(K m)^{1/2}},\frac{z}{(K m)^{1/2}}\right\}, \nonumber
\end{align}
where $\partial^3 \ell_j(\tilde \theta)/\partial \theta^3$ is a 3-dimensional array and $\tilde\theta$ satisfies $\|\tilde\theta - \hat\theta_j\|\leq z/(K m)^{1/2}$. Since $\hat \theta_j\to \theta_0$ in $P_{\theta_0}$-probability, we have $\|\hat \theta_j - \theta_0\|<\delta_0/3$ for all large $m$ with $P_{\theta_0}$-probability approaching 1, and we choose $\delta_1\leq \delta_0/3$ such that $\|\tilde \theta-\theta_0\|<\delta_0$ for all large $m$ given $z\in A_2$. For every fixed $z\in A_2$, $R_m(z)$ in \eqref{wA2} converges to zero as $m\to\infty$ in $P_{\theta_0}$-probability, which implies that on $z\in A_2$, $g_m(z)\to 0$ in $P_{\theta_0}$-probability.  Moreover, by Assumption 3, $R_m(z)$ can be further bounded by
\begin{align*}
 |R_m(z)| & \leq \frac{d^3}{6} \left\|\frac{z}{(K m)^{1/2}}\right\|^3 \sum_{i=1}^m M(X_{ij}) \\
 &\leq \frac{d^3\delta_1}{6K} \|z\|^2  \frac{1}{m}\sum_{i=1}^m M(X_{ij}) \to \frac{d^3\delta_1}{6K} \|z\|^2 E_{P_{\theta_0}} M(X_{11}),
\end{align*}
where the last convergence is almost surely in $P_{\theta_0}$ by the strong law of large numbers.
Therefore, we can choose $\delta_1$ as
$$\delta_1 = \min\left[\frac{\delta_0}{3}, \frac{3 \min_{\theta\in B_{\delta_0}(\theta_0)} \lambda_1\{I(\theta)\}}{4d^3E_{P_{\theta_0}} M(X_{11})}\right],$$
where $\lambda_1(A)$ denotes the smallest eigenvalue of a generic matrix $A$. Assumption 4 indicates that $\min_{\theta\in B_{\delta_0}(\theta_0)} \lambda_1\{I(\theta)\}$ is bounded below by a constant. Thus, in \eqref{wA2}, the choice of $\delta_1$ implies that for every $z\in A_2$, for all large $m$ with $P_{\theta_0}$-probability approaching 1,
\begin{align*}
 |R_m(z)|& \leq  \frac{1}{4K} z^\T I(\hat\theta_j) z,\\
 \exp\{K w(z)\} &\leq \exp \left[K\left\{-\frac{1}{2K} z^\T I(\hat\theta_j)z +|R_m(z)|  \right\}\right]\\
\leq{}& \exp \left\{ - \frac{1}{4}z^\T I(\hat\theta_j)z \right\} \leq \exp \left\{ - \frac{1}{8}z^\T I(\theta_0)z \right\}.
\end{align*}
Therefore for $z\in A_2$, for all large $m$ with $P_{\theta_0}$-probability approaching 1,
\begin{align*}
|g_m(z)| & \leq  \left(1+\|z\|^2\right) \left[ \exp \left\{- \frac{1}{8}z^\T I(\theta_0)z \right\} \pi\left\{ \hat \theta_j +\frac{z}{(K m)^{1/2}}\right\} + \exp\left\{-\frac{1}{2}z^\T I(\theta_0)z\right\} \pi(\theta_0) \right] \\
\leq{}& \sup_{\theta\in \Theta} \pi(\theta) \times  2\left(1+\|z\|^2\right)\exp \left\{ - \frac{1}{8}z^\T I(\theta_0)z \right\}.
\end{align*}
Hence $\int_{A_2} |g_m(z)|\ud z < +\infty$, since $\sup_{\theta\in \Theta} \pi(\theta) <\infty$ by Assumption 6. We can choose the constant $\delta_2$ sufficiently large, such that $\int_{A_2} |g_m(z)|\ud z$ is arbitrarily small in $P_{\theta_0}$-probability.\\

For the third integral in \eqref{gm}, we fix a constant $\delta_2>0$ and can use the similar Taylor series expansion above, and notice that when $\|z\|<\delta_2$, as $m\to \infty$,
\begin{align}\label{a3r}
\sup_{\|z\|<\delta_2}K|R_m(z)| & \leq \frac{Kd^3\delta_2^3}{6(Km)^{3/2}} \sum_{i=1}^m M(X_{ij})
 \leq \frac{d^3\delta_2^3}{6(K m)^{1/2}} \times \frac{1}{m}\sum_{i=1}^m M(X_{ij}) \to 0,
\end{align}
where the last convergence is almost surely in $P_{\theta_0}$.  It follows from
\eqref{wA2}, \eqref{a3r}, the weak consistency of $\tilde \theta_j$ and the continuity of $I(\theta)$ in $B_{\delta_0}(\theta_0)$ that as $m\to\infty$ in $P_{\theta_0}$-probability,
\begin{align}\label{a3w}
& \sup_{\|z\|<\delta_2} \left|Kw(z) - \frac{1}{2}z^\T I(\theta_0) z \right|\leq  \frac{\delta_2^2}{2} \left\|I(\hat \theta_j) - I(\theta_0)\right\| + \sup_{\|z\|<\delta_2}K|R_m(z)| \to 0.
\end{align}
By the continuity of $\pi(\theta)$ in Assumption 6 and the weak consistency of $\tilde \theta_j$, we also have that
\begin{align}\label{a3pi}
& \sup_{\|z\|<\delta_2} \left|\pi\left\{\hat \theta_j +\frac{t}{(K m)^{1/2}}\right\} - \pi(\theta_0)\right| \to 0,
\end{align}
as $m\to\infty$ in $P_{\theta_0}$-probability.
Therefore, \eqref{a3w} and \eqref{a3pi} together imply that as $m\to\infty$ in $P_{\theta_0}$-probability,
\begin{align*}
& \sup_{\|z\|<\delta_2} \left|e^{Kw(z)}\pi\left\{\hat \theta_j +\frac{t}{(K m)^{1/2}}\right\} - e^{-\frac{1}{2}z^\T I(\theta_0)z}\pi(\theta_0)\right| \to 0.
\end{align*}
Hence by the definition of $g_m(z)$, as $m\to\infty$ in $P_{\theta_0}$-probability,
\begin{align*}
&\int_{A_3} |g_m(z)|\ud z \\
&\leq \int_{\|z\|\leq \delta_2}(1+\|z\|^2)\ud z \times  \sup_{\|z\|<\delta_2} \left|e^{Kw(z)}\pi\left\{ \hat \theta_j +\frac{t}{(K m)^{1/2}}\right\} - e^{-\frac{1}{2}z^\T I(\theta_0)z}\pi(\theta_0)\right| \to 0.
\end{align*}


This has proved that the right-hand side of \eqref{gm} converges to zero in $P_{\theta_0}$-probability, and also completes the proof of \eqref{astv2}.
\vspace{.4cm}

\noindent {Step 3:} Show the convergence in $L_1$ as $m\to \infty$. It is clear from the derivation of \eqref{tv2exp} that
\begin{align}
& TV_2\left[\Pi_{m,t}(t\mid X_j),\Phi\left\{t; 0,I^{-1}(\theta_0)\right\}\right] \nonumber \\
={}& \int_{\mathcal{T}} \left(1+\|z\|^2\right) \left| \frac{ e^{K w(z)} \pi\left\{\hat \theta_j +\frac{z}{(K m)^{1/2}}\right\}}{C_m} - \frac{1}{(2\pi)^{d/2} \left\{\dett{I(\theta_0)}\right\}^{-1/2} }\exp\left\{-\frac{1}{2}z^\T I(\theta_0)z\right\} \right| \ud z \nonumber \\
\leq{}& \int_{\Theta} \left\{1+\left\|n^{1/2}(\theta-\hat\theta_j)\right\|^2\right\} \pi(\theta|X_j) \ud \theta + \int_{\mathcal{R}^d} \frac{\left(1+\|z\|^2\right)}{(2\pi)^{d/2} \left\{\dett{I(\theta_0)}\right\}^{-1/2} }\exp\left\{-\frac{1}{2}z^\T I(\theta_0)z\right\} \ud z \nonumber\\
= {} & 1 + E_{\Pi_m(\theta|X_j)} Km\left\|\theta-\hat\theta_j\right\|^2 + \int_{\mathcal{R}^d} \frac{\left(1+\|z\|^2\right)}{(2\pi)^{d/2} \left\{\dett{I(\theta_0)}\right\}^{-1/2} }\exp\left\{-\frac{1}{2}z^\T I(\theta_0)z\right\} \ud z  \nonumber
\end{align}
In this display, the last term is a finite constant. The middle term is $\psi(X_j)$ defined in Assumption 7. According to Assumption 7, for any fixed $j$, $\left\{\psi(X_j):m\geq m_0, K\geq 1\right\}$ is uniformly integrable under $P_{\theta_0}$. Now since $ TV_2\left[\Pi_{m,t}(t\mid X_j),\Phi\left\{t; 0,I^{-1}(\theta_0)\right\}\right]$ is upper bounded by $\psi(X_j)+C$ for all $m,K$ and some constant $C>0$, we obtain that $\left\{TV_2\left[\Pi_{m,t}(t\mid X_j),\Phi\left\{t; 0,I^{-1}(\theta_0)\right\}\right]: m\geq m_0, K\geq 1\right\}$ is also uniformly integrable. This uniform integrability together with the convergence in $P_{\theta_0}$-probability from Step 2 implies the $L_1$ convergence of $TV_2\left[\Pi_{m,t}(t\mid X_j),\Phi\left\{t; 0,I^{-1}(\theta_0)\right\}\right]$ to zero.
\hfill $\blacksquare$

\vspace{1cm}

Similar to the $W_2$ distance, for any $l \geq 1$, we can define the Wasserstein-$l$ ($W_l$) distance: for any two measures $\nu_1,\nu_2$ on $\Theta$, their $W_l$ distance is defined as
\begin{align}
W_l(\nu_1,\nu_2) = \left(\inf_{\gamma \in \Gamma(\nu_1,\nu_2)}\int_{\Theta\times \Theta} \|\theta_1-\theta_2\|^l \ud \gamma( \nu_1,\nu_2 )\right)^{1/l}, \nonumber
\end{align}
where $\Gamma(\nu_1,\nu_2)$ is the set of all probability measures on $\Theta\times \Theta$ with marginals $\nu_1$ and $\nu_2$, respectively. The $W_l$ distance on the space $\Xi$ can be similarly defined. The $W_l$ distance between two univariate distributions $F_1$ and $F_2$ is the same as the $L_l$ distance between their quantile functions (see Lemma 8.2 of \citep{BikFre81}):
\begin{align*}
W_l(F_1,F_2) = \left[ \int_0^1 \left\{ F_1^{-1}(u)-F_2^{-1}(u)\right\}^l \ud u\right]^{1/l}.
\end{align*}

\begin{lemma}\label{w2ineq}
Let $\hat \xi_j=a^\T\hat\theta_j + b$. Then for any $l\geq 1$,
\begin{align*}
W_l\left(\overline \Pi_n(\xi\mid X),\Phi\left[\xi; \overline \xi, \left\{ nI_{\xi}(\theta_0)\right\}^{-1} \right] \right)
\leq \frac{1}{K} \sum_{j=1}^K W_l\left (\Pi_m(\xi\mid X_j),\Phi\left[\xi ; \hat \xi_j,  \left\{ n I_{\xi}(\theta_0)\right\}^{-1} \right]\right).
\end{align*}
\end{lemma}

\noindent {\bf Proof of Lemma~3:}\\
We use $\Phi(\cdot)$ and $\Phi^{-1}(\cdot)$ to denote the cumulative distribution function and the quantile function of standard normal distribution $\mathcal{N}(0,1)$. From \citep{AguCar11}, the univariate Wasserstein-2 barycenter satisfies that for any $u\in (0,1)$,
\begin{align*}
\overline \Pi_n^{-1} (u\mid X) = \frac{1}{K} \sum_{j=1}^K \Pi_m^{-1} \left(u\mid X_j \right).
\end{align*}
Therefore,
\begin{align}\label{w21}
& W_l\left( \overline \Pi_n(\xi\mid X),\Phi\left[\xi ; \overline \xi, \left\{nI_{\xi}(\theta_0)\right\}^{-1} \right]  \right) \nonumber  \\
={}& \left(\int_0^1 \left| \overline \Pi_n^{-1}(u\mid X) - \Phi^{-1}\left[u ; \overline \xi, \left\{nI_{\xi}(\theta_0)\right\}^{-1} \right] \right|^l \ud u \right)^{1/l} \nonumber \\
={}& \left[\int_0^1 \left| \frac{1}{K} \sum_{j=1}^K \Pi_m^{-1} \left(u\mid X_j \right) - \overline \xi -   \left\{nI_{\xi}(\theta_0)\right\}^{-1/2}  \Phi^{-1}(u) \right|^l \ud u \right]^{1/l} \nonumber \\
={}& \left(\int_0^1 \left| \frac{1}{K} \sum_{j=1}^K \left[\Pi_m^{-1} \left(u\mid X_j \right) -   \hat  \xi_j -   \left\{nI_{\xi}(\theta_0)\right\}^{-1/2}  \Phi^{-1}(u) \right]\right|^l \ud u \right)^{1/l}.
\end{align}
Define
\begin{align}\label{rj}
r_j(u) =\Pi_m^{-1} \left(u\mid X_j \right) -    \hat  \xi_j -   \left\{ nI_{\xi}(\theta_0)\right\}^{-1/2}  \Phi^{-1}(u).
\end{align}
then
$$W_l \left(\Pi_m(\xi\mid X_j),\Phi\left[\xi ;  \hat  \xi_j,  \left\{n I_{\xi}(\theta_0)\right\}^{-1} \right]\right) = \left\{\int_0^1 \left|r_j(u)\right|^l\ud u\right\}^{1/l}.$$
Since $l\geq 1$, we apply Minkowski inequality to the right-hand side of \eqref{w21} and obtain that
\begin{align}\label{w2rj}
&W_l\left( \overline \Pi_n(\xi\mid X),\Phi\left[\xi ; \overline \xi, \left\{ nI_{\xi}(\theta_0)\right\}^{-1} \right]  \right) =  \left\{\int_0^1 \left|\frac{1}{K} \sum_{j=1}^K r_j(u) \right|^l \ud u  \right\}^{1/l} \\
&\leq  \frac{1}{K} \sum_{j=1}^K  \left\{\int_0^1 \left| r_j(u) \right|^l \ud u  \right\}^{1/l}
= \frac{1}{K} \sum_{j=1}^K W_l \left(\Pi_m(\xi\mid X_j),\Phi\left[\xi ;  \hat  \xi_j,  \left\{n I_{\xi}(\theta_0)\right\}^{-1} \right]\right), \nonumber
\end{align}
which concludes the proof.
\hfill $\blacksquare$

\vspace{1cm}

\begin{lemma}\label{mlediff}
Suppose Assumptions 1--7 hold. Then
$$\left|\overline\xi - \hat\xi \right|= o_p\left(m^{-1/2}\right),$$
where $o_p$ is in $P_{\theta_0}$ probability.
Furthermore, if $\hat\theta_1$ is an unbiased estimator of $\theta_0$, then
\begin{align*}
& \left|\overline\xi -\hat\xi\right| = o_p\left(n^{-1/2}\right).
\end{align*}
\end{lemma}

\noindent {\bf Proof of Lemma~4:}\\
Because of the linearity $\xi=a^\T\theta+b$, it suffices to show
\begin{align}
\label{diff0}& \left\|\hat\theta -\overline \theta\right\| = o_p\left(m^{-1/2}\right),
\end{align}
under Assumptions 1--7, and
\begin{align}
\label{diff} & \left\|\hat\theta -\overline \theta\right\| = o_p\left(n^{-1/2}\right),
\end{align}
with the further assumption that $\hat\theta_1$ is an unbiased estimator of $\theta_0$.\\

We use the first order Taylor expansion of $\ell'_j(\hat \theta_j)$ ($j=1,\ldots,K$) and $\hat\theta$ around $\theta_0$:
\begin{align*}
& 0=\ell_j'(\hat\theta_j) = \ell_j'(\theta_0) + \ell_j''(\tilde \theta_j)(\hat \theta_j -\theta_0), \nonumber \\
& 0=\ell'(\hat\theta) = \ell'(\theta_0) + \ell''(\tilde \theta)(\hat \theta -\theta_0),
\end{align*}
where $\tilde\theta_j$ is between $\hat\theta_j$ and $\theta_0$, $\tilde \theta$ is between $\hat\theta$ and $\theta_0$, and $\ell''(\theta) = \sum_{j=1}^K \ell_j''(\theta)$. These expansions lead to
\begin{align}
\label{tay1} \hat \theta_j & = \theta_0 + \frac{1}{m}I^{-1}(\theta_0)\ell_j'(\theta_0) + Z_j \frac{\ell_j'(\theta_0)}{m}, \\
Z_j &\equiv  \left\{-\frac{1}{m}\ell_j''(\tilde \theta_j)\right\}^{-1} - I^{-1}(\theta_0), \nonumber \\
\hat \theta & = \theta_0 + \frac{1}{n}I^{-1}(\theta_0)\ell'(\theta_0) + Z \frac{\ell'(\theta_0)}{n},\nonumber \\
Z &\equiv  \left\{-\frac{1}{n}\ell''(\tilde \theta_j)\right\}^{-1} - I^{-1}(\theta_0).\nonumber
\end{align}
Therefore by the equality $\ell'(\theta_0)=\sum_{j=1}^K \ell'_j(\theta_0)$, the difference between $\overline \theta$ and $\hat \theta$ is
\begin{align}\label{oh1}
\overline \theta - \hat  \theta & = \frac{1}{K} \sum_{j=1}^K Z_j \frac{\ell_j'(\theta_0)}{m} - Z \frac{\ell'(\theta_0)}{n}.
\end{align}
For the second term in \eqref{oh1}, by the central limit theorem $n^{-1/2}\ell'(\theta_0)$ converges in distribution to $\mathcal{N}(0,I(\theta_0))$, so $\left\|\ell'(\theta_0)/n\right\|=O_p(n^{-1/2})$. $Z$ converges in $P_{\theta_0}$-probability to zero given the consistency of $\hat\theta$ to $\theta_0$ in Lemma 2, so $\|Z\|=o_p(1)$. Therefore by the Slutsky's theorem,
\begin{align}\label{z21}
\left\|Z \frac{\ell'(\theta_0)}{n} \right\| = o_p\left(n^{-1/2}\right).
\end{align}

Next we show the first term in \eqref{oh1} is of order $o_p(m^{-1/2})$ under Assumptions 1--7, and is of order $o_p(n^{-1/2})$ if furthermore $E_{P_{\theta_0}}\hat\theta_1=\theta_0$.

Let $W_j = Z_j  \ell_j'(\theta_0)/m^{1/2}$. Then $\{W_j:j=1,\ldots,K\}$ are independent and identically distributed random vectors and the first term in \eqref{oh1} is $\sum_{j=1}^K W_j/(Km^{1/2})$. Since $Z_j\to 0$ in $P_{\theta_0}$-probability as $m\to\infty$, and $m^{-1/2}\ell_j'(\theta_0)=O_p(1)$ as $m\to \infty$, by the Slutsky's theorem again, $W_j\to 0$ in $P_{\theta_0}$-probability. Furthermore, we will show at the end of this proof that $E_{P_{\theta_0}}(\|W_1\|^2)\to 0$ as $m\to \infty$. Assuming this is true, by the Markov's inequality,  for any $c>0$,
\begin{align*}
& P_{\theta_0}\left(\left\|\frac{1}{K}\sum_{j=1}^K \frac{W_j}{m^{1/2}}\right\|\geq cm^{-1/2}\right) \leq
\frac{m E_{P_{\theta_0}}\left\|\frac{1}{K}\sum_{j=1}^K \frac{W_j}{m^{1/2}}\right\|^2} {c^2}\\
\leq {}&\frac{1}{c^2K} \sum_{j=1}^K E_{P_{\theta_0}} \|W_j\|^2 = \frac{E_{P_{\theta_0}} \|W_1\|^2}{c^2} \to 0.
\end{align*}
Hence, $\left\|\sum_{j=1}^K W_j/(Km^{1/2})\right\|=o_p\left(m^{-1/2}\right)$. This together with \eqref{oh1} and \eqref{z21} leads to \eqref{diff0}.\\

If we further assume unbiasedness $E_{P_{\theta_0}}\hat\theta_1=\theta_0$, then from \eqref{tay1} we can obtain that
\begin{align*}
E_{P_{\theta_0}}W_j& = m^{-1/2} E_{P_{\theta_0}}\left\{Z_j  \ell_j'(\theta_0)\right\} \\
& = m^{1/2} E_{P_{\theta_0}}\left\{\hat\theta_j - \theta_0 -m^{-1}I^{-1}(\theta_0)\ell_j'(\theta_0)  \right\} \\
& = m^{1/2} E_{P_{\theta_0}}\left(\hat\theta_j - \theta_0\right) - m^{-1/2} I^{-1}(\theta_0) E_{P_{\theta_0}} \ell_j'(\theta_0) \\
& =0,
\end{align*}
for all $j=1,\ldots,K$. In other words, $W_j$'s are centered at zero. Since $X_j$'s ($j=1,\ldots,K$) are all independent and $W_j$ only depends on $X_j$, we have $E_{P_{\theta_0}} W_{j_1}^\T W_{j_2}=0$ for any $j_1\neq j_2$.

We can again apply Markov's inequality to the first term in \eqref{oh1} and obtain that for any constant $c>0$,
\begin{align}
& P_{\theta_0}\left( \left\|\frac{1}{K} \sum_{j=1}^K \frac{W_j}{m^{1/2}}\right\| > cn^{-1/2}\right)
= {} P_{\theta_0}\left( \left\|\frac{1}{K} \sum_{j=1}^K W_j\right\| > cK^{-1/2} \right)  \nonumber \\
\leq {}& \frac{K E_{P_{\theta_0}}\left\|\frac{1}{K}\sum_{j=1}^K W_j\right\|^2 }{c^2}  \nonumber\\
={}& \frac{K }{K^2c^2}E_{P_{\theta_0}}\left(\sum_{j=1}^K \|W_j\|^2 + \sum_{j_1\neq j_2} W_{j_1}^\T W_{j_2}\right)  \nonumber\\
={}& \frac{E_{P_{\theta_0}}(\|W_1\|^2)}{c^2}. \nonumber
\end{align}
Therefore, assuming that $E_{P_{\theta_0}}\hat\theta_1=\theta_0$ and $E_{P_{\theta_0}}(\|W_1\|^2)\to 0$ as $m\to \infty$, which will be proven below, the display above implies that $\left\|\sum_{j=1}^K W_j/(Km^{1/2})\right\|=o_p\left(n^{-1/2}\right)$. This together with \eqref{oh1} and \eqref{z21} leads to \eqref{diff}.\\

\noindent Proof of $E_{P_{\theta_0}}(\|W_1\|^2)\to 0$ as $m\to \infty$:\\
By Assumption 4, we let $\underline \lambda>0$ be a constant lower bound of the eigenvalues of $-\ell_1''(\theta)/m$ for all $\theta\in \Theta$, all $X_1$ and all sufficiently large $m$. Then we have
\begin{align}\label{z12}
\left\|\left\{-\frac{1}{m}\ell''_1(\tilde \theta_1)\right\}^{-1}\right\|
\leq d^{1/2}\underline\lambda ^{-1}, ~~ \left\|I(\theta_0)^{-1}\right\|
\leq d^{1/2} \underline\lambda ^{-1},
\end{align}
where $d$ is the dimension of $\theta$. We have used the property of the Frobenius norm: for a generic $d\times d$ symmetric positive definite matrix $A$, $\|A^{-1}\|\leq d^{1/2} \overline \lambda(A^{-1}) =  d^{1/2} \left\{\underline \lambda(A)\right\}^{-1}$, where $\overline\lambda(A)$ and $\underline\lambda(A)$ denotes the largest and the smallest eigenvalues of the matrix $A$, respectively. Furthermore, the envelop function condition in Assumption 3 implies that
\begin{align}\label{z13}
\left\|-\frac{1}{m}\ell''_1(\tilde \theta_1)\right\|^2
\leq \frac{d^2}{m}\sum_{i=1}^m  M(X_{i1})^2.
\end{align}
It follows from \eqref{z12} and \eqref{z13} that for all large $m$,
\begin{align*}
\left\|Z_1\right\|^2 & = \left\|\left\{-\frac{1}{m}\frac{\partial^2 \ell_1(\tilde \theta_1)}{\partial \theta \partial\theta^\T}\right\}^{-1}\left\{-\frac{1}{m}\frac{\partial^2 \ell_1(\tilde \theta_1)}{\partial \theta \partial\theta^\T} -I(\theta_0)\right\}I^{-1}(\theta_0) \right\|^2 \\
& \leq \frac{1}{2} d \underline\lambda ^{-2} \left\{\frac{d^2}{m}\sum_{i=1}^m  M(X_{i1})^2 +\|I(\theta_0)\|^2\right\}
d \underline\lambda ^{-2}\\
&\leq c_1\frac{1}{m}\sum_{i=1}^m  M(X_{i1})^2 +c_2,
\end{align*}
where $c_1,c_2$ are positive constants that only depend on $d,\underline\lambda,\|I(\theta_0)\|^2$.

Now define $V_1 = \left\{c_1\sum_{i=1}^m  M(X_{i1})^2/m+c_2\right\} \left\|m^{-1/2}\ell_1'(\theta_0)\right\|^2$. Then we have $\|W_1\|^2 \leq V_1 $. We are going to show that $E_{P_{\theta_0}}(V_1)<\infty$ and then apply the dominated convergence theorem to $\|W_1\|^2$ and conclude that $E_{P_{\theta_0}}\|W_1\|^2\to 0$ since we already have $W_1\to 0$ in $P_{\theta_0}$-probability. To see why $E_{P_{\theta_0}}(V_1)<\infty$, we first apply the Cauchy-Schwarz inequality to $V_1$ and obtain that
\begin{align}\label{v1}
E_{P_{\theta_0}}(V_1) & =E_{P_{\theta_0}} \left\{ c_1\frac{1}{m}\sum_{i=1}^m  M(X_{i1})^2+c_2\right\} \left\|m^{-1/2}\ell_1'(\theta_0)\right\|^2 \nonumber \\
& \leq \left[ E_{P_{\theta_0}} \left\{c_1\frac{1}{m}\sum_{i=1}^m  M(X_{i1})^2+c_2\right\}^2\right] ^{1/2}  \left( E_{P_{\theta_0}}\left\|m^{-1/2}\ell_1'(\theta_0)\right\|^4\right)^{1/2}.
\end{align}
Due to Assumption 3, the first term in \eqref{v1} is bounded by
\begin{align*}
& E_{P_{\theta_0}} \left\{ c_1\frac{1}{m}\sum_{i=1}^m  M(X_{i1})^2+c_2\right\} ^2 \\
\leq{}& 2c_1^2 E_{P_{\theta_0}} \left\{\frac{1}{m}\sum_{i=1}^m  M(X_{i1})^4\right\} +2c_2^2 = 2c_1^2 E\left\{M(X)^4\right\} +2c_2^2 <\infty.
\end{align*}
Now recall that $\ell_1'(\theta_0) = \sum_{i=1}^m p'(X_{i1}\mid \theta_0)/p(X_{i1}\mid \theta_0)$. Denote the $l$th component in the random vector $p'(X_{i1}\mid \theta_0)/p(X_{i1}\mid \theta_0)$ as $U_{il}$, such that $p'(X_{i1}\mid \theta_0)/p(X_{i1}\mid \theta_0) =(U_{i1},...,U_{id})^\T$. Then $U_{i_1l}$ and $U_{i_2l}$ are independent if $i_1\neq i_2$, due to the independence between $X_{ij}$'s. By $E_{P_{\theta_0}}\left\{p'(X\mid \theta_0)/p(X\mid \theta_0)\right\}=0$ in Assumption 3, we have $E_{P_{\theta_0}} U_{il}=0$ for all $i=1,\ldots,m$ and $l=1,\ldots,d$. From Assumption 3 we have for all $l=1,\ldots,d$, $E_{P_{\theta_0}} U_{1l}^4 \leq E_{P_{\theta_0}} M(X)^4<\infty $.
Therefore, the second term in \eqref{v1} can be bounded as
\begin{align*}
& E_{P_{\theta_0}}\left\|m^{-1/2}\ell_1'(\theta_0)\right\|^4 = \frac{1}{m^2} E_{P_{\theta_0}} \left\{\sum_{l=1}^d
\left( \sum_{i=1}^m U_{il}\right)^2\right\}^2 \leq  \frac{d}{m^2} E_{P_{\theta_0}} \sum_{l=1}^d
\left( \sum_{i=1}^m U_{il}\right)^4 \\
& = \frac{d}{m^2}\sum_{l=1}^d  \left\{\sum_{i=1}^m E_{P_{\theta_0}} U_{il}^4 + 3\sum_{i_1\neq i_2}\left(E_{P_{\theta_0}} U_{i_1 l}^2\right)\left(E_{P_{\theta_0}} U_{i_2 l}^2\right)  \right\} \\
& = \frac{d}{m^2} \sum_{l=1}^d \left\{m E_{P_{\theta_0}} U_{1l}^4 + 3m(m-1) \left(E_{P_{\theta_0}} U_{1l}^2\right)^2 \right\}\\
& \leq \frac{d}{m^2} \sum_{l=1}^d \left\{m E_{P_{\theta_0}} U_{1l}^4 + 3m(m-1) E_{P_{\theta_0}} U_{1l}^4\right\} \leq 3d \sum_{l=1}^d E_{P_{\theta_0}} U_{1l}^4 <\infty.
\end{align*}
Thus we have shown that both terms on the right-hand side of \eqref{v1} are finite. Therefore, $E_{P_{\theta_0}}(V_1)<\infty$ and by the dominated convergence theorem, $E_{P_{\theta_0}}\|W_1\|^2\to 0$.
\hfill $\blacksquare$

\subsection{Proof of Theorem 1}
\noindent {\bf Proof of Theorem 1(i):}\\
Since $\xi=a^\T \theta+b$, we can derive the following for subset posteriors in terms of $\xi$ using a change of variable from $\theta$ to $\xi$ in \eqref{tj} of Lemma 2:
\begin{align*}
& \lim_{m\to\infty} E_{P_{\theta_0}} TV_2\left[\Pi_{m,t}(t\mid X_j),\Phi\left\{ t; 0,I_{\xi}^{-1}(\theta_0)\right\} \right] =0,
\end{align*}
where $t= n^{1/2} (\xi-\hat\xi_j)$ is now the local parameter for the $j$th subset. From the relation between norms $W_2$ and $TV_2$ in Lemma 1, this directly implies
$$\lim_{m\to\infty}  E_{P_{\theta_0}} W_2^2\left[ \Pi_{m,t}(t\mid X_j),\Phi\left\{ t; 0,I_{\xi}^{-1}(\theta_0)\right\} \right] =0.$$
We further use the rescaling property of the $W_2$ distance and obtain the equivalent form in terms of the original parameter $\xi$:
\begin{align}\label{rootn2}
\lim_{m\to\infty}  n E_{P_{\theta_0}} W_2^2\left(\Pi_{m}\left(\xi\mid X_j\right),\Phi\left[\xi; \hat\xi_j,\left\{ nI_{\xi}\right(\theta_0)\}^{-1}\right]\right) =0.
\end{align}

From Lemma 3, we have that for any constant $c>0$, as $m\to \infty$,
\begin{align}
& P_{\theta_0}\left\{ W_2\left(\overline \Pi_{n}\left(\xi \mid X\right),\Phi\left[\xi; \overline \xi,\left\{ n I_{\xi}(\theta_0)\right\}^{-1}\right]\right) \geq cn^{-1/2} \right\} \nonumber \\
\overset{(i)}{\leq}{}& P_{\theta_0}\left\{ \frac{1}{K} \sum_{j=1}^K W_2\left (\Pi_m(\xi\mid X_j),\Phi\left[\xi ; \hat \xi_j,  \left\{n I_{\xi}(\theta_0)\right\}^{-1} \right]\right) \geq cn^{-1/2} \right\} \nonumber \\
\overset{(ii)}{\leq}{}& \frac{n}{c^2} {E}_{P_{\theta_0}}\left\{\frac{1}{K}\sum_{j=1}^K W_2 \left(\Pi_m(\xi\mid X_j),\Phi\left[\xi ; \hat \xi_j,  \left\{n I_{\xi}(\theta_0)\right\}^{-1} \right]\right)\right\}^2 \nonumber \\
\overset{(iii)}{\leq}{}& \frac{n}{c^2 K} \sum_{j=1}^K {E}_{P_{\theta_0}} W_2^2 \left (\Pi_m(\xi\mid X_j),\Phi\left[\xi ; \hat \xi_j,  \left\{ n I_{\xi}(\theta_0)\right\}^{-1} \right]\right) \nonumber \\
\leq{}& \frac{n}{c^2 } E_{P_{\theta_0}}W_2^2\left (\Pi_m(\xi\mid X_1),\Phi\left[\xi ; \hat \xi_1,  \left\{ n I_{\xi}(\theta_0)\right\}^{-1} \right]\right) \overset{(iv)}{\to} 0, \nonumber
\end{align}
where (i) follows from Lemma 3 with $l=2$, (ii) uses Markov's inequality, (iii) comes from the relation between $l_1$ norm and $l_2$ norm, and (iv) follows from \eqref{rootn2}. This result indicates that
\begin{align}\label{f11}
& W_2\left(\overline \Pi_{n}\left(\xi \mid X\right),\Phi\left[\xi; \overline \xi,\left\{ n I_{\xi}(\theta_0)\right\}^{-1}\right]\right) = o_p\left(n^{-1/2}\right),
\end{align}
which shows the first relation in Part (i) of Theorem 1. The second relation in Theorem 1~(i)
\begin{align}\label{f12}
& W_2\left(\Pi_n\left(\xi \mid X\right), \Phi\left[\xi; \hat \xi,\left\{ n I_{\xi}(\theta_0)\right\}^{-1}\right]\right) = o_p\left(n^{-1/2}\right).
\end{align}
follows from a similar argument using \eqref{tn} in Lemma 2 for the overall posterior.

From Lemma 4, we have $\left|\overline\xi - \hat\xi \right|=o_p\left(m^{-1/2}\right)$. Therefore,
\begin{align}\label{f13}
& W_2\left(\Phi\left[\xi; \overline \xi,\left\{n I_{\xi}(\theta_0)\right\}^{-1}\right], \Phi\left[\xi; \hat \xi,\left\{ n I_{\xi}(\theta_0)\right\}^{-1}\right]\right) \leq \left|\overline\xi - \hat\xi \right|=o_p\left(m^{-1/2}\right),
\end{align}
where the first inequality follows because of the definition of $W_2$ distance and the same variance shared by the two normal distributions.

Finally, by \eqref{f11}, \eqref{f12}, \eqref{f13} and the triangular inequality, we have
\begin{align*}
& W_2\left\{ \overline \Pi_{n}\left(\xi \mid X\right), \Pi_{n}\left(\xi \mid X\right)\right\} \\
\leq {} & W_2\left(\overline \Pi_{n}\left(\xi \mid X\right),\Phi\left[\xi; \overline \xi,\left\{ n I_{\xi}(\theta_0)\right\}^{-1}\right]\right)   +  W_2\left(\Phi\left[\xi; \overline \xi,\left\{n I_{\xi}(\theta_0)\right\}^{-1}\right], \Phi\left[\xi; \hat \xi,\left\{n I_{\xi}(\theta_0)\right\}^{-1}\right]\right) \\
& + W_2\left(\Phi\left[\xi; \hat \xi,\left\{n I_{\xi}(\theta_0)\right\}^{-1}\right],\Pi_{n}\left(\xi \mid X\right)\right) \\
\leq {}& o_p\left(n^{-1/2}\right) + o_p\left(m^{-1/2}\right) + o_p\left(n^{-1/2}\right) =o_p\left(m^{-1/2}\right),
\end{align*}
which is equivalent to the third relation in Part (i).
\hfill $\blacksquare$
\vspace{.5cm}

\noindent {\bf Proof of Theorem 1(ii):}\\
If $\hat\theta_1$ is an unbiased estimator of $\theta_0$, then by Lemma 4 and the definition of $W_2$ distance, it follows that
\begin{align}\label{f21}
& W_2\left(\Phi\left[\xi; \overline \xi,\left\{n I_{\xi}(\theta_0)\right\}^{-1}\right], \Phi\left[\xi; \hat \xi,\left\{ n I_{\xi}(\theta_0)\right\}^{-1}\right]\right) \leq \left|\overline \xi-\hat \xi\right| = o_p\left(n^{-1/2}\right).
\end{align}
Applying the triangular inequality to \eqref{f11}, \eqref{f12} and \eqref{f21}, we obtain that as $m\to\infty$,
\begin{align*}
& W_2\left\{ \overline \Pi_{n}\left(\xi \mid X\right), \Pi_{n}\left(\xi \mid X\right)\right\} \\
\leq {} & W_2\left(\overline \Pi_{n}\left(\xi \mid X\right),\Phi\left[\xi; \overline \xi,\left\{ n I_{\xi}(\theta_0)\right\}^{-1}\right]\right)  +  W_2\left(\Phi\left[\xi; \overline \xi,\left\{n I_{\xi}(\theta_0)\right\}^{-1}\right], \Phi\left[\xi; \hat \xi,\left\{n I_{\xi}(\theta_0)\right\}^{-1}\right]\right) \\
& + W_2\left(\Phi\left[\xi; \hat \xi,\left\{n I_{\xi}(\theta_0)\right\}^{-1}\right],\Pi_{n}\left(\xi \mid X\right)\right) \\
={}& o_p\left(n^{-1/2}\right)+o_p\left(n^{-1/2}\right)+o_p\left(n^{-1/2}\right) =o_p\left(n^{-1/2}\right).
\end{align*}
Thus the conclusion of Part (ii) follows.
\hfill $\blacksquare$

\subsection{Proof of Theorem 2}
\noindent {\bf Proof of Theorem 2(i):}\\
\cite{AguCar11} have shown that the barycenter $\overline \Pi_{n}(\xi\mid X)$ is related to the $K$ subset posteriors $\Pi_m(\xi\mid X_j)$ ($j=1,\ldots,K$) through the quantile function:
\begin{align*}
\overline \Pi_n^{-1} (u\mid X) = \frac{1}{K} \sum_{j=1}^K \Pi_m^{-1} \left(u\mid X_j \right),
\end{align*}
for any $u\in (0,1)$. Also the expectation of a generic univariate distribution $F$ can be calculated through its quantile functions:
if a random variable $Y$ has the cumulative distribution function $F$, $E_F(Y)=\int_0^1 F^{-1}(u)\ud u$. Therefore
\begin{align*}
& \bias\left\{\overline \Pi_{n}(\xi\mid X)\right\} = E_{\overline \Pi_{n}(\xi\mid X)}(\xi)-\xi_0 \\
& = \int_0^1 \overline \Pi_{n}^{-1}(u\mid X) \ud u -\xi_0  = \int_0^1 \frac{1}{K}\sum_{j=1}^K\Pi_m^{-1}(u\mid X_j) - \xi_0 \\
& \overset{(i)}{=} \frac{1}{K}\sum_{j=1}^K \int_0^1 \left[\hat\xi_j +\left\{nI_{\xi}(\theta_0)\right\}^{-1/2}  \Phi^{-1}(u) +r_j(u)\right] \ud u -\xi_0 \\
& \overset{(ii)}{=} \frac{1}{K}\sum_{j=1}^K \hat\xi_j +  \frac{1}{K}\sum_{j=1}^K \int_0^1 r_j(u)\ud u -\xi_0 = \overline \xi -\xi_0 +  \frac{1}{K}\sum_{j=1}^K \int_0^1 r_j(u)\ud u,
\end{align*}
where (i) follows from the definition of $r_j(u)$ in \eqref{rj}, and (ii) makes use of the fact $\int_0^1\Phi^{-1}(u)\ud u=0$. $|\hat\xi-\xi_0|=O_p(n^{-1/2})$ from the central limit theorem. It remains to be shown that
\begin{align}\label{rj1}
\left|\frac{1}{K}\sum_{j=1}^K \int_0^1 r_j(u)\ud u\right| =o_p\left(n^{-1/2}\right).
\end{align}
To see why this is true, we notice that we have derived the following relation in the proof of Theorem 1:
\begin{align*}
\frac{1}{K}\sum_{j=1}^K W_2\left (\Pi_m(\xi\mid X_j),\Phi\left[\xi ; \hat \xi_j,  \left\{n I_{\xi}(\theta_0)\right\}^{-1} \right]\right) = o_p\left(n^{-1/2}\right).
\end{align*}
But according to the definition of $r_j(u)$ in \eqref{rj}, by Cauchy-Schwarz inequality,
\begin{align*}
& \left|\frac{1}{K}\sum_{j=1}^K \int_0^1 r_j(u)\ud u\right| \leq \frac{1}{K}\sum_{j=1}^K \left\{ \int_0^1 r_j^2(u)\ud u\right\}^{1/2} \\
& = \frac{1}{K}\sum_{j=1}^K W_2\left (\Pi_m(\xi\mid X_j),\Phi\left[\xi ; \hat \xi_j,  \left\{ n I_{\xi}(\theta_0)\right\}^{-1} \right]\right) = o_p\left(n^{-1/2}\right),
\end{align*}
which proves \eqref{rj1}.

On the other hand, for the bias of the overall posterior $\Pi_{n}(\xi\mid X)$, we follow a similar argument as above and obtain that
\begin{align*}
& \bias\left\{ \Pi_{n}(\xi\mid X)\right\} = E_{ \Pi_{n}(\xi\mid X)}(\xi)-\xi_0 = \int_0^1  \Pi_{n}^{-1}(u\mid X) \ud u -\xi_0 \\
& = \int_0^1 \left[\hat\xi + \left\{nI_{\xi}(\theta_0)\right\}^{-1/2}\right]  \Phi^{-1}(u) -\xi_0 + \int_0^1 r(u) \ud u
 = \hat\xi - \xi_0 + \int_0^1 r(u) \ud u,
\end{align*}
where $r(u) = \Pi_{n}^{-1}(u|X) - \hat\xi -  \left\{nI_{\xi}(\theta_0)\right\}^{-1/2} \Phi^{-1}(u)$. Moreover we have
\begin{align*}
& \left|\int_0^1 r(u) \ud u\right| \leq \left(\int_0^1 \left[\Pi_{n}^{-1}(u\mid X) - \hat\xi -  \left\{nI_{\xi}(\theta_0)\right\}^{-1/2}\Phi^{-1}(u)\right]^2 \ud u\right)^{1/2} \\
& = W_2\left(\Pi_{n}(\xi\mid X), \Phi\left[\xi;\hat\xi,\left\{nI_{\xi}(\theta_0)\right\}^{-1}\right]\right) = o_p\left(n^{-1/2}\right)
\end{align*}
by Theorem 1. This completes the proof of Part (i).
\hfill $\blacksquare$

\vspace{.5cm}

\noindent {\bf Proof of Theorem 2(ii):}\\
Similar to the expectation, the variance of a generic univariate distribution $F$ can be calculated through its quantile functions: if $Y\sim F$,
\begin{align*}
{\var}(Y)&=E(Y^2)-(E Y)^2 = \int_{-\infty}^{\infty} y^2 \ud F(y) - \left\{\int_{-\infty}^{\infty} y \ud F(y)\right\}^2\\
& = \int_0^1 \left\{F^{-1}(u)\right\}^2 \ud u - \left\{\int_0^1F^{-1}(u) \ud u \right\}^2.
\end{align*}
Therefore,
\begin{align}
& \var\left\{\overline \Pi_{n}(\xi\mid X)\right\} = \int_0^1 \left\{\overline \Pi_{n}^{-1}(u\mid X)\right\}^2 \ud u - \left\{\int_0^1\overline \Pi_{n}^{-1}(u\mid X) \ud u \right\}^2 \nonumber \\
& = \int_0^1 \left\{\frac{1}{K} \sum_{j=1}^K \Pi_{m}^{-1}(u\mid X)\right\}^2 \ud u - \left\{\int_0^1\frac{1}{K} \sum_{j=1}^K \Pi_{m}^{-1}(u\mid X) \ud u \right\}^2 \nonumber \\
& = \int_0^1 \left[\overline \xi + \left\{nI_{\xi}(\theta_0)\right\}^{-1/2}\Phi^{-1}(u) + \frac{1}{K} \sum_{j=1}^K r_j(u)\right]^2 \ud u \nonumber \\
& - \left(\int_0^1\left[\overline \xi + \left\{nI_{\xi}(\theta_0)\right\}^{-1/2}\Phi^{-1}(u) + \frac{1}{K} \sum_{j=1}^K r_j(u)\right]\ud u \right)^2 \nonumber \\
& = \frac{1}{nI_{\xi}(\theta_0)}\int_0^1 \left\{\Phi^{-1}(u)\right\}^2 \ud u  + \int_0^1 \left\{\frac{1}{K}\sum_{j=1}^K r_j(u)\right\}^2 \ud u -
\left\{\frac{1}{K}\sum_{j=1}^K \int_0^1 r_j(u)\ud u\right\}^2  \nonumber  \\
& + 2\left\{nI_{\xi}(\theta_0)\right\}^{-1/2} \int_0^1 \Phi^{-1}(u) \frac{1}{K}\sum_{j=1}^K r_j(u) \ud u \nonumber
\end{align}
where we have used the fact $\int_0^1 \Phi^{-1}(u)\ud u=0$ and $\int_0^1 \left(\Phi^{-1}(u)\right)^2 \ud u=1$. It remains to be shown that the other three terms in the display above are of order $o_p(n^{-1})$.

From \eqref{w2rj} (with $l=2$) and the conclusion of Theorem 1, we have
\begin{align*}
& \int_0^1 \left\{\frac{1}{K}\sum_{j=1}^K r_j(u)\right\}^2 \ud u = W_2^2\left( \overline \Pi_n(\xi\mid X),\Phi\left[\xi ; \overline \xi, \left\{nI_{\xi}(\theta_0)\right\}^{-1} \right]  \right) = o_p\left(n^{-1}\right).
\end{align*}
By Cauchy-Schwarz inequality, we have
\begin{align*}
& \left\{\frac{1}{K}\sum_{j=1}^K \int_0^1 r_j(u)\ud u\right\}^2 = \left\{ \int_0^1 \frac{1}{K}\sum_{j=1}^K r_j(u)\ud u\right\}^2
\leq \int_0^1 \left\{\frac{1}{K}\sum_{j=1}^K r_j(u)\right\}^2 \ud u = o_p\left(n^{-1}\right).
\end{align*}

Again by Cauchy-Schwarz inequality, we have
\begin{align*}
&  \left|2\left\{nI_{\xi}(\theta_0)\right\}^{-1/2} \int_0^1 \Phi^{-1}(u)\times \frac{1}{K}\sum_{j=1}^K r_j(u) \ud u \right| \\
&\leq 2\left\{nI_{\xi}(\theta_0)\right\}^{-1/2} \left[\int_0^1 \left\{\Phi^{-1}(u)\right\}^2 \ud u\right]^{1/2}
\left[\int_0^1 \left\{\frac{1}{K}\sum_{j=1}^K r_j(u)\right\}^2 \ud u\right]^{1/2} \\
& = O\left(n^{-1/2}\right)\times o_p\left(n^{-1/2}\right) = o_p\left(n^{-1}\right).
\end{align*}
Therefore, we have shown that $\var\left\{\overline \Pi_{n}(\xi\mid X)\right\} =  \left\{nI_{\xi}(\theta_0)\right\}^{-1}+ o_p\left(n^{-1}\right)$.

For the variance of $\Pi_n(\xi\mid X)$, we use the same definition of $r(u)$ as in Part (i) and derive that
\begin{align}
& \var\left\{\Pi_{n}(\xi\mid X)\right\} = \int_0^1 \left\{ \Pi_{n}^{-1}(u\mid X)\right\}^2 \ud u - \left\{\int_0^1 \Pi_{n}^{-1}(u\mid X) \ud u \right\}^2 \nonumber \\
& = \int_0^1 \left[\hat \xi + \left\{nI_{\xi}(\theta_0)\right\}^{-1/2}\Phi^{-1}(u) + r(u)\right]^2 \ud u \nonumber \\
& - \left(\int_0^1\left[\hat \xi + \left\{nI_{\xi}(\theta_0)\right\}^{-1/2}\Phi^{-1}(u) +r(u)\right]\ud u \right)^2 \nonumber \\
& = \frac{1}{nI_{\xi}(\theta_0)} + \int_0^1 r(u)^2 \ud u -\left\{\int_0^1 r(u) \ud u \right\}^2+ 2\left\{nI_{\xi}(\theta_0)\right\}^{-1/2}\int_0^1 \Phi^{-1}(u)r(u)\ud u.  \nonumber
\end{align}
Based on the conclusion of Theorem 1 and Cauchy-Schwarz inequality, we have
\begin{align*}
& \int_0^1 r(u)^2 \ud u = W_2^2 \left(\Pi_n(\xi\mid X),\Phi\left[\xi ; \hat \xi, \left\{nI_{\xi}(\theta_0)\right\}^{-1} \right]  \right) = o_p\left(n^{-1}\right),\\
& \left\{\int_0^1 r(u) \ud u \right\}^2 \leq \int_0^1 r(u)^2 \ud u = o_p\left(n^{-1}\right),
\end{align*}
and also
\begin{align*}
&  \left|2\left\{nI_{\xi}(\theta_0)\right\}^{-1/2}\int_0^1 \Phi^{-1}(u)r(u)\ud u \right| \\
&\leq 2\left\{nI_{\xi}(\theta_0)\right\}^{-1/2} \left[\int_0^1 \left\{\Phi^{-1}(u)\right\}^2 \ud u\right]^{1/2}
\left\{\int_0^1 r(u)^2 \ud u\right\}^{1/2} \\
& = O\left(n^{-1/2}\right) o_p\left(n^{-1/2}\right) = o_p\left(n^{-1}\right),
\end{align*}
which proves $\var\left\{\Pi_{n}(\xi\mid X)\right\} =  \left\{nI_{\xi}(\theta_0)\right\}^{-1} + o_p\left(n^{-1}\right)$.
\hfill $\blacksquare$
\vspace{.5cm}

\noindent {\bf Proof of Theorem 2(iii):}\\
The convergence in $W_2$ distance implies weak convergence. Therefore, it follows from Theorem 1 that in $P_{\theta_0}$ probability, both $\overline \Pi_{n,s}(s\mid X)$ and $\Pi_{n,s}(s\mid X)$ converge in distribution to normal distributions as $m\to\infty$. The weak convergence also implies the convergence of quantile functions at any continuous point. Since both $\overline \Pi_{n}\left(\xi\mid X\right)$ and $\Pi_{n}\left(\xi\mid X\right)$ are continuous distributions with posterior densities, their quantiles also converge pointwise to the quantiles of their limiting normal distributions. For any fixed $u\in (0,1)$, as $m\to\infty$, Theorem 1 implies that for $s=n^{1/2}(\xi-\overline\xi)$,
\begin{align*}
& \left|\overline \Pi_{n,s}^{-1}(u\mid X) - \Phi^{-1} \{u;0,I_{\xi}^{-1}(\theta_0)\}\right| = o_p(1).
\end{align*}
We can make this convergence uniform over all quantiles $u\in [u_1,u_2]\subset (0,1)$. Divide $[u_1,u_2]$ into $L$ equally spaced subintervals $[u_{(j)},u_{(j+1)}]$ for $j=0,\ldots,L-1$ and $u_{(j)}=u_1+j(u_2-u_1)/L$. For any $\epsilon>0$, since $\Phi^{-1} \{u;0,I_{\xi}^{-1}(\theta_0)\}$ is uniformly continuous on $[u_1,u_2]$, we can pick $L$ sufficiently large such that
$$\Phi^{-1} \{u_{(j+1)};0,I_{\xi}^{-1}(\theta_0)\} - \Phi^{-1} \{u_{(j)};0,I_{\xi}^{-1}(\theta_0)\} < \epsilon/2,$$
for all $j=0,\ldots,L-1$. Furthermore, because $\Phi^{-1}(\cdot)$ is continuous everywhere, we can find a sufficiently large $n_0$, such that for all $n>n_0$, all $j=0,\ldots,L-1$ with the $L$ chosen above,
$$\left|\overline \Pi_{n,s}^{-1}(u_{(j)}\mid X) - \Phi^{-1} \{u_{(j)};0,I_{\xi}^{-1}(\theta_0)\}\right| < \epsilon/2.$$
For any $u\in [u_1,u_2]$, we can find a $j_0\in \{0,\ldots,L-1\}$ such that $u\in [u_{(j_0)},u_{(j_0+1)}]$. Therefore using the monotonicity of quantile functions,
\begin{align*}
&\overline \Pi_{n,s}^{-1}(u\mid X) - \Phi^{-1} \{u;0,I_{\xi}^{-1}(\theta_0)\}
 \leq \overline \Pi_{n,s}^{-1}(u_{(j_0+1)}\mid X) - \Phi^{-1} \{u;0,I_{\xi}^{-1}(\theta_0) \}\\
\leq{}& \overline \Pi_{n,s}^{-1}(u_{(j_0+1)}\mid X) - \Phi^{-1} \{u_{(j_0+1)};0,I_{\xi}^{-1}(\theta_0)\} \\
& + \Phi^{-1} \{u_{(j_0+1)};0,I_{\xi}^{-1}(\theta_0)\} - \Phi^{-1} \{u_{(j_0)};0,I_{\xi}^{-1}(\theta_0)\} \\
 <{}& \epsilon/2 + \epsilon/2 < \epsilon.
\end{align*}
and
\begin{align*}
&\overline \Pi_{n,s}^{-1}(u\mid X) - \Phi^{-1} \{u;0,I_{\xi}^{-1}(\theta_0)\}
 \geq \overline \Pi_{n,s}^{-1}(u_{(j_0)}\mid X) - \Phi^{-1} \{u;0,I_{\xi}^{-1}(\theta_0) \}\\
\geq{}& \overline \Pi_{n,s}^{-1}(u_{(j_0)}\mid X) - \Phi^{-1} \{u_{(j_0)};0,I_{\xi}^{-1}(\theta_0)\} \\
& + \Phi^{-1} \{u_{(j_0)};0,I_{\xi}^{-1}(\theta_0)\} - \Phi^{-1} \{u_{(j_0+1)};0,I_{\xi}^{-1}(\theta_0)\} \\
 > {}& -\epsilon/2 - \epsilon/2 > -\epsilon.
\end{align*}
Therefore, we have shown that
\begin{align*}
\sup_{u\in [u_1,u_2]} \left|\overline \Pi_{n,s}^{-1}(u\mid X) - \Phi^{-1} \{u;0,I_{\xi}^{-1}(\theta_0)\}\right| = o_p(1),
\end{align*}
which implies that for the quantiles in terms of $\xi$,
\begin{align*}
& \sup_{u\in [u_1,u_2]} \left| \overline \Pi_{n}^{-1}(u\mid X) - \overline \xi - \left\{ nI_{\xi}(\theta_0)\right\} ^{-1/2}\Phi^{-1}(u) \right| = o_p\left(n^{-1/2}\right).
\end{align*}
Similarly for the overall posterior
\begin{align*}
&  \sup_{u\in [u_1,u_2]} \left| \Pi_{n}^{-1}(u\mid X) - \hat \xi - \left\{ nI_{\xi}(\theta_0)\right\} ^{-1/2}\Phi^{-1}(u) \right| = o_p\left(n^{-1/2}\right).
\end{align*}
Therefore, by the triangular inequality,
\begin{align}\label{q12comb}
& \sup_{u\in [u_1,u_2]} \left| \overline \Pi_{n}^{-1}(u\mid X) - \Pi_{n}^{-1}(u\mid X) \right | \leq \left|\overline\xi - \hat\xi\right| + o_p\left(n^{-1/2}\right).
\end{align}
By plugging in the order $\left|\overline\xi - \hat\xi\right|=o_p(m^{-1/2})$ from the proof of Theorem 1, we have
\begin{align*}
&\sup_{u\in [u_1,u_2]} \left| \overline \Pi_{n}^{-1}(u\mid X) - \Pi_{n}^{-1}(u\mid X) \right| =  o_p\left(m^{-1/2}\right).
\end{align*}
If we further assume that $\hat\theta_1$ is an unbiased estimator of $\theta_0$, then Lemma 4 says that $\left|\overline\xi - \hat\xi \right|=o_p\left(n^{-1/2}\right)$. Therefore, using the results from Part (i), we have
$$\bias\left\{\overline \Pi_{n}(\xi\mid X)\right\} - \bias\left\{\Pi_{n}(\xi\mid X)\right\}
= \overline\xi - \hat\xi +o_p \left(n^{-1/2}\right) = o_p \left(n^{-1/2}\right).$$
Then \eqref{q12comb} leads to
$$ \sup_{u\in [u_1,u_2]} \left| \overline \Pi_{n}^{-1}(u\mid X) - \Pi_{n}^{-1}(u\mid X) \right | \leq o_p\left(n^{-1/2}\right) + o_p \left(n^{-1/2}\right)= o_p\left(n^{-1/2}\right),$$
which completes the proof.
\hfill $\blacksquare$
\vspace{1cm}

\section{Theoretical Results for the Posterior Monte Carlo Errors}
In practice, the credible intervals are calculated from the averages of empirical quantiles from subset posterior samples. In Algorithm 1, suppose that for each $j=1,\ldots,K$, $\Pi^\circ_j(\theta)$ and $\kappa_j(\theta,\theta')$ for $\theta,\theta'\in \Theta$ are the initial distribution and the transition kernel for the Markov chain of the $j$th subset posterior. $\{\theta_{1j},\ldots,\theta_{Tj}\}$ with sample size $T$ are drawn sequentially with $\theta_{1j}\sim \Pi^\circ_j(\cdot)$ and $\theta_{l+1,j}\sim \kappa_j(\theta_{lj},\cdot)$ for $l=1,\ldots, T-1$. $\xi_{lj}=a^\T \theta_{lj}+b$ for $l=1,\ldots,T$ and $j=1,\ldots,K$. Let $\widehat \Pi_m(\xi \mid X_j)$ be the empirical distribution of $\{\xi_{1j},\ldots,\xi_{Tj}\}$ for $j=1,\ldots,K$. Let $\widehat \Pi_n(\xi\mid X)$ be the Wasserstein barycenter of $\widehat \Pi_m(\xi \mid X_1),\ldots, \widehat \Pi_m(\xi \mid X_K)$, which can be calculated through its quantile function $\widehat \Pi^{-1}_n(u \mid X) = \sum_{j=1}^K \widehat \Pi_m^{-1} (u \mid X_j)/K$ for all $u\in (0,1)$. Let $L_2\{\Pi_m(\cdot \mid X_j)\}$ for $j=1,\ldots,K$ be the $L_2$ space of functions on $\Theta$ such that for any $f\in L_2\{\Pi_m(\cdot \mid X_j)\}$, $\|f(\theta)\|^2_{L_2,j}=E_{\Pi_m(\cdot \mid X_j)}f^2(\theta) < \infty$ almost surely in $P_{\theta_0}$. We need three additional assumptions as follows.

\begin{assumption}\label{a8}
The $j$th subset posterior $\Pi_m(\theta \mid X_j)$ is the unique stationary distribution that satisfies the balance condition $\pi_m(\theta'\mid X_j) = \int_{\Theta} \pi_m(\theta\mid X_j)\kappa_j(\theta,\theta') \ud \theta$ for any $\theta,\theta'\in \Theta$, where $\pi_m(\theta\mid X_j)$ is the density of $\Pi_m(\theta \mid X_j)$. Furthermore, the Markov chain of each subset posterior is reversible with the detailed balance condition $\pi_m(\theta \mid X_j)\kappa_j(\theta,\theta') = \pi_m(\theta'\mid X_j) \kappa_j(\theta',\theta)$ for any $\theta,\theta'\in \Theta$ and all $j=1,\ldots,K$.
\end{assumption}
\begin{assumption}\label{a9}
$\max_{1\leq j\leq K}E_{\Pi_m(\cdot \mid X_j)} \|\theta\|^7$ is upper bounded by a constant almost surely in $P_{\theta_0}$. $~\max_{1\leq j\leq K} E_{\Pi_m(\cdot\mid X_j)} \{\pi^\circ_j( \theta) / \pi_m( \theta\mid X_j)\}^3$ is upper bounded by a constant almost surely in $P_{\theta_0}$, where $\pi^\circ_j(\theta)$ is the density of $\Pi^\circ_j(\theta)$ for $j=1,\ldots,K$.
\end{assumption}
\begin{assumption}\label{a10}
Every subset posterior $\Pi_m(\theta\mid X_j)$ ($j=1,\ldots,K$) is $\rho$-mixing: there exists a nonnegative constant sequence $\{\rho_{l}\}_{l\geq 1}$ decreasing to zero and $\sum_{l=1}^\infty \rho_{l}<\infty$, such that almost surely in $P_{\theta_0}$, for any integer $l\geq 1$, any $f\in L_2\{\Pi_m(\cdot \mid X_j)\}$ and all $j=1,\ldots,K$,
$$\left\|E_{\kappa_j^l(\cdot\mid \theta_{1j}=\theta)} f(\theta_{l+1,j}) - E_{\Pi_m(\cdot\mid X_j)} f(\theta)\right\|_{L_2,j} \leq \rho_l \left\|f(\theta) -E_{\Pi_m(\cdot\mid X_j)} f(\theta)\right\|_{L_2,j},$$
where $\theta_{l+1,j}$ is the $l$th draw in the Markov chain with initial draw $\theta_{1j}$, and $E_{\kappa_j^l(\cdot\mid \theta_{1j}=\theta)}$ is the conditional distribution of $\theta_{l+1,j}$ given $\theta_{1j}=\theta$.
\end{assumption}
Then the following theorem accounts for the Monte Carlo error in the empirical version of Wasserstein posterior due to finite sample approximations.
\begin{theorem}\label{cormix}
Suppose Assumptions 1--10 hold. Then for two arbitrary fixed numbers $0<u_1<u_2<1$,
\begin{align*}
& W_2\left\{ \widehat \Pi_{n}\left(\xi \mid  X\right),\Pi_{n}\left(\xi \mid  X\right)\right\} = O_p\left(m^{-1/2}\right) + O_p\left(T^{-1/4}\right); \\
& \bias\left\{\widehat \Pi_{n}(\xi\mid  X)\right\} - \bias\left\{\Pi_{n}(\xi\mid  X)\right\} =  o_p\left(m^{-1/2}\right) + O_p\left(T^{-1/2}\right); \\
& \var\left\{\widehat \Pi_{n}(\xi\mid  X)\right\} - \var\left\{\Pi_{n}(\xi\mid  X)\right\} =o_p(n^{-1}) + O_p\left(T^{-1/2}\right); \\
& \sup_{u\in [u_1,u_2]}\left|\widehat \Pi_n^{-1}(u\mid  X) - \Pi_n^{-1}(u\mid  X)\right| = o_p\left(m^{-1/2}\right) +  O_p\left(T^{-1/2}\right),
\end{align*}
where $O_p$ and $o_p$ are in $P_{\theta_0}$-probability.  Furthermore, if $\hat\theta_1$ is an unbiased estimator of $\theta_0$, then
\begin{align*}
& W_2\left\{ \widehat \Pi_{n}\left(\xi \mid  X\right),\Pi_{n}\left(\xi \mid  X\right)\right\} = O_p\left(n^{-1/2}\right) + O_p\left(T^{-1/4}\right); \\
& \bias\left\{\widehat \Pi_{n}(\xi\mid  X)\right\} - \bias\left\{\Pi_{n}(\xi\mid  X)\right\} =  o_p\left(n^{-1/2}\right) + O_p\left(T^{-1/2}\right); \\
& \sup_{u\in [u_1,u_2]}\left|\widehat \Pi_n^{-1}(u\mid  X) - \Pi_n^{-1}(u\mid  X)\right| = o_p\left(n^{-1/2}\right) +  O_p\left(T^{-1/2}\right).
\end{align*}
\end{theorem}

\noindent {\bf Proof of Theorem 3:}\\
In this proof, we first establish the key relations between the empirical distribution $\widehat \Pi_m(\xi \mid X_j)$ and the exact continuous subset posterior $\Pi_m(\xi \mid X_j)$, using the recent results from \citep{FouGui15}. Given the linear relation $\xi=a^\T \theta+b$ and all the assumptions in Theorem 3,
\begin{align}\label{fgw2}
& E_{\Pi^\circ_j} \left[W_{1+\delta} \left\{\widehat \Pi_m(\xi \mid X_j), \Pi_m(\xi \mid X_j)\right\}\right]^{1+\delta}
\leq C_1 T^{-1/2}
\end{align}
almost surely in $P_{\theta_0}$ for all $j=1,\ldots,K$, where $0\leq \delta\leq 1$, $C_1$ is a constant that only depends on the sequence $\{\rho_l\}_{\l\geq 1}$, the constant upper bound of $\max_{1\leq j\leq K}E_{\Pi_m(\cdot \mid X_j)} \|\theta\|^7$, and the constant upper bound of $\max_{1\leq j\leq K} E_{\Pi_m(\cdot\mid X_j)} \{\pi^\circ_j( \theta) / \pi_m( \theta\mid X_j)\}^3$ in Assumption 9. The expectation in \eqref{fgw2} is taken with respect to $\Pi^\circ_j$ because the first posterior sample $\theta_{1j}$ is drawn from the initial distribution $\Pi^\circ_j$. Given Assumptions 8-10, the inequality \eqref{fgw2} is the consequence of Theorem 15 of \citep{FouGui15} by setting their $d=1,~p=1+\delta,~r=3,~q=7$.


For the empirical Wasserstein barycenter $\widehat \Pi_n(\xi\mid X)$, we can establish a similar inequality to Lemma \ref{w2ineq}: for any $l\geq 1$,
\begin{align}\label{w2hatineq}
W_l\left(\widehat \Pi_n(\xi\mid X),\Phi\left[\xi; \overline \xi, \left\{ nI_{\xi}(\theta_0)\right\}^{-1} \right] \right) \leq \frac{1}{K} \sum_{j=1}^K W_l\left (\widehat \Pi_m(\xi\mid X_j),\Phi\left[\xi ; \hat \xi_j,  \left\{ n I_{\xi}(\theta_0)\right\}^{-1} \right]\right),
\end{align}
where $\hat\xi$ is defined in Lemma \ref{w2ineq}. Therefore, taking $l=2$ in \eqref{w2hatineq}, we obtain that
\begin{align*}
&E_{P_{\theta_0}} E_{\Pi^\circ_1,\ldots,\Pi^\circ_K}W^2_2\left(\widehat \Pi_n(\xi\mid X),\Phi\left[\xi; \overline \xi, \left\{ nI_{\xi}(\theta_0)\right\}^{-1} \right] \right) \\
\overset{(i)}{\leq}{}& \frac{1}{K} \sum_{j=1}^K E_{P_{\theta_0}}E_{\Pi^\circ_j} W^2_2\left(\widehat \Pi_m(\xi\mid X_j),\Phi\left[\xi ; \hat \xi_j,  \left\{ n I_{\xi}(\theta_0)\right\}^{-1} \right]\right) \\
\overset{(ii)}{\leq}{}& \frac{2}{K} \sum_{j=1}^K E_{P_{\theta_0}}E_{\Pi^\circ_j} W^2_2\left\{\widehat \Pi_m(\xi\mid X_j), \Pi_m(\xi\mid X_j)\right\} \\
& + \frac{2}{K} \sum_{j=1}^K E_{P_{\theta_0}} W^2_2\left( \Pi_m(\xi\mid X_j), \Phi\left[\xi ; \hat \xi_j,  \left\{ n I_{\xi}(\theta_0)\right\}^{-1} \right]\right) \\
\overset{(iii)}{=}{}& O(T^{-1/2}) + o(n^{-1}),
\end{align*}
where (i) is from the relation between $l_1$ and $l_2$ norms, (ii) is from the triangular inequality of the $W_2$ distance and $(x_1+x_2)^2\leq 2(x_1^2+x_2^2)$ for $x_1,x_2\in \mathcal{R}$, and (iii) follows from \eqref{rootn2} and \eqref{fgw2} with $\delta=1$. By Markov's inequality, it is clear that $W_2\left(\widehat \Pi_n(\xi\mid X),\Phi\left[\xi; \overline \xi, \left\{ nI_{\xi}(\theta_0)\right\}^{-1} \right] \right) = o_p(n^{-1/2})+O_p(T^{-1/4})$.

We can also take $l=1$ in \eqref{w2hatineq} and obtain that
\begin{align}\label{w1hat1}
& W_1\left(\widehat \Pi_n(\xi\mid X),\Phi\left[\xi; \overline \xi, \left\{ nI_{\xi}(\theta_0)\right\}^{-1} \right] \right)  \nonumber \\
={}& \int_0^1 \left|\frac{1}{K}\sum_{j=1}^K \left[\widehat \Pi_m^{-1} (u\mid X_j) - \hat \xi_j - \{nI_{\xi}(\theta_0)\}^{-1/2} \Phi^{-1}(u)\right] \right| \ud u \nonumber \\
={}& \int_0^1 \left|\frac{1}{K}\sum_{j=1}^K \{\hat r_j(u) + r_j(u)\}\right| \ud u,
\end{align}
where $r_j(u)$ is defined in \eqref{rj} and $\hat r_j(u) = \widehat \Pi_m^{-1} (u\mid X_j) - \Pi_m^{-1} (u\mid X_j)$.

For the bias of $\widehat \Pi_n(\xi\mid X)$, we have
\begin{align}
& \left|\bias \{\widehat \Pi_n(\xi\mid X)\} - \bias \{\overline \Pi_n(\xi\mid X)\}\right|
=  \left|E_{\widehat \Pi_n(\cdot\mid X)}(\xi) - E_{\overline \Pi_n(\xi\mid X)}(\xi)\right| \nonumber \\
& = \left|\int_0^1 \widehat \Pi^{-1}_n(u\mid X) \ud u - \int_0^1 \overline \Pi^{-1}_n(u\mid X) \ud u\right| \leq \int_0^1 \left|\frac{1}{K}\sum_{j=1}^K \hat r_j(u) \right|\ud u \nonumber \\
& \leq \frac{1}{K} \sum_{j=1}^K \int_0^1 \left|\hat r_j(u)\right|\ud u
= \frac{1}{K}\sum_{j=1}^K W_1 \left\{\widehat \Pi_m(\xi \mid X_j), \Pi_m(\xi \mid X_j)\right\}.
\end{align}
By Markov's inequality and \eqref{fgw2} with $\delta=0$, for any $c>0$,
\begin{align*}
& P\left[\frac{1}{K}\sum_{j=1}^K W_1 \left\{\widehat \Pi_m(\xi \mid X_j), \Pi_m(\xi \mid X_j)\right\} >c T^{-1/2} \right] \\
\leq {}& \frac{\frac{1}{K}\sum_{j=1}^K E_{P_{\theta_0}} E_{\Pi^\circ_j} W_1 \left\{\widehat \Pi_m(\xi \mid X_j), \Pi_m(\xi \mid X_j)\right\} }{cT^{-1/2}} \leq \frac{C_1}{c}.
\end{align*}
Therefore, we have shown that
\begin{align}\label{rjhat1}
&\left|\bias \{\widehat \Pi_n(\xi\mid X)\} - \bias \{\overline \Pi_n(\xi\mid X)\}\right|  = O_p(T^{-1/2}), \nonumber \\
& \frac{1}{K} \sum_{j=1}^K \int_0^1 \left|\hat r_j(u)\right|\ud u = O_p(T^{-1/2}).
\end{align}
Together with Theorem 2, we conclude that
\begin{align*}
& \left|\bias \{\widehat \Pi_n(\xi\mid X)\} - \bias \{\Pi_n(\xi\mid X)\}\right| \\
\leq{} & \left|\bias \{\overline \Pi_n(\xi\mid X)\} - \bias \{\Pi_n(\xi\mid X)\}\right| + \left|\bias \{\widehat \Pi_n(\xi\mid X)\} - \bias \{\overline \Pi_n(\xi\mid X)\}\right| \\
={}& O_p(m^{-1/2}) + O_p(T^{-1/2}).
\end{align*}
Furthermore, if $\hat\theta_1$ is unbiased for $\theta$, then
\begin{align*}
& \left|\bias \{\widehat \Pi_n(\xi\mid X)\} - \bias \{\Pi_n(\xi\mid X)\}\right| = O_p(n^{-1/2}) + O_p(T^{-1/2}).
\end{align*}
The results for quantiles can be derived similarly and therefore the proofs are omitted here.

Next we derive the rate for the posterior variance of $\widehat \Pi_n(\xi\mid X)$. Similar to the derivation in the proof of Theorem 2(ii), we can obtain the following equality:
\begin{align}\label{varhat1}
&\var\left\{\widehat \Pi_n(\xi\mid X)\right\} \nonumber \\
&= \frac{1}{nI_{\xi}(\theta_0)}  + \int_0^1 \left[\frac{1}{K}\sum_{j=1}^K \{\hat r_j(u)+r_j(u)\}\right]^2 \ud u -
\left[\frac{1}{K}\sum_{j=1}^K \int_0^1 \{\hat r_j(u)+r_j(u)\}\ud u\right]^2 \nonumber  \\
& + 2\left\{nI_{\xi}(\theta_0)\right\}^{-1/2} \int_0^1 \Phi^{-1}(u) \frac{1}{K}\sum_{j=1}^K \{\hat r_j(u)+r_j(u)\} \ud u.
\end{align}
We bound the last three terms in the display above. It is clear that by Cauchy-Schwarz inequality, the third term is upper bounded by the second term. For the second term, we have
\begin{align}\label{varhat2}
& \int_0^1 \left[\frac{1}{K}\sum_{j=1}^K \{\hat r_j(u)+r_j(u)\}\right]^2 \ud u \nonumber\\
\leq{}&  2\int_0^1 \left\{\frac{1}{K}\sum_{j=1}^K \hat r_j(u)\right\}^2\ud u  +2\int_0^1 \left\{\frac{1}{K}\sum_{j=1}^K r_j(u)\right\}^2\ud u \nonumber\\
\leq{}& 2W_2^2\left\{\widehat \Pi_m(\xi \mid X_j), \overline \Pi_m(\xi \mid X_j)\right\} + 2W_2^2\left( \overline \Pi_n(\xi\mid X),\Phi\left[\xi ; \overline \xi, \left\{nI_{\xi}(\theta_0)\right\}^{-1} \right]  \right) \nonumber\\
={}& O_p(T^{-1/2}) + o_p(n^{-1}),
\end{align}
where the last relation follows from \eqref{f11} and applying Markov's inequality to \eqref{fgw2}.

For the last term in \eqref{varhat1}, we have the following bound:
\begin{align}\label{varhat3}
& \left|2\left\{nI_{\xi}(\theta_0)\right\}^{-1/2} \int_0^1 \Phi^{-1}(u) \frac{1}{K}\sum_{j=1}^K \{\hat r_j(u)+r_j(u)\}\ud u\right| \nonumber \\
\leq{}& 2 \left\{nI_{\xi}(\theta_0)\right\}^{-1/2} \left\{\int_0^1 \Phi^{-1}(u)\left|\frac{1}{K}\sum_{j=1}^K \hat r_j(u)\right|\ud u + \int_0^1 \Phi^{-1}(u)\left|\frac{1}{K}\sum_{j=1}^K r_j(u)\right|\ud u \right\} \nonumber \\
\leq {}& 2 \left\{nI_{\xi}(\theta_0)\right\}^{-1/2} \Bigg( \left[\int_0^1 \left\{\Phi^{-1}(u)\right\}^{1+1/\delta}\ud u\right]^{\delta/(1+\delta)} \Bigg\{\int_0^1 \Big|\frac{1}{K}\sum_{j=1}^K \hat r_j(u)\Big|^{1+\delta}\ud u\Bigg\}^{1/(1+\delta)} \nonumber \\
& + \left[\int_0^1 \left\{\Phi^{-1}(u)\right\}^2\ud u\right]^{1/2} \Bigg\{\int_0^1 \Big|\frac{1}{K}\sum_{j=1}^K r_j(u)\Big|^2\ud u \Bigg\}^{1/2} \Bigg) \nonumber \\
\leq {}& 2 \left\{nI_{\xi}(\theta_0)\right\}^{-1/2}  \Bigg\{ \frac{2^{(1+\delta)/2}}{\sqrt{\pi}}\Gamma\left(\frac{1+\delta}{2}\right) W_{1+\delta}\left\{\widehat \Pi_n(\xi\mid X), \overline \Pi_n(\xi\mid X)\right\} \nonumber \\
& + W_2\left(\overline \Pi_n(\xi \mid X), \Phi\left[\xi ; \overline \xi, \left\{nI_{\xi}(\theta_0)\right\}^{-1} \right]\right) \Bigg\} \nonumber \\
= {}&\frac{2}{\sqrt{\pi}}  \left\{nI_{\xi}(\theta_0)\right\}^{-1/2} 2^{(1+\delta)/2} \Gamma\left(\frac{1+\delta}{2}\right) W_{1+\delta}\left\{\widehat \Pi_n(\xi\mid X), \overline \Pi_n(\xi\mid X)\right\} + o_p(n^{-1}),
\end{align}
where in the second inequality we used the H\"{o}lder's inequality and the Cauchy-Schwarz inequality, and the last step is from Theorem 1. By \eqref{fgw2}, almost surely in $P_{\theta_0}$,
\begin{align}\label{varhat4}
& E_{\Pi^\circ_1,\ldots,\Pi^\circ_K} W_{1+\delta}\left\{\widehat \Pi_n(\xi\mid X), \overline \Pi_n(\xi\mid X)\right\}
\leq \frac{1}{K}\sum_{j=1}^K E_{\Pi^\circ_j} W_{1+\delta} \left\{\widehat \Pi_m(\xi \mid X_j), \Pi_m(\xi \mid X_j)\right\} \nonumber \\
&\overset{(i)}{\leq} \frac{1}{K}\sum_{j=1}^K \left(E_{\Pi^\circ_j} \left[W_{1+\delta} \left\{\widehat \Pi_m(\xi \mid X_j), \Pi_m(\xi \mid X_j)\right\}\right]^{1+\delta}\right)^{1/(1+\delta)}
\leq C_1 T^{-1/\{2(1+\delta)\}},
\end{align}
where (i) is from $0\leq \delta\leq 1$ and Jensen's inequality. Now we set $\delta=\min\{1,\log n/(2\log T)\}$ and derive from \eqref{varhat4} that
\begin{align}
& E_{\Pi^\circ_1,\ldots,\Pi^\circ_K} \frac{2}{\sqrt{\pi}}  \left\{nI_{\xi}(\theta_0)\right\}^{-1/2} 2^{(1+\delta)/2} \Gamma\left(\frac{1+\delta}{2}\right) W_{1+\delta}\left\{\widehat \Pi_n(\xi\mid X), \overline \Pi_n(\xi\mid X)\right\} \nonumber \\
\leq{}& \frac{2}{\sqrt{\pi}}  \left\{nI_{\xi}(\theta_0)\right\}^{-1/2} \times 2\Gamma(1) \times C_1 T^{-1/\{2(1+\delta)\}} \nonumber \\
\leq{}& 4C_1 \left\{I_{\xi}(\theta_0)\right\}^{-1/2} \exp \left\{-\frac{1}{2}\log n - \frac{1}{2(1+\delta)} \log T\right\} \nonumber \\
={}& 4C_1 \left\{I_{\xi}(\theta_0)\right\}^{-1/2} \exp \left\{ -\frac{1}{2}\log T -\frac{1}{2}\log n + \frac{\delta}{2(1+\delta)} \log T \right\} \nonumber \\
\leq&{} 4C_1 \left\{I_{\xi}(\theta_0)\right\}^{-1/2} T^{-1/2} \exp \left( -\frac{1}{2}\log n + \frac{\delta}{2} \log T\right) \nonumber \\
\leq&{} 4C_1 \left\{I_{\xi}(\theta_0)\right\}^{-1/2} T^{-1/2} \exp \left( -\frac{1}{4}\log n \right) = o(T^{-1/2}). \nonumber
\end{align}
Hence, by Markov's inequality, the right-hand side of \eqref{varhat3} can be bounded by
\begin{align}\label{varhat4}
& \left|2\left\{nI_{\xi}(\theta_0)\right\}^{-1/2} \int_0^1 \Phi^{-1}(u) \frac{1}{K}\sum_{j=1}^K \{\hat r_j(u)+r_j(u)\}\ud u\right| = o_p(T^{-1/2}) + o_p(n^{-1}).
\end{align}
Now we combine \eqref{varhat1}, \eqref{varhat2} and \eqref{varhat4} and conclude that
\begin{align*}
&\var\left\{\widehat \Pi_n(\xi\mid X)\right\} = \frac{1}{nI_{\xi}(\theta_0)}  +  O_p(T^{-1/2}) + o_p(n^{-1}) + o_p(T^{-1/2}) + o_p(n^{-1})\\
& = \frac{1}{nI_{\xi}(\theta_0)} + o_p(n^{-1}) + O_p(T^{-1/2}).
\end{align*}
If we compare this with the results in Theorem 2, we obtain that
\begin{align*}
& \var\left\{\widehat \Pi_{n}(\xi\mid  X)\right\} - \var\left\{\Pi_{n}(\xi\mid  X)\right\} = o_p(n^{-1}) + O_p\left(T^{-1/2}\right).
\end{align*}
This concludes the proof of Theorem 3.
\hfill $\blacksquare$

\section{Justification of Assumption 7}
In this section, we verify Assumption 7 for two special examples: the normal linear model and some exponential family distributions. Without loss of generality, all the samples considered in this section refer to the first subset sample $X_1$ in Assumption 7.\\

\noindent {1. Normal Linear Model}\\

We consider the following normal linear model based on independent and identically distributed observations:
\begin{align}\label{nlm}
&y_i =Z_i^\T\beta+\varepsilon_i,~ \varepsilon_i \sim \mathcal{N}\left(0, \sigma^2\right),~ i=1,\ldots,m,
\end{align}
where $\dimm(\beta)=p$ and $\varepsilon_i$'s are independent. We write $y=(y_1,\ldots,y_m)^\T$, $Z=(Z_1,\ldots,Z_m)^\T$, $\varepsilon=(\varepsilon_1,\ldots,\varepsilon_m)^\T$, and the true parameter is $\theta_0=(\beta_0^\T,\sigma_0^2)^\T$. We impose the following conjugate prior on the parameter $\theta=(\beta^\T,\sigma^2)^\T$:
\begin{align} 
&\beta \Big| \sigma^2,\mu^*,\Omega \sim \mathcal{N} \left(\mu^*, \sigma^2\Omega\right), \nonumber\\
&\sigma^2 \Big| a,b  \sim \text{Inverse-Gamma} \left(a/2,b/2\right),\nonumber
\end{align}
where $a>4,b>0$ is to guarantee a finite variance for the prior of $\sigma^2$, and $\Omega$ is a positive definite matrix.  The subset posterior after the stochastic approximation is given by
\begin{align*}
\pi_m\left(\beta,\sigma^2 \Big| y,Z\right) \propto & ~(\sigma^2)^{-Km/2}\exp\left\{-\frac{K(y-Z\beta)^\T(y-Z\beta)}{2\sigma^2}\right\}\times\\
&~ \exp\left\{-\frac{(\beta-\mu^*)^\T \Omega^{-1}(\beta-\mu^*)}{2\sigma^2}\right\}
\times (\sigma^2)^{-a/2-1}\exp\left\{-\frac{b}{2\sigma^2}\right\}
\end{align*}
We have the following proposition, which shows that the $\psi(\cdot)$ function in Assumption 7 is $L_1$-integrable uniformly for all $m$ and $K$, which implies the uniform integrability condition.\\

\begin{proposition}\label{nlma7}
In the normal linear model \eqref{nlm}, assume that $\|\mu^*\|$ is upper bounded by a constant. Assume that the eigenvalues of $\Omega$ and $Z^{\T} Z/m$ are lower and upper bounded by constants for all $m\geq 2$. Assume that the error $\varepsilon_i$ in \eqref{nlm} has finite 4th moment. Let $\widehat\beta$ and $\widehat{\sigma^2} $ be the maximum likelihood estimators of $\beta$ and $\sigma^2$ respectively. Then
\begin{align}
\label{betabound} & \sup_{m\geq 2,K\geq 1 }E_{P_{\theta_0}}E_{\Pi_m(\cdot|y,Z)}Km\|\beta-\widehat\beta\|^2 <+\infty,\\
\label{sigmabound} & \sup_{m\geq 2,K\geq 1 }E_{P_{\theta_0}}E_{\Pi_m(\cdot|y,Z)}Km\|\sigma^2-\widehat{\sigma^2}\|^2<+\infty.
\end{align}
\end{proposition}

\noindent {\bf Proof of Proposition \ref{nlma7}:}\\
Let $\|\beta_0\|,\|\mu^*\|\leq c_1<+\infty$. Let the eigenvalues of $\Omega$ and $Z^\T Z/m$ be lower bounded by $c_2>0$ and upper bounded by $c_3>0$. Let $E(\varepsilon_i^4)=c_4<+\infty$.
The subset posterior distributions of $\beta$ and $\sigma^2$ are given by
\begin{align*}
&\beta\Big | y,Z, \mu^*, \Omega,a,b~ \sim ~ \text{Multi-}t_{a+Km+p}\left\{\beta^*, \frac{b^*}{a+Km}\left(KZ^\T Z+\Omega^{-1}\right)^{-1}\right\},\\
&\sigma^2 \Big| y,Z, \mu^*, \Omega,a,b ~\sim ~  \text{Inverse-Gamma} \left(\frac{a+Km}{2},\frac{b^*}{2}\right),\\
& \beta^* = \left(KZ^\T Z+\Omega^{-1}\right)^{-1}\left( KZ^\T y+\Omega^{-1}\mu^*\right),\\
& b^*= b+ \mu^{*\T}\Omega^{-1}\mu^*+ Ky^\T \left\{I_m-KZ\left(KZ^\T Z+\Omega^{-1}\right)^{-1}Z^\T\right\}y,
\end{align*}
where $\text{Multi-}t_{\nu}(\mu,\Sigma)$ denotes the multivariate-t distribution with mean $\mu$, variance matrix $\Sigma$, and $\nu$ degrees of freedom.

The maximum likelihood estimators of $\beta$ and $\sigma^2$ are given by
\begin{align*}
\widehat\beta &= (Z^\T Z)^{-1}Z^\T y, \\
\widehat{\sigma^2} &= m^{-1}\|y-Z^\T\beta\|^2 = m^{-1}y^\T \left\{I_m-Z(Z^\T Z)^{-1}Z^\T\right\}y.
\end{align*}
We first prove \eqref{betabound}. It is clear that
\begin{align}\label{betab0}
E_{P_{\theta_0}}E_{\Pi_m(\cdot|y,Z)}Km\|\beta-\widehat{\beta}\|^2 =
KmE_{P_{\theta_0}}\tr\left\{{\var}_{\pi_m(\cdot|y,Z)}(\beta)\right\}  +
KmE_{P_{\theta_0}}\|E_{\Pi_m(\cdot|y,Z)}\beta - \widehat{\beta}\|^2,
\end{align}
where $\tr(A)$ denotes the trace of a generic square matrix $A$. The posterior variance of $\beta$ can be bounded as
\begin{align}\label{betab1}
& Km E_{P_{\theta_0}}\tr\left\{{\var}_{\pi_m(\cdot|y,Z)}(\beta)\right\} \nonumber \\
&= Km \frac{a+Km+p}{a+Km+p-2} \times\tr\left\{ E_{P_{\theta_0}}\frac{b^*}{a+Km} \left(KZ^\T Z+\Omega^{-1}\right)^{-1}\right\} \nonumber \\
&\leq 2\tr\left\{ E_{P_{\theta_0}}\left(b+c_1^2c_2^{-1} + Ky^\T y \right) \left(KZ^\T Z+\Omega^{-1}\right)^{-1}\right\}  \nonumber\\
&\leq 2 \tr\left\{ E_{P_{\theta_0}}\left(b+c_1^2c_2^{-1} + Kmc_1^2 c_2 + Km \sigma_0^2\right) \left(Kmc_2I_p+c_3^{-1}I_p\right)^{-1}\right\} \nonumber \\
&= 2p\frac{Km(c_1^2 c_2 + \sigma_0^2)+b+c_1^2c_2^{-1}}{Kmc_2+c_3^{-1}} \to \frac{2p(c_1^2 c_2 + \sigma_0^2)}{c_2}  ~\text{ as } m\to \infty.
\end{align}
The second term in \eqref{betab0} can be bounded as
\begin{align}\label{betab2}
&KmE_{P_{\theta_0}}\|E_{\Pi_m(\cdot|y,Z)}\beta - \widehat{\beta}\|^2 \nonumber \\
&= KmE_{P_{\theta_0}}\left\|\left\{\left(KZ^\T Z+\Omega^{-1}\right)^{-1}-\left(KZ^\T Z\right)^{-1}\right\} (KZ^\T y) + \left(KZ^\T Z+\Omega^{-1}\right)^{-1}\Omega^{-1}\mu^{*}\right\|^2 \nonumber \\
&\leq 2Km E_{P_{\theta_0}}\left\|\left\{\left(KZ^\T Z+\Omega^{-1}\right)^{-1}-\left(KZ^\T Z\right)^{-1}\right\} (KZ^\T y)\right\|^2 \nonumber \\
& ~~ + 2Km \left\|\left(KZ^\T Z+\Omega^{-1}\right)^{-1}\Omega^{-1}\mu^{*}\right\|^2 \nonumber \\
&\leq  2Km E_{P_{\theta_0}}  \left\|\left(KZ^\T Z+\Omega^{-1}\right)^{-1}\Omega^{-1}(Z^\T Z)^{-1}(Z^\T y)\right\|^2 \nonumber \\
& ~~ + 2Km \left\|\left(Kmc_2I_p+c_3^{-1}I_p\right)^{-1}c_2^{-1}c_1\right\|^2 \nonumber \\
&\leq  2Km E_{P_{\theta_0}}  \left\|\left(Kmc_2+c_3^{-1}\right)^{-1}c_2^{-1}(Z^\T Z)^{-1}(Z^\T y)\right\|^2 + \frac{2c_1^2}{Kmc_2^4}  \nonumber \\
&\leq \frac{2Km}{(Kmc_2+c_3^{-1})^2c_2^2} \left\{\left\|\beta_0\right\|^2 + E_{P_{\theta_0}}\left\|(Z^\T Z)^{-1}(Z^\T\varepsilon)\right\|^2 \right\} + \frac{2c_1^2}{Kmc_2^4}  \nonumber \\
&\leq \frac{2Km}{(Kmc_2+c_3^{-1})^2c_2^2} \left(c_1^2 + c_2^{-2}c_3 \sigma_0^2 \right) + \frac{2c_1^2}{Kmc_2^4} \to 0 ~\text{ as } m\to \infty.
\end{align}
Since \eqref{betab1} and \eqref{betab2} have finite limits as $m\to\infty$, they are both bounded by constants, regardless of the value of $K$. They together with \eqref{betab0} lead to \eqref{betabound}.

Next we prove \eqref{sigmabound}. We have the similar decomposition
\begin{align}\label{sigma0}
E_{P_{\theta_0}}E_{\Pi_m(\cdot|y,Z)}Km\|\sigma^2-\widehat{\sigma^2}\|^2 =~&
KmE_{P_{\theta_0}}{\var}_{\pi_m(\cdot|y,Z)}(\sigma^2)  \nonumber \\
&+ KmE_{P_{\theta_0}}\|E_{\Pi_m(\cdot|y,Z)}\sigma^2  - \widehat{\sigma^2}\|^2.
\end{align}
We show an useful bound for the square of $y^\T y$:
\begin{align}\label{yb2}
&E_{P_{\theta_0}}\left(y^\T y\right)^2  = E_{P_{\theta_0}}\left(\left\|Z\beta_0+\varepsilon\right\|^2\right)^2 \nonumber\\
&\leq 4E_{P_{\theta_0}} \left(\left\|Z\beta_0\right\|^2+\left\|\varepsilon\right\|^2\right)^2 \leq 4E_{P_{\theta_0}} \left\{\beta_0^\T (Z^\T Z)\beta_0 + \left\|\varepsilon\right\|^2 \right\}^2 \nonumber\\
&\leq 4E_{P_{\theta_0}} \left(mc_1^2c_3 + \left\|\varepsilon\right\|^2 \right)^2 \leq 8m^2c_1^4c_3^2 + 8E_{P_{\theta_0}} \left(\sum_{i=1}^m\varepsilon_i^2\right)^2 \nonumber \\
&\leq 8m^2c_1^4c_3^2 + 8m E_{P_{\theta_0}} \sum_{i=1}^m\varepsilon_i^4 \leq 8m^2(c_1^4c_3^2 +c_4).
\end{align}
By using \eqref{yb2}, the first term in \eqref{sigma0} can be bounded as
\begin{align}\label{sigma1}
&Km E_{P_{\theta_0}}{\var}_{\pi_m(\cdot|y,Z)}(\sigma^2) \leq Km E_{P_{\theta_0}}\frac{ b^{*2}/4}{\left\{(Km+a)/2-2\right\}^3} \nonumber \\
& \leq \frac{2}{(Km)^2} E_{P_{\theta_0}} \left[b+ \mu^{*\T}\Omega^{-1}\mu^*+ Ky^\T \left\{I_m-KZ\left(KZ^\T Z+\Omega^{-1}\right)^{-1}Z^\T\right\}y\right]^2 \nonumber \\
& \leq \frac{2}{(Km)^2} E_{P_{\theta_0}} \left(b+ c_1^2 c_2^{-1} + K y^\T y \right)^2   \nonumber \\
& \leq \frac{4(b+ c_1^2 c_2^{-1})^2}{(Km)^2} + \frac{4}{m^2}  E_{P_{\theta_0}} \left(y^\T y\right)^2 \nonumber \\
& \leq \frac{4(b+ c_1^2 c_2^{-1})^2}{(Km)^2} + 32(c_1^4c_3^2 +c_4) \to 32(c_1^4c_3^2 +c_4) ~\text{ as } m\to \infty.
\end{align}
And the second term in \eqref{sigma0} can be bounded as
\begin{align}\label{sigma2}
&Km E_{P_{\theta_0}}\left\|E_{\Pi_m(\cdot|y,Z)}\sigma^2  - \widehat{\sigma^2}\right\|^2 = Km E_{P_{\theta_0}}\left\|\frac{b^*/2}{(a+Km)/2-1}  - \widehat{\sigma^2}\right\|^2 \nonumber \\
& = Km E_{P_{\theta_0}}\Bigg\| \frac{b+\mu^{*\T}\Omega^{-1}\mu^*}{Km+a-2} - \frac{(a-2)y^\T\left\{I_m-Z(Z^\T Z)^{-1}Z^\T\right\}y}{(Km+a-2)m} \nonumber \\
& + \frac{Ky^\T\left\{Z(Z^\T Z)^{-1}Z^\T-KZ\left(KZ^\T Z+\Omega^{-1}\right)^{-1}Z^\T \right\}y}{Km+a-2} \Bigg\|^2 \nonumber \\
& \leq \frac{3Km\left(b+\mu^{*\T}\Omega^{-1}\mu^*\right)^2}{(Km+a-2)^2} + \frac{3(a-2)}{(Km+a-2)^2} E_{P_{\theta_0}} \left[m^{-1}y^\T\left\{I_m-Z(Z^\T Z)^{-1}Z^\T\right\}y\right]^2 \nonumber \\
& + \frac{3Km}{(Km+a-2)^2} E_{P_{\theta_0}}\left\{ y^\T Z (Z^\T Z+\Omega^{-1}/K)^{-1}\Omega^{-1}(Z^\T Z)^{-1}Z^\T y\right\}^2 \nonumber \\
& \leq \frac{3(b+c_1^2c_2^{-1})^2}{Km} +  \frac{3(a-2)}{(Km+a-2)^2} E_{P_{\theta_0}} \left(m^{-1}y^\T y\right)^2 \nonumber \\
& + \frac{3}{Km (mc_2+c_3^{-1}/K)^2 c_2^2 mc_2^2} E_{P_{\theta_0}}\left(y^\T Z Z^\T y\right)^2 \nonumber \\
& \leq \frac{3(b+c_1^2c_2^{-1})^2}{Km} +  \frac{24(a-2)(c_1^4c_3^2 + c_4)}{(Km+a-2)^2} + \frac{3pmc_3}{Km^4c_2^6} E_{P_{\theta_0}}\left(y^\T y\right)^2  \nonumber \\
& \leq \frac{3(b+c_1^2c_2^{-1})^2}{Km} +  \frac{24(a-2)(c_1^4c_3^2 + c_4)}{(Km+a-2)^2} + \frac{24p c_3(c_1^4c_3^2 + c_4)}{Kmc_2^6} \to 0 ~\text{ as } m\to \infty,
\end{align}
where we have used the relation $\overline \lambda(ZZ^\T) \leq \tr(ZZ^\T) = \tr(Z^\T Z)\leq p\overline\lambda(Z^\T Z) \leq pmc_3$, and $\overline\lambda(A)$ denotes the largest eigenvalue of a generic matrix $A$.  Since \eqref{sigma1} and \eqref{sigma2} have finite limits as $m\to\infty$, they are both bounded by constants, regardless of the value of $K$. They together with \eqref{sigma0} lead to \eqref{sigmabound}.
\hfill $\blacksquare$

\vspace{.5cm}
\noindent {2. Some Exponential Family Models}\\

In this section, we verify Assumption 7 for the following three commonly used exponential family distributions: Poisson, exponential, and binomial.
\begin{proposition}\label{3expex}
(i) Suppose the data $y_i$ ($i=1,\ldots,m$) are independent and identically distributed as $\text{Poisson}(\theta)$ with the probability mass function $p(y|\theta) = \theta^y e^{-\theta}/y!$ and the true parameter $\theta_0$. Suppose the prior on $\theta$ is $\text{Gamma}(a,b)$ for some constants $a>0,b>0$. Let $\widehat \theta = \sum_{i=1}^my_i/m$ be the maximum likelihood estimator of $\theta$. Then
\begin{align*}
& \sup_{m\geq 1,K\geq 1} E_{P_{\theta_0}}E_{\Pi_m(\cdot|y)}Km\left|\theta-\widehat\theta\right|^2 <+\infty;
\end{align*}
\noindent (ii) Suppose the data $y_i$ ($i=1,\ldots,m$) are independent and identically distributed as $\text{Exp}(\theta)$ with the probability density function $p(y|\theta) =\theta e^{-\theta y}$ and the true parameter $\theta_0$. Suppose the prior on $\theta$ is $\text{Gamma}(a,b)$ for some constants $a>0,b>0$. Let $\widehat \theta = m/\sum_{i=1}^my_i$ be the maximum likelihood estimator of $\theta$. Then
\begin{align*}
& \sup_{m\geq 3,K\geq 1} E_{P_{\theta_0}}E_{\Pi_m(\cdot|y)}Km\left|\theta-\widehat\theta\right|^2 <+\infty;
\end{align*}
\noindent (iii) Suppose the data $y_i$ ($i=1,\ldots,m$) are $\{0,1\}$ binary data independent and identically distributed as $\text{Bernoulli}(\theta)$ with the probability density function $p(y|\theta) =\theta^{y}(1-\theta)^{1-y}$ and the true parameter $\theta_0 \in (0,1)$. Suppose the prior on $\theta$ is $\text{Beta}(a,b)$ for some constants $a>0,b>0$. Let $\widehat \theta = \sum_{i=1}^my_i/m$ be the maximum likelihood estimator of $\theta$. Then
\begin{align*}
& \sup_{m\geq 1,K\geq 1} E_{P_{\theta_0}}E_{\Pi_m(\cdot|y)}Km\left|\theta-\widehat\theta\right|^2 <+\infty;
\end{align*}
\end{proposition}

\noindent \textbf{Proof of Proposition \ref{3expex}:}\\
(i) The subset posterior distribution of $\theta$ is $\text{Gamma}(K\sum_{i=1}^my_i+a,Km+b)$. Therefore
\begin{align}
& E_{P_{\theta_0}}E_{\Pi_m(\cdot|y)}Km\left|\theta-\widehat\theta\right|^2 \nonumber \\
={}& E_{P_{\theta_0}}\left\{ Km\left|E_{\Pi_m(\cdot|y)}(\theta)-\widehat\theta\right|^2 + Km{\var}_{\pi_m(\cdot|y)}(\theta) \right\} \nonumber \\
={}& E_{P_{\theta_0}}\left\{ Km\left|\frac{K\sum_{i=1}^my_i+a}{Km+b}-\frac{\sum_{i=1}^my_i}{m}\right|^2 +
\frac{Km\left(K\sum_{i=1}^my_i+a\right)}{(Km+b)^2}  \right\} \nonumber \\
={}& E_{P_{\theta_0}}\left\{ \frac{K(b\sum_{i=1}^m y_i - am)^2}{m(Km+b)^2} +
\frac{Km\left(K\sum_{i=1}^my_i+a\right)}{(Km+b)^2}  \right\} \nonumber \\
\leq{}&  E_{P_{\theta_0}}\left\{ \frac{2Kb^2m\sum_{i=1}^m y_i^2+2Km^2a^2}{m(Km+b)^2} +
\frac{Km\left(K\sum_{i=1}^my_i+a\right)}{(Km+b)^2}  \right\} \nonumber \\
={}& \frac{2Km^2b^2(\theta_0^2+\theta_0)+2Km^2a^2}{m(Km+b)^2} +
\frac{Km\left(Km\theta_0+a\right)}{(Km+b)^2} \to \theta_0 \quad \text{ as } m\to \infty. \nonumber
\end{align}
Hence, the conclusion holds.\\

\noindent (ii) The subset posterior distribution of $\theta$ is $\text{Gamma}(Km+a,K\sum_{i=1}^my_i+b)$, and notice that $W\equiv 1/\sum_{i=1}^m y_i$ follows $\text{Inverse-Gamma}(m,\theta_0)$ with $E(W)=\theta_0/(m-1)$, $E(W^2)=\theta_0^2/\{(m-1)(m-2)\}$, $E(W^3)= \theta_0^3/\{(m-1)(m-2)(m-3)\}$. Therefore
\begin{align}
& E_{P_{\theta_0}}E_{\Pi_m(\cdot|y)}Km\left|\theta-\widehat\theta\right|^2 \nonumber \\
={}& E_{P_{\theta_0}}\left\{ Km\left|E_{\Pi_m(\cdot|y)}(\theta)-\widehat\theta\right|^2 + Km{\var}_{\pi_m(\cdot|y)}(\theta) \right\} \nonumber \\
={}& E_{P_{\theta_0}}\left\{ Km\left|\frac{Km+a}{K\sum_{i=1}^my_i+b}-\frac{m}{\sum_{i=1}^my_i}\right|^2 +
\frac{Km\left(Km+a\right)}{(K\sum_{i=1}^my_i+b)^2}  \right\} \nonumber \\
={}& E_{P_{\theta_0}}\left\{ \frac{K(a\sum_{i=1}^m y_i - bm)^2}{(\sum_{i=1}^m y_i)(K\sum_{i=1}^m y_i+b)^2} +
\frac{Km\left(Km+a\right)}{(K\sum_{i=1}^my_i+b)^2}  \right\} \nonumber \\
\leq{}&  E_{P_{\theta_0}}\left\{ \frac{2a^2(\sum_{i=1}^m y_i)^2+2b^2m^2}{K(\sum_{i=1}^m y_i)^3} +
\frac{m\left(Km+a\right)}{K\left(\sum_{i=1}^my_i\right)^2}  \right\} \nonumber \\
\leq{}& \frac{2a^2\theta_0}{K(m-1)} + \frac{2b^2m^2\theta_0^3}{K(m-1)(m-2)(m-3)} + \frac{m\left(Km+a\right)\theta_0^2}{K(m-1)(m-2)} \to \theta_0^2 \quad \text{ as } m\to \infty. \nonumber
\end{align}
Therefore, the conclusion holds.\\

\noindent (iii) The subset posterior distribution of $\theta$ is $\text{Beta}\left\{K\sum_{i=1}^m y_i+a,K\sum_{i=1}^m(1-y_i)+b\right\}$. Therefore
\begin{align}
& E_{P_{\theta_0}}E_{\Pi_m(\cdot|y)}Km\left|\theta-\widehat\theta\right|^2 \nonumber \\
={}& E_{P_{\theta_0}}\left\{ Km\left|E_{\Pi_m(\cdot|y)}(\theta)-\widehat\theta\right|^2 + Km{\var}_{\pi_m(\cdot|y)}(\theta) \right\} \nonumber \\
={}& E_{P_{\theta_0}}\left[ Km\left|\frac{K\sum_{i=1}^m y_i+a}{Km+a+b}-\frac{\sum_{i=1}^my_i}{m}\right|^2 +
\frac{Km\left(K\sum_{i=1}^m y_i+a\right)\left\{K\sum_{i=1}^m(1-y_i)+b\right\}}{(Km+a+b)^2(Km+a+b+1)}  \right] \nonumber \\
\leq{}& E_{P_{\theta_0}}\Bigg[\frac{2Kma^2}{(Km+a+b)^2} + \frac{2Km(a+b)^2\left(\sum_{i=1}^m y_i\right)^2}{m^2(Km+a+b)^2}\nonumber\\
& +
\frac{Km\left(K\sum_{i=1}^m y_i+a\right)\left\{K\sum_{i=1}^m(1-y_i)+b\right\}}{(Km+a+b)^2(Km+a+b+1)}  \Bigg] \nonumber \\
={}&  \frac{2Kma^2}{(Km+a+b)^2} + \frac{2Km(a+b)^2\left\{ m^2\theta_0^2 + m\theta_0(1-\theta_0)\right\} }{m^2(Km+a+b)^2}\nonumber \\
&+ \frac{Km\left\{ K^2(m^2-m)\theta_0(1-\theta_0) + Kam(1-\theta_0) + Kbm\theta_0 + ab\right\} }{(Km+a+b)^2(Km+a+b+1)}  \nonumber \\
& \to \theta_0(1-\theta_0) \quad \text{ as } m\to \infty.\nonumber
\end{align}
Therefore, the conclusion holds.
\hfill $\blacksquare$

\section{Data Analysis}
\subsection{Simulated data analysis: Linear model with varying dimension}
\label{para-app}


The prior distributions of $\beta$ and $\sigma$ are specified as follows:
$$\beta \sim \text{generalized double Pareto}(\alpha, \eta),  \;
  \sigma \sim \text{Half-}t(\nu, A).$$
The prior density of $\beta=(\beta_1,\ldots,\beta_p)^\T$ given $\alpha$ and $\eta$ is given by
$$\pi(\beta\mid \alpha,\eta) = \prod_{j=1}^p \frac{\alpha}{2\eta}\left(1+\frac{|\beta_j|}{\eta}\right)^{-(\alpha+1)}.$$
The prior mean and variance of $\beta$ are set to be 0 and $2\eta^2 (\alpha-1)^{-1} (\alpha -2)^{-1}$. $\alpha$ and $\eta$ have independent hyperpriors with densities $\pi(\alpha)=1/(1+\alpha)^2$ and $\pi(\eta)=1/(1+\eta)^2$. The Half-$t$ prior has a convenient parameter expanded form in terms of Inverse-Gamma($a$, $b$) distribution, where $a$ and $b$ are shape and scale parameters: if $\sigma^2 \mid \rho \sim $ Inverse-Gamma($\nu/2$, $\nu/\rho$) and $\rho \sim $ Inverse-Gamma($1/2$, $1/A^2$), then $\sigma \sim $ Half-$t$($\nu$, $A$). We fixed the hyperparameters $\nu$ and $A$ at recommended default values 2 and $100$. We used griddy Gibbs for generating samples of $\alpha$ and $\eta$ from their posterior distribution; see Section 3 in \citep{ArmDunLee13} for details. The Gibbs sampler in \citep{ArmDunLee13} is modified by changing the sample size, $n$, in their sampler to $m K$, where $m$ is sample size for the subset and $K$ is the number of subsets.

Let $\mathcal N(\hat m_1,\hat V_1), \ldots, \mathcal N(\hat m_K, \hat V_K)$ represent the asymptotic approximations of $K$ subset posteriors, then \citep{AguCar11} has shown that their barycenter in Wasserstein-2 space is also Gausssian with mean $m^*$ and covariance matrix $V^*$, where
\begin{align*}
  m^* = K^{-1}\sum_{j=1}^K \hat m_j \quad \text{and}\quad  V^* \text{ satisfies } \sum_{j=1}^K \left(  V^{*^{1/2}} \hat V_j V^{*^{1/2}} \right)^{1/2} = KV^*.
\end{align*}
Therefore, we use the formula above to calculate the $W_2$ barycenter of $K$ normal approximations to the $K$ subset posteriors.
Given $\hat V_1,\ldots,\hat V_K$, we can find $V^*$ efficiently using fixed-point iteration.


\vspace{.75cm}

Although the priors of $\beta$ and $\sigma$ specified above are heavy-tailed with infinite second moments, in the following proposition and its proof, we verify that every subset posterior after conditioning on the first $m_0$ observations has finite second moment in both $\beta$ and $\sigma$, for some fixed integer $m_0$.

\begin{proposition}\label{2momentcheck}
Suppose the form of a linear model and its priors are specified in Section 4.1 of the main paper with fixed $\nu>0$ and $A>0$. Assume that in the model $X$ and $\epsilon$ are independent. Let $\tilde y$ and $\tilde X$ be the response vector and the design matrix of the first $m_0$ observations ($m_0\geq 1$). Suppose that the true parameters are $\theta_0=(\beta_0^\T, \sigma_0)^\T$ with $\sigma_0>0$. Assume that the eigenvalues of $\tilde X^\T \tilde X$ are bounded from above and below by positive constants almost surely. Then the posterior distribution of $\theta=(\beta^\T,\sigma)^\T$ conditional on $\tilde y$ and $\tilde X$ has finite second moment almost surely in $P_{\theta_0}$, if $m_0$ satisfies $m_0\geq p+4$.
\end{proposition}

\noindent {\bf Proof of Proposition \ref{2momentcheck}:}\\
For convenience we define the quadratic term $S(\beta,\tilde y,\tilde X)=(\tilde y - \tilde X\beta)^\T (\tilde y - \tilde X\beta)$, which has the decomposition $S(\beta,\tilde y,\tilde X)=\tilde \epsilon^\T (I_{m_0}-\tilde H) \tilde \epsilon + (\beta-\tilde \beta)^\T \tilde X^\T \tilde X (\beta-\tilde\beta)$ with $\tilde \epsilon=\tilde y-\tilde X\beta_0$, $\tilde H= \tilde X(\tilde X^\T \tilde X)^{-1} \tilde X^\T$, $I_{m_0}$ being the $m_0$-dimensional identity matrix, and $\tilde \beta=(\tilde X^\T\tilde X)^{-1}\tilde X^\T \tilde y$. Since $m_0\geq p+4$ and $\tilde X^\T\tilde X$ is nonsingular, $I_{m_0}-\tilde H$ is idempotent with rank $m_0-p>0$. Since $\sigma_0>0$, the residual sum of squares $\tilde \epsilon^\T (I_{m_0}-\tilde H) \tilde \epsilon$ is then almost surely positive. Let the smallest eigenvalue of $\tilde X^\T \tilde X$ be lower bounded by $c_1>0$. Then $S(\beta,\tilde y,\tilde X)\geq \tilde \epsilon^\T (I_{m_0}-\tilde H) \tilde \epsilon + c_1\|\beta-\tilde\beta\|^2$.

The subset posterior of the model parameter $\theta=(\beta^\T,\sigma)^\T$ given only $\tilde y, \tilde X$ has the following expression
\begin{align}\label{postbetasigma}
&\pi_{m_0}(\beta,\sigma\mid \tilde y,\tilde X, \nu, A) \nonumber \\
= {}& \frac{(2\pi)^{-Km_0/2} \sigma^{-Km_0} \exp\left\{-\frac{K}{2\sigma^2} S(\beta,\tilde y,\tilde X)\right\} \pi(\beta)\pi(\sigma\mid \nu, A)}
{\iint (2\pi)^{-Km_0/2} \sigma^{-Km_0} \exp\left\{-\frac{K}{2\sigma^2} S(\beta,\tilde y,\tilde X)\right\} \pi(\beta)\pi(\sigma\mid \nu, A) \ud \beta \ud \sigma} \nonumber \\
= {}& \frac{\sigma^{-Km_0} \exp\left\{-\frac{K}{2\sigma^2} S(\beta,\tilde y,\tilde X)\right\} \left\{1+\nu^{-1}\left(\sigma/A\right)^2\right\}^{-(\nu+1)/2} \pi(\beta)}
{\int\left[ \int_0^\infty \sigma^{-Km_0} \exp\left\{-\frac{K}{2\sigma^2} S(\beta,\tilde y,\tilde X)\right\} \left\{1+\nu^{-1}\left(\sigma/A\right)^2\right\}^{-(\nu+1)/2} \ud \sigma \right]\pi(\beta)\ud \beta}
\end{align}
where the likelihood has been raised to the power of $K$ according to our stochastic approximation. In the following, we bound $E_{\Pi_{m_0}(\cdot \mid \tilde y,\tilde X, \nu, A)}\|\beta\|^2$ and $E_{\Pi_{m_0}(\cdot \mid \tilde y,\tilde X, \nu, A)}(\sigma^2)$ respectively.
\vspace{.3cm}

\noindent Step 1: Show that $E_{\Pi_{m_0}(\cdot \mid \tilde y,\tilde X, \nu, A)}\|\beta\|^2$ is finite almost surely in $P_{\theta_0}$.
\vspace{.2cm}

In the following, we use \eqref{postbetasigma} to calculate $E_{\Pi_{m_0}(\cdot \mid \tilde y,\tilde X, \nu, A)}\|\beta\|^2$ and bound its numerator and denominator respectively. For the numerator part, we have
\begin{align}\label{betanumeratorA11}
& \int_0^\infty \int_{\mathcal{R}^p} \|\beta\|^2 \sigma^{-Km_0} \exp\left\{-\frac{K}{2\sigma^2} S(\beta,\tilde y,\tilde X)\right\} \left\{1+\nu^{-1}\left(\sigma/A\right)^2\right\}^{-(\nu+1)/2} \pi(\beta) \ud \beta \ud \sigma \nonumber \\
\leq {}& (A\nu^{1/2})^{\nu+1} \int_{\mathcal{R}^p} \|\beta\|^2\pi(\beta)\left[ \int_0^\infty \sigma^{-(Km_0+\nu+1)} \exp\left\{-\frac{K}{2\sigma^2}S(\beta,\tilde y,\tilde X) \right\}  \ud \sigma\right] \ud \beta \nonumber \\
\leq {}&  2^{(Km_0+\nu)/2-1} K^{-(Km_0+\nu)/2}(A\nu^{1/2})^{\nu+1}\Gamma\left(\frac{Km_0+\nu}{2}\right) \nonumber \\
&~~ \times \int_{\mathcal{R}^p} \frac{\|\beta\|^2\times \pi(\beta) }{\left\{\tilde \epsilon^\T (I_{m_0}-\tilde H) \tilde \epsilon + c_1\|\beta-\tilde\beta\|^2\right\}^{(Km_0+\nu)/2}} \ud \beta.
\end{align}
The last integral of \eqref{betanumeratorA11} can be further bounded by
\begin{align}\label{betanumeratorA12}
& \int_{\mathcal{R}^p} \frac{\|\beta\|^2\times \pi(\beta) }{\left\{\tilde \epsilon^\T (I_{m_0}-\tilde H) \tilde \epsilon + c_1\|\beta-\tilde\beta\|^2\right\}^{(Km_0+\nu)/2}} \ud \beta \nonumber \\
\leq{}& \int_{\mathcal{R}^p} \frac{2\left(\|\beta-\tilde\beta\|^2+\|\tilde\beta\|^2\right)\times \pi(\beta) }{\left\{\tilde \epsilon^\T (I_{m_0}-\tilde H) \tilde \epsilon + c_1\|\beta-\tilde\beta\|^2\right\}^{(Km_0+\nu)/2}} \ud \beta \nonumber \\
\leq{}& \int_{\mathcal{R}^p} 2c_1^{-1} \left\{\tilde \epsilon^\T (I_{m_0}-\tilde H) \tilde \epsilon + c_1\|\beta-\tilde\beta\|^2\right\}^{-(Km_0+\nu)/2+1} \pi(\beta)\ud \beta  \nonumber \\
&~~ + \int_{\mathcal{R}^p}2\|\tilde\beta\|^2\left\{\tilde \epsilon^\T (I_{m_0}-\tilde H) \tilde \epsilon\right\}^{-(Km_0+\nu)/2} \pi(\beta)\ud \beta  \nonumber \\
\leq{}& 2\left\{c_1^{-1}\tilde \epsilon^\T (I_{m_0}-\tilde H) \tilde \epsilon + \|\tilde\beta\|^2\right\}  \left\{\tilde \epsilon^\T (I_{m_0}-\tilde H) \tilde \epsilon \right\}^{-(Km_0+\nu)/2}.
\end{align}

Next we provide a lower bound for the denominator of \eqref{postbetasigma}. The integral of $\sigma$ can be lower bounded by using a change of variable $u=KS(\beta,\tilde y,\tilde X)/(2\sigma^2)$:
\begin{align}\label{sigmadenom}
& \int_0^\infty \sigma^{-Km_0} \exp\left\{-\frac{K}{2\sigma^2} S(\beta,\tilde y,\tilde X)\right\} \left\{1+\nu^{-1}\left(\sigma/A\right)^2\right\}^{-(\nu+1)/2} \ud \sigma  \nonumber \\
={}& \frac{1}{2}\left\{KS(\beta,\tilde y,\tilde X)/2\right\}^{-(Km_0-1)/2} \int_0^\infty \left\{1+KS(\beta,\tilde y,\tilde X)/(2A^2\nu u)\right\}^{-(\nu+1)/2}u^{(Km_0-3)/2} e^{-u} \ud u  \nonumber \\
\geq{}& \frac{1}{2}\left\{KS(\beta,\tilde y,\tilde X)/2\right\}^{-(Km_0-1)/2} \left\{1+KS(\beta,\tilde y,\tilde X)/(2A^2\nu)\right\}^{-(\nu+1)/2} \int_1^\infty u^{(Km_0-3)/2} e^{-u} \ud u   \nonumber  \\
\geq{}& \frac{1}{2}\left\{KS(\beta,\tilde y,\tilde X)/2\right\}^{-(Km_0-1)/2} \left\{1+KS(\beta,\tilde y,\tilde X)/(2A^2\nu)\right\}^{-(\nu+1)/2}\nonumber\\ &\times e^{-1}\Gamma\left(\frac{Km_0-1}{2}\right)\left(\frac{Km_0+1}{Km_0-1}\right)^{(Km_0-3)/2},
\end{align}
where we have used the fact $Km_0\geq 4$ and the lower bound for the incomplete gamma function $\int_1^\infty u^{s-1}e^{-u}\ud u\geq e^{-1}\Gamma(s) (1+1/s)^{s-1}$ for $s\geq 1$. Now to evaluate the denominator of \eqref{postbetasigma}, we need to integrate the lower bound in \eqref{sigmadenom} with respect to $\beta$. Consider the set $A_3=\{\beta\in \mathcal{R}^p:~ \|\beta\|\leq 1\}$. Clearly the prior of $\beta$ has positive probability mass on $A_3$. Define the constant $c_3=\int_{A_3} \pi(\beta)\ud \beta = \int_{A_3} \iint \pi(\beta\mid \alpha,\eta) \pi(\alpha)\pi(\eta)\ud \alpha \ud \eta \ud \beta >0$ which only depends on the dimension $p$. Let the largest eigenvalue of $\tilde X^\T \tilde X$ be upper bounded by $c_2>0$. Then on $A_3$, $S(\beta,\tilde y,\tilde X)\leq c_2\|\beta-\tilde \beta\|^2 + \tilde \epsilon^\T (I_{m_0}-\tilde H) \tilde \epsilon \leq 2c_2(\|\tilde\beta\|^2+1) + \tilde \epsilon^\T (I_{m_0}-\tilde H) \tilde \epsilon$. This and \eqref{sigmadenom} imply that the denominator of \eqref{postbetasigma} can be lower bounded by
\begin{align}\label{bsdenom}
& \int\left[ \int_0^\infty \sigma^{-Km_0} \exp\left\{-\frac{K}{2\sigma^2} S(\beta,\tilde y,\tilde X)\right\} \left\{1+\nu^{-1}\left(\sigma/A\right)^2\right\}^{-(\nu+1)/2} \ud \sigma \right]\pi(\beta)\ud \beta \nonumber \\
\geq{}&  \frac{c_3}{2e}\Gamma\left(\frac{Km_0-1}{2}\right)
\left[\left\{2c_2(\|\tilde\beta\|^2+1) + \tilde \epsilon^\T (I_{m_0}-\tilde H) \tilde \epsilon \right\}K/2\right]^{-(Km_0-1)/2} \nonumber \\
& ~\times \left[1+\left\{2c_2(\|\tilde\beta\|^2+1) + \tilde \epsilon^\T (I_{m_0}-\tilde H) \tilde \epsilon \right\}K/(2A^2\nu)\right]^{-(\nu+1)/2} \nonumber \\
\geq{}& 2^{(Km_0-3)/2}e^{-1}c_3(A\nu^{1/2})^{\nu+1}K^{-(Km_0+\nu)/2}\Gamma\left(\frac{Km_0-1}{2}\right)\nonumber \\
& ~\times \left\{2c_2(\|\tilde\beta\|^2+1) + \tilde \epsilon^\T (I_{m_0}-\tilde H) \tilde \epsilon \right\}^{-(Km_0+\nu)/2},
\end{align}
where the last inequality follows if we choose $c_2>A\nu^2/K$. \\

We can combine \eqref{betanumeratorA11}, \eqref{betanumeratorA12}, \eqref{bsdenom} and obtain that
\begin{align}\label{boundbeta2}
&E_{\Pi_{m_0}(\cdot \mid \tilde y,\tilde X, \nu, A)}\|\beta\|^2  \nonumber \\
\leq{}& c_4 K^{(\nu+1)/2} \left\{c_1^{-1} \tilde \epsilon^\T (I_{m_0}-\tilde H) \tilde \epsilon + \|\tilde\beta\|^2\right\} \left\{\tilde \epsilon^\T (I_{m_0}-\tilde H) \tilde \epsilon\right\}^{-(Km_0+\nu)/2} \nonumber \\
& ~\times \left\{2c_2(\|\tilde\beta\|^2+1) + \tilde \epsilon^\T (I_{m_0}-\tilde H) \tilde \epsilon \right\}^{(Km_0+\nu)/2}
\end{align}
for some constant $c_4>0$ that only depends on $m_0,p,\nu,A,c_1,c_2,c_3$. Conditional on $\tilde y,\tilde X$, both $\|\tilde\beta\|^2$ and $\tilde \epsilon^\T (I_{m_0}-\tilde H) \tilde \epsilon$ are almost surely positive constants. Therefore, we have proved that
$E_{\Pi_{m_0}(\cdot \mid \tilde y,\tilde X, \nu, A)}\|\beta\|^2 <\infty$ almost surely in $P_{\theta_0}$.
\vspace{.3cm}

\noindent Step 2: Show that $E_{\Pi_{m_0}(\cdot \mid \tilde y,\tilde X, \nu, A)}(\sigma^2)$ is finite almost surely in $P_{\theta_0}$.

To calculate $E_{\Pi_{m_0}(\cdot \mid \tilde y,\tilde X, \nu, A)}\sigma^2$, we integrate $\sigma^2$ with respect to the posterior density of \eqref{postbetasigma}. We start with upper bounding the numerator:
\begin{align}\label{sigmanumerator}
& \int \int_0^\infty \sigma^2\times \sigma^{-Km_0} \exp\left\{-\frac{K}{2\sigma^2} S(\beta,\tilde y,\tilde X)\right\} \left\{1+\nu^{-1}\left(\sigma/A\right)^2\right\}^{-(\nu+1)/2} \pi(\beta)\ud \sigma \ud\beta \nonumber \\
\leq{}& \int (A\nu^{1/2})^{(\nu+1)/2} \left[\int_0^\infty \sigma^{-(Km_0+\nu-1)} \exp\left\{-\frac{K}{2\sigma^2} S(\beta,\tilde y,\tilde X)\right\} \ud \sigma\right] \pi(\beta) \ud\beta \nonumber \\
\leq{}& \int \frac{1}{2}(A\nu^{1/2})^{(\nu+1)/2} \left\{KS(\beta,\tilde y,\tilde X)/2\right\}^{-(Km_0+\nu)/2+1} \Gamma\left(\frac{Km_0+\nu}{2}-1\right) \pi(\beta)\ud \beta \nonumber \\
\leq{}& 2^{(Km_0+\nu)/2-2} (A\nu^{1/2})^{(\nu+1)/2} K^{-(Km_0+\nu)/2+1} \Gamma\left(\frac{Km_0+\nu}{2}-1\right)  \nonumber \\
&\times \left\{\tilde \epsilon^\T (I_{m_0}-\tilde H) \tilde \epsilon\right\}^{-(Km_0+\nu)/2+1},
\end{align}
where we have used the fact that $(Km_0+\nu)/2\geq 2$ and $S(\beta,\tilde y,\tilde X)\geq \tilde \epsilon^\T (I_{m_0}-\tilde H) \tilde \epsilon$. If we combine \eqref{bsdenom} and \eqref{sigmanumerator}, then we can obtain that
\begin{align}\label{boundsigma}
E_{\Pi_{m_0}(\cdot \mid \tilde y,\tilde X, \nu, A)}(\sigma^2) & \leq c_5 K^{(\nu+1)/2} \left\{\tilde \epsilon^\T (I_{m_0}-\tilde H) \tilde \epsilon\right\}^{-(Km_0+\nu)/2+1} \nonumber \\
&\times \left\{2c_2(\|\tilde\beta\|^2+1) + \tilde \epsilon^\T (I_{m_0}-\tilde H) \tilde \epsilon \right\}^{(Km_0+\nu+1)/2},
\end{align}
for some constant $c_5>0$ that only depends on $m_0,p,\nu,A,c_2,c_3$. Conditional on $\tilde y,\tilde X$, both $\|\tilde\beta\|^2$ and $\tilde \epsilon^\T (I_{m_0}-\tilde H) \tilde \epsilon$ are almost surely positive constants. Therefore, we have proved that
$E_{\Pi_{m_0}(\cdot \mid \tilde y,\tilde X, \nu, A)}(\sigma^2) <\infty$ almost surely in $P_{\theta_0}$.
\hfill $\blacksquare$

\begin{remark}
The second moment of the prior has only appeared in \eqref{prior2moment} in the proof of Lemma \ref{subsetbound}. We need the right-hand side of \eqref{prior2moment} to go to zero as $n\to \infty$. Since the gdp prior and half-$t$ prior are both heavy tailed, Proposition \ref{2momentcheck} proposes to replace the prior in \eqref{prior2moment} by the posterior conditional on the first $m_0$ observations in the linear model example. It is straightforward to check that the finite upper bounds for $E_{\Pi_{m_0}(\cdot \mid \tilde y,\tilde X, \nu, A)}\|\beta\|^2$ and $E_{\Pi_{m_0}(\cdot \mid \tilde y,\tilde X, \nu, A)}(\sigma^2)$ in \eqref{boundbeta2} and \eqref{boundsigma} increase at most exponentially fast in $K$. Even if $K\to\infty$, we can see that the exponential term $\exp(-Km\epsilon_1)$ in the right-hand side of \eqref{prior2moment} decays faster than any exponential rate in $K$ since $m\to\infty$. Therefore, the conclusions of Lemma \ref{subsetbound} and all subsequent theorems remain valid conditional on the first $m_0$ observations.
\end{remark}


\subsection{Simulated data analysis: Linear mixed effects model}
\label{stoc-lme}

Stochastic approximation for subset posteriors can be easily implemented in Stan. The sampling model for linear mixed effects models implies that likelihood of $\beta$ and $\Sigma$ is
\begin{align*}
  L(\beta, \Sigma) = \prod_{i=1}^s \int_{\mathcal{R}^q} p(y_i \mid X_i,Z_i, \beta, u_i) p(u_i \mid \Sigma)~\ud u_i = \prod_{i=1}^s \phi(y_i \mid X_i \beta,  Z_i \Sigma Z_i^\T),
\end{align*}
where $\phi( \cdot \mid \mu, \Sigma)$ is the multivariate normal density with mean $\mu$ and covariance matrix $\Sigma$. The likelihood after stochastic approximation is
\begin{align}
  L_{K}(\beta, \Sigma) = \left\{ L(\beta, \Sigma) \right\}^K = \prod_{i=1}^s \left\{ \phi(y_i \mid X_i \beta,  Z_i \Sigma Z_i^\T) \right\}^K \label{llk}.
\end{align}
 The generative model is completed by imposing default priors for $\beta$ and $\Sigma$ in Stan. We take advantage of the \texttt{increment\_log\_prob} function in Stan to specify that
\begin{align}
  y_i \mid  \beta, \Sigma\sim f_K(y_i \mid \beta, \Sigma),\nonumber
\end{align}
where $f_K$ is the density that leads to the term for $y_i$ in the likelihood $L_{K}(\beta, \Sigma)$ in \eqref{llk}. In general $f_K$ would be analytically intractable, but in the present case it corresponds to $\{\phi( \cdot \mid \mu, \Sigma)\}^K$.
The computation time of different methods is summarized in Figure \ref{fig:time1}.
%
\begin{table}[ht]
\caption{\it 90\% credible intervals for fixed effects in simulated data analysis. The upper and lower bounds are averaged over 10 replications. MLE, maximum likelihood estimator; MCMC, Markov chain Monte Carlo based on the full data; VB, variational Bayes; CMC, consensus Monte Carlo; SDP, semiparametric density product; WASP, the algorithm in \citet{Srietal15}; PIE, our posterior interval estimation algorithm.}
{
\begin{tabular}{r|llll}
  \hline
  & $\beta_1$ & $\beta_2$ & $\beta_3$ & $\beta_4$ \\
  \hline
  MLE & (-1.01, -1.00) & (0.99, 1.01) & (-1.01, -0.99) & (1.00, 1.01) \\
  MCMC & (-1.01, -1.00) & (0.99, 1.01) & (-1.01, -0.99) & (1.00, 1.01) \\
  VB & (-1.01, -1.00) & (0.99, 1.01) & (-1.01, -1.00) & (1.00, 1.01) \\
  CMC & (-1.01, -0.99) & (0.99, 1.01) & (-1.01, -0.99) & (1.00, 1.01) \\
  SDP & (-1.01, -0.99) & (0.99, 1.01) & (-1.01, -0.99) & (1.00, 1.01) \\
  WASP & (-1.01, -0.99) & (0.99, 1.01) & (-1.01, -0.99) & (1.00, 1.01) \\
  PIE & (-1.01, -0.99) & (0.99, 1.01) & (-1.01, -0.99) & (1.00, 1.01) \\
  \hline
\end{tabular}
}
\label{tab:sim-fix-ci}
\end{table}

\begin{table}[ht]
\caption{\it Accuracy of approximate posteriors for fixed effects in simulated data analysis. The standard deviation of accuracy across 10 replications is in parentheses. MLE, maximum likelihood estimator; VB, variational Bayes; CMC, consensus Monte Carlo; SDP, semiparametric density product; WASP, the algorithm in \citet{Srietal15}; PIE, our posterior interval estimation algorithm.}
{
\begin{tabular}{r|cccc}
  \hline
  & $\beta_1$ & $\beta_2$ & $\beta_3$ & $\beta_4$ \\
  \hline
  MLE & 0.96 (0.01) & 0.96 (0.01) & 0.96 (0.01) & 0.96 (0.01) \\
  VB & 0.86 (0.11) & 0.87 (0.10) & 0.86 (0.10) & 0.87 (0.10) \\
  CMC & 0.96 (0.01) & 0.96 (0.01) & 0.95 (0.02) & 0.96 (0.01) \\
  SDP & 0.96 (0.01) & 0.94 (0.02) & 0.95 (0.03) & 0.95 (0.02) \\
  WASP & 0.95 (0.02) & 0.95 (0.02) & 0.96 (0.01) & 0.94 (0.02) \\
  PIE & 0.95 (0.01) & 0.95 (0.01) & 0.94 (0.01) & 0.94 (0.02) \\
   \hline
\end{tabular}
}
\label{tab:sim-fix-acc}
\end{table}

\subsection{Real data analysis: United States natality data}
\label{sec:us-natality-data}

\begin{figure}[H]
  \subfloat[Linear mixed effects model]{
    \includegraphics[scale=0.28]{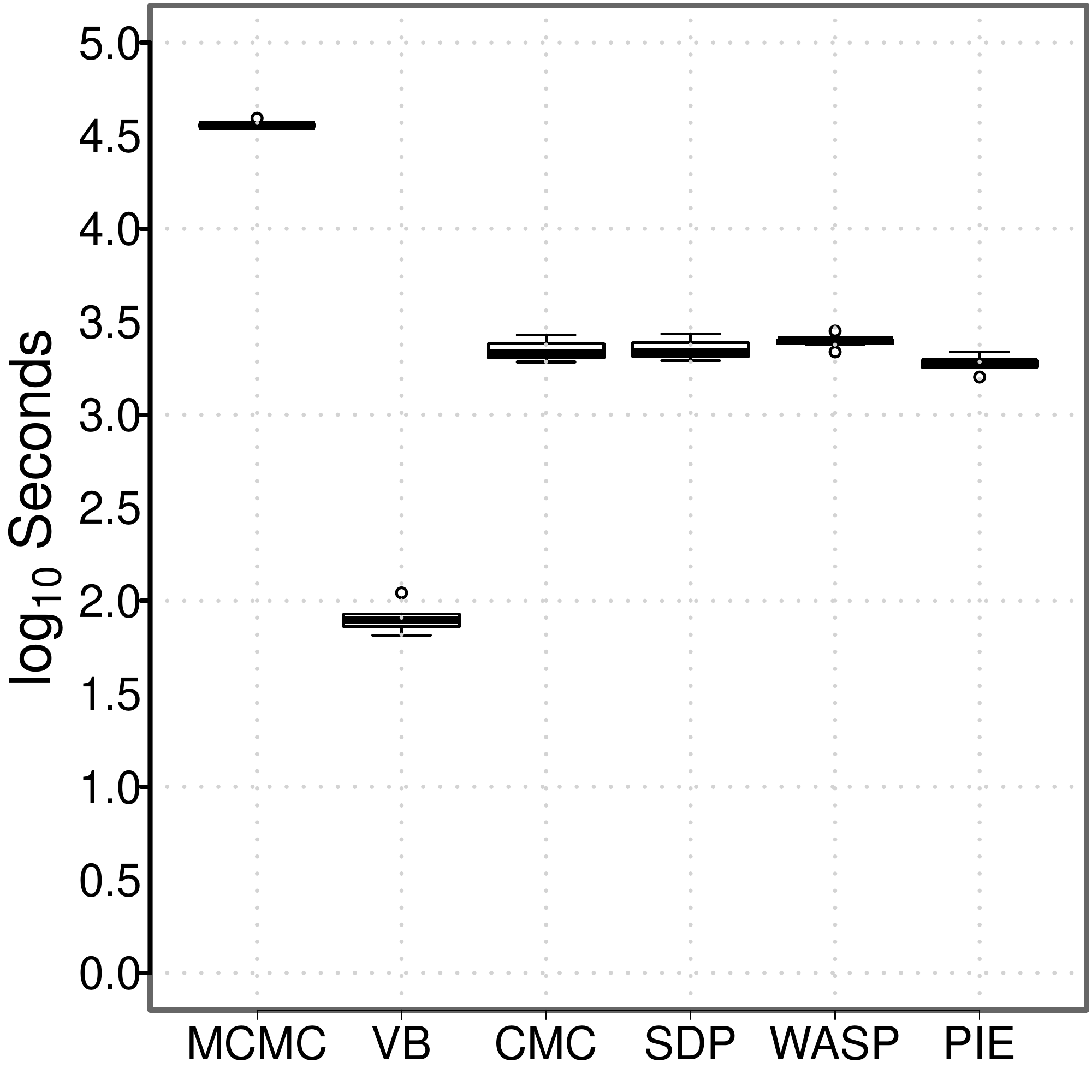}
    \label{fig:time1}}
  \subfloat[United States natality data]{
    \includegraphics[scale=0.28]{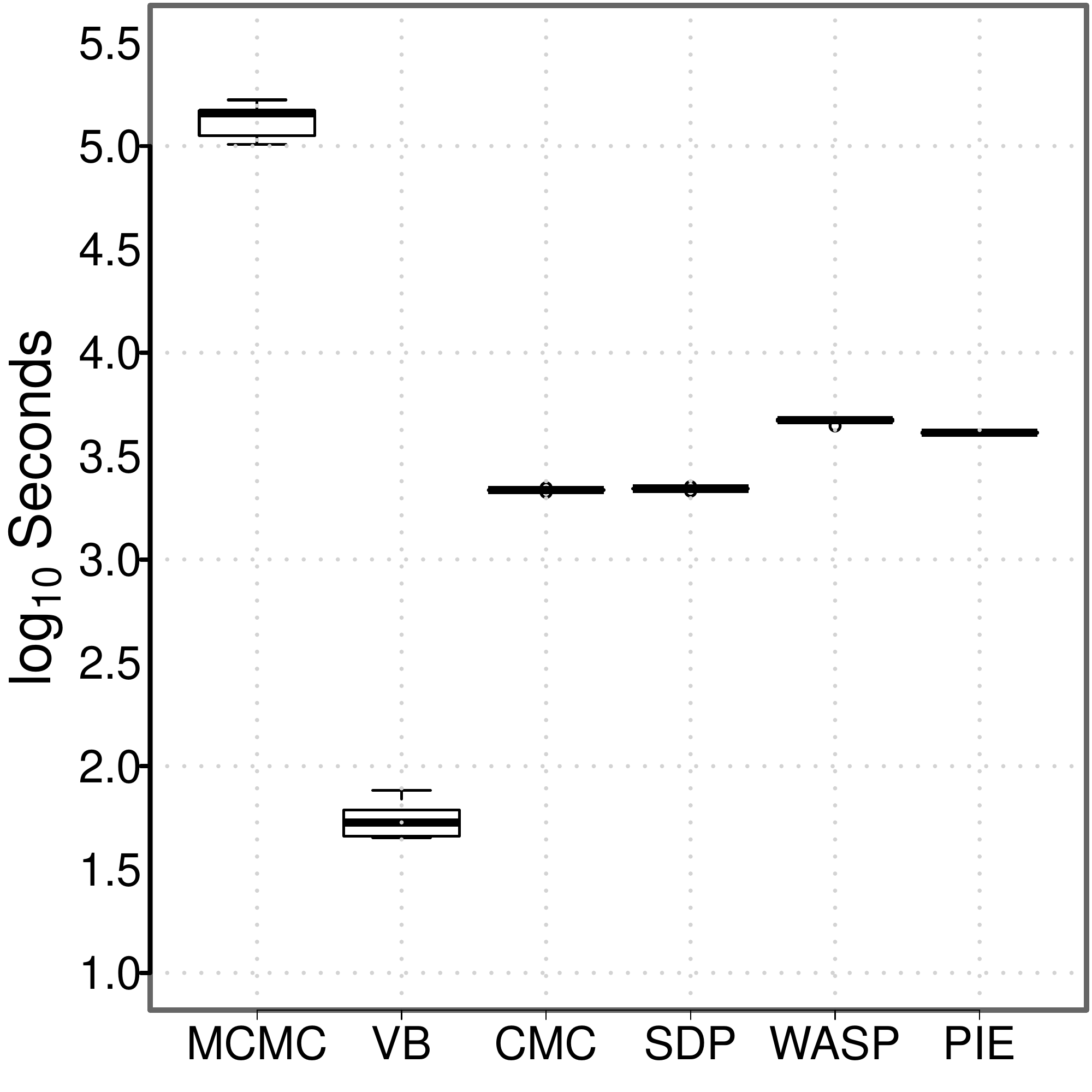}
    \label{fig:time3}}
  \captionsetup{font=small}
  \caption{Computation time for the methods used in simulated and real data analysis. MCMC, Markov chain Monte Carlo based on the full data; VB, variational Bayes; CMC, consensus Monte Carlo; SDP, semiparametric density product; WASP, the algorithm in \citet{Srietal15}; PIE, our posterior interval estimation algorithm.}
  \label{fig:times}
\end{figure}

We selected thirteen variables from the United States natality data summarized in Table \ref{tab:abe} and analyzed in \citep{Abe2006} and \citep{LeeWan16}. These data are available at \url{http://qed.econ.queensu.ca/jae/datasets/abrevaya001}. The computation time of different methods is summarized in Figure \ref{fig:time3}.
\begin{table}[ht]
  \caption{\it Variables used in the United States natality data}
 { \begin{tabular}{ll}
    \hline
    Variable & Description\\
    \hline
\texttt{dmage}     &        age of mother in years\\
\texttt{nlbnl}     &        number of live births now living\\
\texttt{gestat}    &        length of gestation in weeks\\
\texttt{male}      &        indicator variable for baby gender\\
\texttt{married}   &        indicator variable for marital status\\
\texttt{hsgrad}    &        high-school graduate indicator\\
\texttt{agesq}     &        age of mother squared\\
\texttt{black}     &        indicator variable for black race\\
\texttt{novisit}   &        indicator of no prenatal care visit\\
\texttt{adeqcode2} &        indicator that Kessner index  2\\
    \texttt{adeqcode3} &        indicator that Kessner index  3\\
    \texttt{pretri2}   &        indicator that first prenatal visit occurred in 2nd trimester\\
    \texttt{pretri3}   &        indicator that first prenatal visit occurred in 3nd trimester\\
    \hline
  \end{tabular}}
  \label{tab:abe}
\end{table}

\begin{table}[H]
\caption{\it 90\% credible intervals for fixed effects in United States natality data analysis. The upper and lower bounds are averaged over 10 folds of cross-validation. MLE, maximum likelihood estimator; MCMC, Markov chain Monte Carlo based on the full data; VB, variational Bayes; CMC, consensus Monte Carlo; SDP, semiparametric density product; WASP, the algorithm in \citet{Srietal15}; PIE, our posterior interval estimation algorithm.}
{\begin{tabular}{r|ccccc}
      \hline
  & \texttt{Intercept} & \texttt{dmage} & \texttt{nlbnl} & \texttt{gestat} & \texttt{male} \\
      \hline
  MLE & (-1.31, -0.18) & (0.09, 0.57) & (0.02, 0.08) & (0.22, 0.25) & (0.25, 0.32)  \\
  MCMC & (-0.48, 0.11) & (-0.04, 0.26) & (0.02, 0.08) & (0.22, 0.24) & (0.24, 0.31)  \\
  VB & (-1.96, -0.94) & (0.17, 0.56) & (0.01, 0.06) & (0.25, 0.27) & (0.24, 0.31)  \\
  CMC & (-1.13, -0.04) & (0.06, 0.51) & (0.02, 0.08) & (0.22, 0.24) & (0.24, 0.32) \\
  SDP  & (-0.97, -0.10) & (0.05, 0.42) & (0.02, 0.08) & (0.22, 0.24) & (0.25, 0.32) \\
  WASP & (-0.93, 0.03) & (0.04, 0.45) & (0.02, 0.08) & (0.22, 0.24) & (0.25, 0.31)  \\
  PIE & (-0.93, 0.04) & (0.03, 0.45) & (0.02, 0.08) & (0.22, 0.24) & (0.25, 0.31)  \\
   \hline
  &  \texttt{married} & \texttt{hsgrad} & \texttt{agesq} & \texttt{black} & \texttt{ageqcode2} \\
   \hline
  MLE & (-0.03, 0.06) & (0.02, 0.12) & (0.00, 0.00) & (-0.46, -0.3) & (-0.24, -0.11)  \\
  MCMC & (-0.03, 0.07) & (0.03, 0.13) & (0.00, 0.00) & (-0.44, -0.28) & (-0.22, -0.1) \\
  VB & (-0.02, 0.06) & (0.02, 0.11) & (0.00, 0.00) & (-0.43, -0.29) & (-0.26, -0.13)  \\
  CMC & (-0.03, 0.07) & (0.02, 0.12) & (0.00, 0.00) & (-0.45, -0.28) & (-0.24, -0.11) \\
  SDP & (-0.03, 0.07) & (0.02, 0.12) & (0.00, 0.00) & (-0.45, -0.30) & (-0.23, -0.11) \\
   WASP & (-0.02, 0.07) & (0.03, 0.13) & (0.00, 0.00) & (-0.45, -0.28) & (-0.23, -0.11)  \\
  PIE & (-0.02, 0.07) & (0.03, 0.13)& (0.00, 0.00) & (-0.44, -0.28) & (-0.23, -0.11) \\
  \hline
   & \texttt{ageqcode3} & \texttt{novisit} & \texttt{petri2} & \texttt{pertri3} & \\
   \hline
  MLE  & (-0.42, -0.22) & (-0.12, 0.16) & (0.03, 0.16) & (0.1, 0.33) & \\
  MCMC  & (-0.37, -0.18) & (-0.14, 0.11) & (0.01, 0.14) & (0.06, 0.28)& \\
  VB  & (-0.43, -0.22) & (-0.12, 0.18) & (0.04, 0.18) & (0.09, 0.34) & \\
  CMC & (-0.39, -0.19) & (-0.13, 0.17) & (0.02, 0.16) & (0.08, 0.32) & \\
  SDP & (-0.38, -0.20) & (-0.13, 0.13) & (0.02, 0.15) & (0.09, 0.30) & \\
  WASP  & (-0.39, -0.2) & (-0.14, 0.15) & (0.02, 0.15) & (0.08, 0.31)& \\
  PIE  & (-0.39, -0.2) & (-0.14, 0.15) & (0.02, 0.15) & (0.08, 0.31) & \\
  \hline
    \end{tabular}
  }
\label{tab:abe-fix-ci}
\end{table}

\begin{table}[H]
\caption{\it Accuracy of approximate posteriors for fixed effects in US natality data analysis. The standard deviation of accuracy across 10 replications is in parentheses. MLE, maximum likelihood estimator; VB, variational Bayes; CMC, consensus Monte Carlo; SDP, semiparametric density product; WASP, the algorithm in \citet{Srietal15}; PIE, our posterior interval estimation algorithm.}
{\begin{tabular}{r|ccccc}
      \hline
      & \texttt{Intercept} & \texttt{dmage} & \texttt{nlbnl} & \texttt{gestat} & \texttt{male}\\
      \hline
      MLE  & 0.28 (0.06) & 0.35 (0.06) & 0.93 (0.03) & 0.77 (0.06) & 0.93 (0.02) \\
      VB   & 0.02 (0.01) & 0.22 (0.04) & 0.77 (0.06) & 0.03 (0.01) & 0.91 (0.05) \\
      CMC  & 0.40 (0.10) & 0.42 (0.09) & 0.91 (0.02) & 0.83 (0.11) & 0.95 (0.03) \\
      SDP   & 0.39 (0.23) & 0.47 (0.19) & 0.91 (0.07) & 0.79 (0.16) & 0.91 (0.05) \\
      WASP & 0.55 (0.12) & 0.53 (0.10) & 0.90 (0.04) & 0.87 (0.11) & 0.91 (0.06) \\
      PIE  & 0.55 (0.12) & 0.52 (0.11) & 0.91 (0.05) & 0.87 (0.11) & 0.92 (0.05) \\
      \hline
      &  \texttt{married} & \texttt{hsgrad} & \texttt{agesq} & \texttt{black} & \texttt{ageqcode2} \\
      \hline
      MLE & 0.92 (0.02) & 0.90 (0.02) & 0.36 (0.05) & 0.84 (0.01) & 0.83 (0.03) \\
      VB & 0.91 (0.03) & 0.80 (0.03) & 0.23 (0.04) & 0.91 (0.02) & 0.66 (0.06)  \\
      CMC & 0.90 (0.06) & 0.82 (0.08) & 0.44 (0.10) & 0.88 (0.10) & 0.85 (0.06) \\
      SDP & 0.90 (0.06) & 0.87 (0.08) & 0.48 (0.19) & 0.86 (0.08) & 0.87 (0.10) \\
      WASP  & 0.92 (0.05) & 0.92 (0.06) & 0.54 (0.11) & 0.87 (0.08) & 0.88 (0.08) \\
      PIE & 0.92 (0.06) & 0.92 (0.06) & 0.54 (0.11) & 0.87 (0.08) & 0.88 (0.08) \\
      \hline
      &  \texttt{ageqcode3} & \texttt{novisit} & \texttt{petri2} & \texttt{pertri3}\\
      \hline
      MLE & 0.74 (0.04) & 0.81 (0.04) & 0.82 (0.02) & 0.73 (0.03)  \\
      VB  & 0.67 (0.04) & 0.77 (0.06) & 0.67 (0.04) & 0.72 (0.05) \\
      CMC & 0.86 (0.08) & 0.80 (0.08) & 0.84 (0.10) & 0.83 (0.08) \\
      SDP & 0.87 (0.09) & 0.89 (0.07) & 0.88 (0.09) & 0.84 (0.10) \\
      WASP &  0.84 (0.11) & 0.84 (0.09) & 0.89 (0.11) & 0.83 (0.06) \\
      PIE & 0.84 (0.11) & 0.84 (0.08) & 0.89 (0.11) & 0.82 (0.07) \\
      \hline
    \end{tabular}
  }
  \label{tab:abe-fix-acc}
\end{table}

\end{document}